\documentclass[%
 amsmath,amssymb,
 twocolumn,
 aps,
]{revtex4-2}

\usepackage{xr-hyper} 
\usepackage{hyperref} 
\makeatletter

\usepackage{xfrac}  
\usepackage[compat=1.1.0]{tikz-feynman}
\usepackage{comment}
\usepackage{amsmath}
\usepackage{color}

\usepackage[english]{babel}

\usepackage{graphicx}
\usepackage{dcolumn}
\usepackage{bm}

\newcommand{\ket}[1]{\left|#1\right>}

\newcommand{\braket}[1]{\left<#1\right>}

\begin{document}


\title{Bounds on Phonon-mediated Hydrodynamic Transport in a Type-I Weyl Semimetal}
\author{Joan Bernabeu}%
\affiliation{Departamento de F\'isica de la Materia Condensada, Universidad Aut\'onoma de Madrid, Cantoblanco, E-28049 Madrid, Spain}
\email{joan.bernabeu@uam.es}

\author{Alberto Cortijo}
\affiliation{Departamento de F\'isica de la Materia Condensada, Universidad Aut\'onoma de Madrid, Cantoblanco, E-28049 Madrid, Spain}
\affiliation{Condensed Matter Physics Center (IFIMAC), Cantoblanco, E-28049 Madrid, Spain}
\email{alberto.cortijo@uam.es}
%





\date{\today}

\begin{abstract}
We analyze from a microscopic point of view the thermoelectric transport properties of a type-I Weyl semimetal driven by electron-electron interactions mediated by virtual phonons, particularly the feasibility of entering a hydrodynamic regime. Considering also the effects of of impurities and the absorption/emission of real phonons, electric and thermal conductivities are obtained. At temperatures $T$ above the Bloch-Gr\"uneisen temperature $T_\textrm{BG}$, virtual phonons behave similarly to real phonons, but the Lorenz ratio is modified by a constant prefactor dependent on the Fermi surface geometry. For temperatures below $T_\textrm{BG}$, virtual phonons induce a $T^2$ dependence in the resistivities, opening a window where momentum-conserving interactions could dominate transport signatures. We find that the onset of such a hydrodynamic regime requires impurity scattering whose inverse relaxation time is one or two orders of magnitude smaller than realistic values, depending on the electron-phonon coupling strength.

\end{abstract}

\maketitle

\section{Introduction}
Despite having been originally postulated decades ago \cite{Gurzhi63,Gurzhi68}, electron hydrodynamics has become of interest in recent years in light of experimental observations hinting such behavior. In contrast to the conventional diffusive dynamics of electrons in a Drude-like regime, the hydrodynamic regime in an electron system is characterized qualitatively by momentum-conserving electron-electron interactions providing the shortest scattering time. 

Recent experiments \cite{moll2016evidence, gooth2018thermal, jaoui2018departure, fu2020largely, vool2021imaging, osterhoudt2021evidence} have focused  in part on measuring the thermoelectric transport signatures within this hydrodynamic regime, particularly the electric conductivity $\sigma$ and the thermal conductivity $\kappa$, where in principle each one can be proportional to different relaxation times, $\tau_\sigma$ and $\tau_\kappa$ respectively. In systems where electric and thermal conductivities have the same characteristic scattering time,  the Lorenz ratio $L \equiv \kappa/(T\sigma)$ acquires a constant value $L_0$ independent of the temperature $T$ (and any particular characteristic scale in the system). This is known as the Wiedemann-Franz (WF) law and is the prevalent case for most metals. For systems where $\tau_\sigma$ and $\tau_\kappa$ are different, the Lorenz ratio $L \propto \tau_\kappa/\tau_\sigma$ no longer satisfies the WF law. Electronic systems in the hydrodynamic regime fall into the latter category as $\tau_{\sigma, \textrm{e-e}} \ne \tau_{\kappa, \textrm{e-e}}$ since the local electric current is relaxed only by processes that do not conserve the flow of the quasiparticles.

The violation the WF law observed in several materials, particularly some type-II Weyl semimetals (WSM) such as WP\textsubscript{2} \cite{gooth2018thermal,jaoui2018departure} and Co\textsubscript{2}MnAl \cite{robinson2021large}, has been suggested as a signature of such hydrodynamic regime, with phonon-mediated electron-electron interactions being proposed as the culprit for such behavior. This suggestion is supported by several studies from first-principles calculations \cite{Coulter18,vool2021imaging,osterhoudt2021evidence,PhysRevMaterials.5.L091202} in several type-II WSMs that appear to show regions where the smallest relaxation time is the one associated to phonon-mediated scattering. Anisotropy of the Fermi surface in these type-II WSMs has been suggested \cite{Coulter18} as the driving factor enhancing this mechanism. It is of note that in these works, Coulomb scattering  is seen to be the least effective scattering processes for the studied materials. 

Regarding the WF law, it is imperative to bear in mind that momentum-relaxing interactions between electrons and (real) phonons by themselves might lead to its violation without entering in a hydrodynamic regime. For instance, in three dimensional Fermi liquids where phonon absorption and emission by electrons are the dominant scattering processes at low temperatures, one expects $\tau_{\sigma, \textrm{e-e}} \propto T^{-5}$ whereas $\tau_{\kappa, \textrm{e-e}} \propto T^{-3}$, leading to a clearly nonconstant $L \propto T^2$, but no hydrodynamic regime, as phonon absorption and emission processes fail to conserve electron momentum \cite{Lavasani19}. Similar results can be expected in the bulk of a Weyl semimetal, with some deviations due to surface effects \cite{buccheri22}. Nevertheless, most of the theoretical approaches to the problem still consider absorption and emission of real phonons as the primary scattering mechanism for electrons, but with the phonon gas slightly out of equilibrium and thus conserving the total momentum of the electron-phonon fluid in absence of umklapp scattering \cite{levchenko2020transport, Huang21}. Experiments in WP$_2$ show however that the violation of the WF law is stronger than expected by electron-phonon scattering alone \cite{gooth2018thermal}, with the contribution of electron-electron scattering being noticeably different from that of other metals \cite{jaoui2018departure}.

In the present work we consider the role of electron-electron interactions in thermoelectric transport with virtual phonons playing the role of mediators instead of the conventional Coulomb interaction in type-I Weyl SMs. These are clearly different systems to the type-II SMs where potential signatures of hydrodynamics have been observed exclusively thus far. Evidence of hydrodynamics has also been reported in PdCoO$_2$, which is not a WSM \cite{moll2016evidence}). Nevertheless, there has been extensive theoretical work based on the premise that a particular instance of a type-I SM may indeed harbor (relativistic) hydrodynamic behavior \cite{son2009hydrodynamics,isachenkov2011chiral,lucas2016hydrodynamic,galitski2018dynamo,Gorbar18,gorbar2018hydrodynamic,sukhachov2018collective, Sukhachov22,Zhu22,matus2022skin}. It is thus of interest to ask whether a system where processes derived from electron-phonon interactions dominate over Coulomb scattering. In the process, we elucidate the general behavior of these scattering processes, whose microscopic treatment is seemingly absent in the literature, and how it entangles with the features of a type-I Weyl SM, finding their imprint on thermoelectric transport and in particual, the WF law.

Let us summarize our results. First of all, there is a characteristic temperature scale, the Bloch-Gr\"uneisen temperature, $T_\textrm{BG}$, delimiting two temperature regimes. Such temperature scale also appears in models that only consider phonon emission/absorption \cite{Lavasani19,Huang21}. In this work we assume that the Debye temperature $T_\textrm{D} > T_\textrm{BG}$, as is the case for metals with small Fermi momenta $p_f$. 

Above $T_\textrm{BG}$, virtual phonon effects are non-neglible when the phonon linewidth $\Gamma$ is dominated by the effects of the phonon-electron interaction. In this regime electronic scattering times are equal to those of phonon emission and absorption as the virtual phonons act like on-shell real phonons, with quantum interference effects becoming negligible, but the Lorenz ratio acquires a constant value different from $L_0$. When there are stronger decay channels for phonons than those provided by the interaction with electrons, then the effects of electron scattering in this temperature regime are suppressed and $L = L_0$.

For temperatures below $T_\textrm{BG}$, phonons no longer act on-shell as they are not thermally activated, and the quantum fermionic inteference between different scattering channels becomes non-neglible. For a simple quadratic band model, this would lead to the effects of electron scattering by virtual phonons being suppressed with respect to the phonon emission/absorption and scattering off impurities. Fundamentally, the wavefunction overlaps associated with the scattering of the Weyl quasiparticles make the contribution of phonon-mediated electron scattering non-negligible. This results in $\tau_\textrm{e-e} \propto T^{-2}$ relaxation times, opening up a parameter window for which electron scattering can dominate over scattering with real phonons for small enough temperatures. Nevertheless, we predict that the onset of a hydrodynamic regime is foiled for realistic electron-phonon coupling strengths and impurity scattering, the latter of which is expected to start dominating at higher temperatures of the order $T\sim 0.1 T_\textrm{BG}$.

\section{The model}
\subsection{Particles}
We consider the electron dynamics in a WSM described by the following two-band Hamiltonian \cite{burkov11,halasz12}
\begin{equation}\label{two_band_Hamiltonian}
    H_0 = \int_{\bm{p}}\psi_{\bm{p}}^{\dagger} [\bm{\sigma}\cdot\bm{d}(\bm{p})]\psi_{\bm{p}},
\end{equation}
where $\psi_{\bm{p}}$ is a two-component spinor, and $\int_{\bm{p}} \equiv \int \frac{d^3p}{(2\pi)^3}$. A simple diagonalization of the Hamiltonian shows that there are positive and negative energy eigenstates whose energies are given by $\epsilon_{\pm, \bm{p}} = \pm d(\bm{p}) \equiv \sqrt{|\bm{d}(\bm{p})|^2}$. In particular, we set
\begin{equation}\label{weyl_semimetal_d}
    \bm{d}(\bm{p}) = v_0\begin{pmatrix}
        p_x, & p_y, &\frac{1}{2b}\left(b^2-p_z^2\right)
    \end{pmatrix},
\end{equation}
so that the Hamiltonian has two isotropic Weyl nodes (valleys) around $\bm{b}_\pm = (0,0,\pm b)$. For small momenta around these nodes, $\delta \bm{p}_\pm \equiv \bm{p} - \bm{b}_\pm $, one obtains the low energy Hamiltonians, one for each chirality ($\pm\equiv L/R$),
\begin{equation}\label{Weyl_Hamiltonian}
    \mathcal{H}_{L,R}(\delta \bm{p}_\pm) = \delta p_{\pm,x}\sigma_x + \delta p_{\pm,y}\sigma_y \pm \delta p_{\pm,z} \sigma_z,
\end{equation}
where we have set the Fermi velocity $v_0 \equiv 1$, as is also done throughout the rest of this work. The nodes are thus separated by a distance $2b$. Importantly, we assume that the chemical potential is the same for both valleys, and much larger than the temperature $T$. Without loss of generality, we assume that the chemical potential is positive, meaning that the occupied valence band has no impact in the transport properties and can be neglected. In addition we also assume that $p_f < b \equiv|\bm{b}|$ as well in order to have a well defined chirality for each valley.

Longitudinal acoustic phonons have on-shell isotropic energies $\omega_{\bm{q}} =c|\bm{q}|\equiv c q$, where $c$ is the (dimensionless) speed of sound measured in units of $v_0$. For realistic WSM it makes sense to assume $c \sim 10^{-2} \ll 1$ \cite{peng2016high, Huang21, buccheri22,cheng2023first}. In the present work phonons interact with electrons through a deformation potential term \cite{Lavasani19} (neglecting umklapp processes):
\begin{equation}\label{deformation_potential}
    H_{\textrm{e-ph}} = \int_{\bm{p},\bm{q}}g_{\bm{p}+\bm{q},\bm{p}}^{\bm{q}}\left(a_{\bm{q}}  + a_{\bm{q}}^{\dagger}\right)\psi^{\dagger}_{\bm{p}+\bm{q}}\psi_{\bm{p}}.
\end{equation}
The operator $a_{\bm{q}}$ ($a_{\bm{q}}^{\dagger}$) destroy (create) a phonon with momentum $\bm{q}$, and the electron-phonon coupling is given by
\begin{equation}\label{electron_phonon_coupling}
    g^{\bm{q}}_{\bm{p}',\bm{p}} = \sqrt{\frac{U}{2\omega_{\bm{q}}}}\left(\bm{p}'-\bm{p}\right)\cdot\hat{\bm{q}},
\end{equation}
where $U$ is some constant regulating the strength of the electron-phonon interaction that has dimension of $\textrm{energy}^{-2}$ after setting $\hbar = v_0 = 1$ and $\hat{\bm{q}}=\bm{q}/|\bm{q}|$.

When projecting the Hamiltonian term (\ref{deformation_potential}) onto the conduction band states of (\ref{weyl_semimetal_d}), the effective coupling  (\ref{electron_phonon_coupling}) reads
\begin{equation}\label{effective_electron_phonon_coupling}
    g_{\bm{p}',\bm{p}}^{\textrm{eff}\ \bm{q}} = g_{\bm{p}',\bm{p}}^{\bm{q}}\braket{+\bm{p}'|+\bm{p}},
\end{equation}
where the state $\ket{+\bm{p}}$ is the positive energy eigenstate of the Hamiltonian (see Appendix section \ref{appendix_hamiltonian_and_scattering_amplitudes}). Note that because of overlap term $\braket{+\bm{p}'|+\bm{p}}$ the effective fermion-phonon coupling is now sensitive to chirality flips in the scattering processes involving different valleys, as can be checked by expressing $\bm{p}'$ and $\bm{p}$ in terms of $\delta\bm{p}_{\pm'}$ and $\delta\bm{p}_{\pm}$. In the rest of this work, the label $\bm{p}$ will be used instead of $\delta \bm{p}_\pm$ to refer to the momenta with respect to the Weyl nodes. For chirality-violating processes where confusion could arise, the momentum around the right-handed node will simply be altered as $\bm{p} \rightarrow \bm{p} + 2\bm{b}$, setting it to the reference frame of the left-handed node. 

\begin{figure}
    \includegraphics[width = 0.45\textwidth]{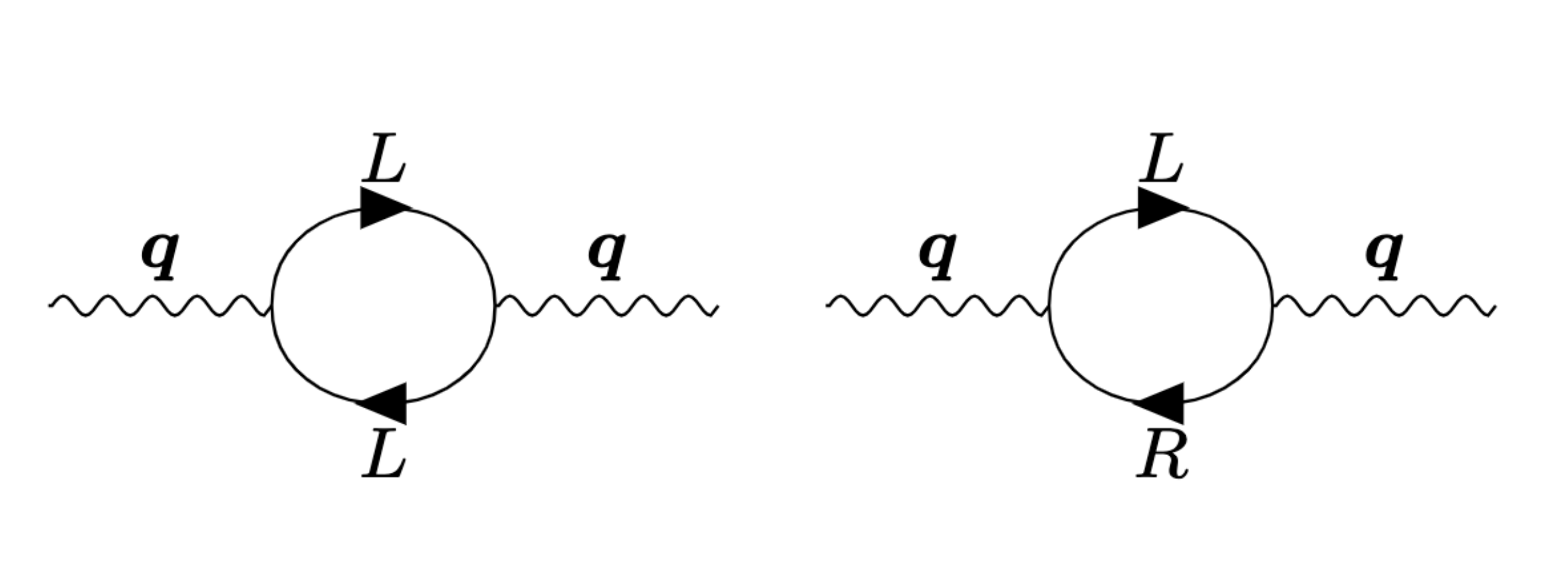}
    \caption{Intravalley and intervalley contributions to the phonon linewidth from the electron-phonon interaction. The bubble diagram with two $R$ fermions provides the same contribution as the the first diagram.}
    \label{fig:linewidth_diagrams}
\end{figure}

The phonon propagator is given by
\begin{equation}\label{phonon_propagator}
    \mathcal{D}(\bm{q},\omega)   =  \frac{2\omega_{\bm{q}}}{\omega^2 - \omega_{\bm{q}}^2 +i\omega_{\bm{q}}\Gamma_{\bm{q}}}.
\end{equation}
We have included the effects of interactions through a small but non-vanishing phonon linewidth $\Gamma_{\bm{q}} \ll \omega_{\bm{q}}$. Here we focus on the electron-phonon interaction (\ref{deformation_potential}) as the dominant source of this phonon linewidth,  represented to lowest order in perturbation theory by the diagrams in Fig.(\ref{fig:linewidth_diagrams}), i.e., $\Gamma_{\bm{q}}=\Gamma^{\textrm{e-ph}}_{\bm{q}}$. In general, other effects like phonon-impurity scattering or anharmonic effects might enhance $\Gamma_{\bm{q}}$, so we expect for realistic WSMs, 
\begin{equation}\label{phonon_linewidth_bound}
    \Gamma_{\bm{q}} \ge \Gamma^{\textrm{e-ph}}_{\bm{q}}.
\end{equation}
In particular, one expects for anharmonic interactions to dominate at sufficiently large temperatures. Conversely, for small enough temperatures, particularly in a region close to $T_\textrm{BG}$, anharmonic effects may be suppressed with respect to the absorption/emission of phonons. Hence our assumption of $\Gamma_{\bm{q}}=\Gamma^{\textrm{e-ph}}_{\bm{q}}$ may be sensible below and also for a limited temperature range above $T_\textrm{BG}$.

\twocolumngrid
%
%
%
%
%


\twocolumngrid

\subsection{Boltzmann Equation}

The evolution of the distribution function $f_\lambda(\bm{p},\bm{x}$) for the fermionic quasiparticles with chirality $\lambda = L,R$ is governed by the quantum Boltzmann equation
\begin{equation}\label{boltzmann}
    \partial_t f_{\lambda} + \bm{F}\cdot\nabla_{\bm{p}}f_{\lambda} + \bm{v}_{\lambda}\cdot\nabla_{\bm{x}}f_{\lambda} = -\mathcal{I}[f_\lambda].
\end{equation}
where $\bm{F}$ is the external force acting on the particles and $\bm{v}_{\lambda} \equiv \nabla_{\bm{p}}\epsilon_{\bm{p}} = \hat{\bm{p}}$ is the group velocity for linearly dispersive WSMs. Note that the $\lambda$ dependence of the group velocity is implicitly encoded in the reference point for the momentum $\bm{p}$.
We point out that we are not considering the contribution from the Berry curvature to the velocity, that might lead to anomalous Hall currents, as we are interested in transport responses linear in external applied fields \cite{Stephanov12,Son12,Gorbar18,Sukhachov22}. The collision integral (right hand side of (\ref{boltzmann})) contains terms arising from the scattering of electrons by impurities, absortion/emission of real phonons and electron-electron interactions:
\begin{equation}\label{collision_integral_contributions}
    \mathcal{I}[f_\lambda] \equiv \mathcal{I}_\textrm{imp}[f_\lambda] + \mathcal{I}_\textrm{e-ph}[f_\lambda] + \mathcal{I}_\textrm{e-e}[f_\lambda].
\end{equation}
Electron-electron interactions are the only processes that conserve electron momentum. We do not consider here screened Coulomb interactions, and assume that virtual phonon exchange is the only source of electron-electron interactions.

We also assume that phonons are at thermal equilibrium, following Bose-Einstein distribution,
\begin{equation}
    b_0(\omega) = \frac{1}{e^{\beta\omega} - 1}.
\end{equation}
The electric ($\bm{J}_{\textrm{E}}$) and heat ($\bm{J}_{\textrm{T}}$) currents can be obtained from the solution $f_\lambda$ of (\ref{boltzmann}):
\begin{equation}\label{currents}
    \bm{J}_{\textrm{E,T}} \equiv \sum_{\lambda}\int_{\bm{p}} q_\textrm{E,T}^{\lambda} \bm{v}_{\lambda} f_\lambda,
\end{equation}
where $q_\textrm{E}^\lambda \equiv q_\textrm{E} \equiv e$ and $q_\textrm{T}^\lambda \equiv q_\textrm{T} \equiv (p-p_f)$ are the chirality-independent charges associated to each current, and $p\equiv|\bm{p}|$. \\

In thermal equilibrium, fermions follow the Fermi-Dirac distribution
\begin{equation}\label{Fermi-Dirac}
    f_0\equiv f_{0}(p)= \frac{1}{e^{\beta(p-p_f)} + 1},
\end{equation}
for which the collision integral (\ref{collision_integral_contributions}) and the currents (\ref{currents}) identically vanish. We thus analyze the non-equilibrium perturbative response to small electric fields and thermal gradients. As usual, the nonlinear Boltzmann equation (\ref{boltzmann}) can be linearized by writing
$f_\lambda = f_{0} + \delta f_\lambda$ and keeping only the terms $\delta f_\lambda$ that depend linearly with the external fields. Eq.(\ref{boltzmann}) then becomes 
\begin{equation}\label{linearized_boltzmann_equation}
    \hat{\bm{p}}\cdot\left[e\bm{\mathcal{E}} - \beta\left(p - p_F\right)\bm{\nabla} T\right]\beta f_0[1-f_0]= - \mathcal{I}[f_{0} + \delta f_\lambda],
\end{equation}
where $\bm{\mathcal{E}} \equiv \bm{E} + \frac{1}{e}\bm{\nabla} \mu$ is the electrochemical force and $\bm{\nabla}T$ the temperature gradient. We also assume that the system size is much larger than any other scale in the problem. In absence of external magnetic fields that trigger anomaly-related phenomena, we expect that $\delta f_L=\delta f_R\equiv \delta f$ in Eq.(\ref{linearized_boltzmann_equation}) so in practice one only needs to study the Boltzmann equation for a single chirality, even when chirality-flipping processes are taken into account. Due to the form of the LHS of Eq.(\ref{linearized_boltzmann_equation}), it is convenient to parametrize $\delta f$ as
\begin{eqnarray}
     \delta f = \beta^2 f_{0}[1-f_{0}][\bm{\mathcal{X}}_\textrm{E}(\bm{p})\cdot e\bm{\mathcal{E}}-\bm{\mathcal{X}}_\textrm{T}(\bm{p})\cdot\bm{\nabla} T],
\end{eqnarray}
where $\bm{\mathcal{X}}_\textrm{E,T}$ are dimensionless vector perturbations. Substituting this expression into (\ref{linearized_boltzmann_equation}) and separating the electric and thermal contributions, one formally obtains a vector equation
\begin{equation}\label{linearized_vector_boltzmann_equation}
 \bm{\mathcal{S}}_\textrm{E,T}=  \mathcal{C}\bm{\mathcal{X}}_\textrm{E,T},
\end{equation}
where the source term is defined as
$
    \bm{\mathcal{S}}_\textrm{E,T} \equiv \hat{q}_\textrm{E,T}\hat{\bm{p}}.
$
The quantities $\hat{q}_\textrm{E} \equiv 1$ and $\hat{q}_\textrm{T} \equiv \beta(p-p_f)$ are the dimensionless charges for the sources, and  the operator $\mathcal{C}$ on the RHS is the vectorized version of the collision integral acting on the vector functions $\bm{\mathcal{X}}_{\textrm{E,T}}$.
Although isotropy is technically broken in our model through the internodal vector $\bm{b}$, we assume that the solution is isotropic and hence postulate that $\bm{\mathcal{X}}_{\textrm{E,T}}(\bm{p}) = \mathcal{X}_{\textrm{E,T}}(\epsilon_{\bm{p}})\bm{v} = \mathcal{X}_{\textrm{E,T}}(p)\hat{\bm{p}}$.
For isotropic systems, the currents (\ref{currents}) are related to the external fields through the Onsager transport coefficients
\begin{equation}
    \begin{pmatrix}
        \bm{J}_\textrm{E} \\
        \bm{J}_\textrm{T}
    \end{pmatrix}
    =
    \begin{pmatrix}
        L_\textrm{EE} & L_\textrm{ET} \\
        L_\textrm{TE} & L_\textrm{TT}
    \end{pmatrix}
    \begin{pmatrix}
        e\bm{\mathcal{E}} \\
        \bm{\nabla} T
    \end{pmatrix},
\end{equation}
which for our model read
\begin{equation}
    L_\textrm{AB} =\frac{4\beta^2}{3(2\pi)^2}\int dp\ p^2 f_0[1-f_0]q_\textrm{A}(p) \mathcal{X}_\textrm{B}(p).\\
\end{equation}
Note that an extra factor of $2$ has been introduced to account for the effect of both valleys/chiralities. The electric conductivity as well as the open-circuit and closed-circuit thermal conductivities are respectively defined as
\begin{eqnarray}
    \nonumber
    \sigma \equiv e^2 L_\textrm{EE}, \quad \bar{\kappa} \equiv L_\textrm{TT},  \\
    \label{conductivities}
    \kappa \equiv L_\textrm{TT} - \frac{L_\textrm{TE}L_\textrm{ET}}{L_\textrm{EE}}.
\end{eqnarray}
The thermopower is given by $\alpha \equiv e L_\textrm{TE} = e\beta^{-1}L_\textrm{ET}$, and the latter equality is due to Onsager reciprocity\cite{onsager31}.

\subsection{Variational Approach}
To solve Eqs.(\ref{linearized_vector_boltzmann_equation}), we make use of the standard variational approach \cite{Ziman:100360, arnold2000transport, PhysRevB.78.085416}. Introducing the inner product
\begin{equation}\label{inner_product}
    (\bm{\mathcal{F}},\bm{\mathcal{G}}) \equiv \int \frac{d^3p}{(2\pi)^3}\beta f_0(p)[1-f_0(p)]\bm{\mathcal{F}}(\bm{p})\cdot\bm{\mathcal{G}}(\bm{p}),
\end{equation}

the following functional is defined:
\begin{equation}\label{action_functional}
    \mathcal{Q}_\textrm{E,T}[\bm{\chi}_\textrm{E,T}] \equiv (\bm{\mathcal{X}}_\textrm{E,T}, \bm{\mathcal{S}}_\textrm{E,T}) - \frac{1}{2}(\bm{\mathcal{X}}_\textrm{E,T}, \bm{\mathcal{C}}\bm{\mathcal{X}}_\textrm{E,T}).
\end{equation}
Eqs.(\ref{linearized_vector_boltzmann_equation}) are then obtained by applying the variational principle on this functional $\delta\mathcal{Q}_{E,T}/\delta\bm{\mathcal{X}}_\textrm{E,T} = 0$. The issue of solving the Boltzmann equation (\ref{linearized_boltzmann_equation}) can hence be converted into an $N$-dimensional algebraic problem by expanding $\bm{\mathcal{X}}_\textrm{E,T}$ into an arbitrary functional basis $\left\{ F^n\right\}_{n=0}^{N-1}$, i.e., $\mathcal{X}_\textrm{E,T}(p) = \bm{x}_\textrm{E,T}\cdot \bm{F}(p)$.
Now the Boltzmann equation (\ref{linearized_vector_boltzmann_equation}) becomes a linear equation for the vector of unknown coefficients $\bm{x}_\textrm{E,T}$,
\begin{equation}\label{functional_boltzmann_equation}
    \bm{S}_\textrm{E,T} = \bm{C}\ \bm{x}_\textrm{E,T},
\end{equation}
where the components of the source vector are $S^n_\textrm{E,T} \equiv (F^n\hat{\bm{p}}, \bm{\mathcal{S}}_\textrm{E,T} )$ and the collision matrix components are given by $C^{nm} \equiv (F^n\hat{\bm{p}}, \mathcal{C}[F^m\hat{\bm{p}}])$. The matrix $\mathbf{C}$ is not to be confused with the eletron-phonon coupling $C$ (\ref{electron_phonon_coupling}). If $\bm{C}$ is an invertible operator, Eq.(\ref{functional_boltzmann_equation}) has a unique solution and the transport coefficients will read $L_{AB} \propto \bm{S}_\textrm{A}\bm{C}^{-1}\bm{S}_\textrm{B}$. For the conductivities it implies
\begin{eqnarray}
    \nonumber
    & \sigma = \frac{2e^2\beta}{3}\bm{S}_\textrm{E}^\dagger\bm{C}^{-1}\bm{S}_\textrm{E}, \quad \bar{\kappa} = \frac{2}{3}\bm{S}_\textrm{T}^\dagger\bm{C}^{-1}\bm{S}_\textrm{T}, \\
    \label{conductivities_matrix_form}
    & \alpha = \frac{2e\beta}{3}\bm{S}_\textrm{E}^\dagger\bm{C}^{-1}\bm{S}_\textrm{T} =  \frac{2e\beta}{3}\bm{S}_\textrm{T}^\dagger\bm{C}^{-1}\bm{S}_\textrm{E}.
\end{eqnarray}
Again, the latter equality in (\ref{conductivities_matrix_form}) is just the well-known Onsager reciprocity. In the case at hand, it easily proven using (\ref{functional_boltzmann_equation}) and the hermiticity of $\bm{C}$.

The expressions in (\ref{conductivities_matrix_form}) are meaningful if the collision matrix is invertible as it happens when momentum-relaxing processes are considered. However, in the presence of exclusively momentum-conserving processes (that also conserve chirality, as we will see later) $\bm{C}$ would have a zero-mode associated with momentum conservation and given by the function $F^0(p) = \beta p$. In such a scenario the conductivities (\ref{conductivities_matrix_form}) would be infinite with the electric field/thermal gradient accelerating electrons indefinitely, preventing any steady state. This is indeed the case for the electric conductivity $\sigma$, where a net movement of the electron fluid is needed to carry charge between electrodes, but not necessarily for the thermal conductivity $\kappa$, measured in short-circuit conditions. In this short-circuit scenario, heat can be transferred among electronic degrees of freedom without the need of a net fluid movement, so it is expected to be finite even for a purely momentum-conserving system.

Since $\bm{C}_\textrm{e-e}$ is hermitian, it can be diagonalized through an orthogonal basis $\left\{ F^n\right\}_{n=0}^{N-1}$, where $F^0(p) = \beta p$ is the zero-mode described above. If a small momentum-relaxing perturbation to the total collision matrix of the form $\tau_\textrm{mr}^{-1}\bm{1}$ is introduced so that $\bm{C} = \tau_\textrm{mr}^{-1}\bm{1} + \bm{C}_\textrm{e-e}$, then its inverse is given by
\begin{equation}
    \bm{C}^{-1} =
    \begin{pmatrix}
        \tau_\textrm{mr} & 0 & \dots \\
        0 & (\tau_\textrm{mr}^{-1} + \tau_1^{-1})^{-1} & \dots \\
        \vdots & \vdots & \ddots
    \end{pmatrix},
\end{equation}
where $\tau_n^{-1}$ is the eigenvalue corresponding to the basis function $F^n$. Plugging this into the equation for the electric conductivity (\ref{conductivities_matrix_form}) and taking the momentum conserving limit $\tau_\textrm{mr} \gg \tau_n$ (which is relevant for studying hydrodynamic effects) one obtains 
\begin{equation}\label{electric_conductivity_limit}
    \sigma \propto  \tau_\textrm{mr}(S^0_\textrm{E})^2,
\end{equation}
which in the $\tau_\textrm{mr} \rightarrow \infty$ limit is infinite, as expected. In our case where $p_f \gg T$, one can go further by noticing that at leading order, the zero-mode in the functional basis is related to the source term $\bm{p} \approx p_f \bm{S}_\textrm{E}$. Since the zero-mode $F^0(p)$ is orthogonal to all other basis vectors, $0= P^n \approx S^n_\textrm{E}$ for all $n > 0$. Plugging this into Eq.(\ref{conductivities}), the expression in Eq.(\ref{electric_conductivity_limit}) is automatically recovered, independently of the hierarchy of $\tau_\textrm{mr}$ with respect to $\tau_{n}$.

In the case of $\kappa$, one can plug in Eq.(\ref{conductivities_matrix_form}) into the definition given in Eq.(\ref{conductivities}) and find that $\kappa \propto \sigma^{-1} \textrm{Tr}\left[\left(\bm{\Sigma} \bm{C}^{-1}\right)^2\right]$, where $\Sigma^{ij} \equiv S^i_\textrm{E}S^j_\textrm{T}-S^i_\textrm{T}S^j_\textrm{E}$. Since $\bm{\Sigma}$ is antisymmetric, the only possibly non-convergent term  in the large $\tau_\textrm{mr}$ limit, which is proportional $ [C^{-1, 00}]^2 = \tau_\textrm{mr}^2$, vanishes. In fact, only terms in the trace proportional to $\tau_\textrm{mr}$ survive in the momentum-conserving limit since the electrical conductivity $\sigma \propto\tau_\textrm{mr}$ in the denominator renders their contribution finite. The resulting expression for $\kappa$ in the momentum-conserving limit is
\begin{equation}
    \kappa \propto \sum_{n=1}^{N-1}\tau_n\left(S_\textrm{T}^n-\frac{S_\textrm{T}^0}{S_\textrm{E}^0}S_\textrm{E}^n\right)^2.
\end{equation}
In our model, $\bm{S}_\textrm{T} = \bm{p} - p_f\bm{S}_\textrm{E}$, where $\bm{p}$ is the zero-mode. Using again that $P^n = 0$ for $n>0$, the thermal conductivity can be expressed as
\begin{eqnarray}\label{thermal_conductivity_matrix}
    \kappa =\frac{2}{3}\left(\frac{P^0}{S^0_\textrm{E}}\right)^2 \bar{\bm{S}}_\textrm{E}^\dagger\bar{\bm{C}}^{-1}\bar{\bm{S}}_\textrm{E},
\end{eqnarray}
where the $\bar{\bm{S}}_\textrm{E}$ and $\bar{\bm{C}}$ are the respective projections of $\bm{S}_\textrm{E}$ and $\bm{C}$ into the subspace orthogonal to the zero mode, where $\bar{\bm{C}}$ is invertible. Eq.(\ref{thermal_conductivity_matrix}) is independent of the basis chosen for the subspace orthogonal to the zero mode. Physically speaking, momentum-conserving effects are projected out as they are irrelevant for the short-circuited system where $\kappa$ is measured.
The expression for the thermal conductivity (\ref{thermal_conductivity_matrix}) is valid for any system described by a Dirac Hamiltonian such as graphene. A similarly simple expression can be derived for systems with a quadratic dispersion relation, where $\bm{S}_\textrm{E} \propto \bm{p}$. \\

The discussion in the preceding paragraphs is generic and qualitative. To derive quantitative results for the thermoelectric conductivities, it is necessary to adopt a concrete set of basis functions. For our purposes, it is enough to choose $\left\{F^n\right\}_{n=0}^1$ with
\begin{equation}\label{basis}
    F^0(p) = \beta p, \quad F^1(p) = 1.
\end{equation}
As just argued, $F^0(p)$ is useful to distinguish momentum-relaxing and momentum-conserving effects in the momentum-conserving limit. In our model, we consider just one further basis function $F^1$, from which the perpendicular subspace to $F^0$ needed to calculate $\kappa$ (\ref{thermal_conductivity_matrix}) can be extracted trough the Gram-Schmidt method. In the Fermi liquid regime $p_f \gg T$, other functions are not needed, as they only contribute with subleading terms in the Sommerfeld expansion with the small parameter $(\beta p_f)^{-1}$. This implies that the result (\ref{thermal_conductivity_matrix}) is in fact exact at leading order in $T/p_f$ using only the basis (\ref{basis}). \\


\subsection{Impurities and Thermal Phonons}

Returning to our model, electrons can interact with impurities in the material, real phonons through emission and absorption, and other electrons through virtual phonons (\ref{collision_integral_contributions}). Impurities are considered though a relaxation time approximation (RTA),
\begin{equation}\label{impurities}
    \mathcal{I}_\textrm{imp}[f] \equiv -\frac{\delta f}{\tau_\textrm{imp}},
\end{equation}
i.e., the strength of impurity scattering is accounted for by the parameter $\tau_\textrm{imp}$. In the matrix form for the Boltzmann equation given in (\ref{functional_boltzmann_equation}), this is equivalent to $\bm{C}_\textrm{imp} = \tau_\textrm{imp}^{-1} \bm{1}$. This shows that impurities relax all modes equally, implying that, for impurity-collision dominated systems, $\tau_\sigma = \tau_\kappa$ and non-violation of the WF law.

On the other hand, phonon emission and absorption by electrons is a well-known mechanism \cite{Ziman:100360, Lavasani19} leading to $\tau_\sigma \ne \tau_\kappa$ and an explicit violation of the WF law at small temperatures. At large temperatures however, the WF law is restored. Details on the derivation of these conclusions are given in \cite{SupplementaryMaterial}. We also show that, at temperatures above the Bloch-Gr\"uneisen temperature $T_\textrm{BG}$,
\begin{equation}\label{e-ph_relaxation_high_T}
    \tau_{\sigma/\kappa,\textrm{e-ph}}^{-1} =  \beta^{-1} \frac{2(Up_f^2)}{3(2\pi) c^2}\cdot \begin{cases}
        \frac{5}{2} & \textrm{if $T \gg T_\textrm{BG,W}$}, \\
        1 & \textrm{if $T \ll T_\textrm{BG,W}$},
    \end{cases}
\end{equation}
where the difference in temperature regimes comes from the fact that for $T \ll T_\textrm{BG,W} \equiv 2cb$, chirality-breaking phonon absorption/emission is suppressed as real/thermal phonons do not have enough energy to make the jump between valleys Here we have defined a second Bloch-Grüneisen temperature $T_\textrm{BG,W}$ associated to the distance between the two fermion valleys \cite{buccheri22}. At temperatures $T \ll T_\textrm{BG}$ chirality-breaking processes are of course still suppressed but chirality-conserving ones lead to
\begin{eqnarray}\label{e-ph_relaxation_low_T}
    \tau_{\sigma, \textrm{e-ph}}^{-1} = \beta^{-5}\frac{12g_5U}{(2\pi)c^6p_f^2} = \frac{(\beta p_f)^{-2}}{c_2 c^2}\cdot \tau_{\kappa, \textrm{e-ph}}^{-1}.
\end{eqnarray}
and the WF law is violated when these relaxation times dominate, $L/L_0 = \tau_\kappa/\tau_\sigma \ne 1$. In (\ref{e-ph_relaxation_low_T}), $c_2$ and $g_5$ are numerical constants of order unity, see Appendix section \ref{appendix_numeric_integrals}.

\section{Electron-Electron Collisions}

In this section we consider the effects of normal (momentum-conserving) electron-electron interactions mediated by virtual phonons in transport, the primary focus of this work. We first study the collision integral for intravalley processes $L,L \rightarrow L,L$, and then extend it to the full system which also includes the $R$ node. For intravalley processes the collision integral in the action functional (\ref{action_functional}) is given by
\begin{widetext}
\begin{eqnarray}\nonumber
    &(\bm{\mathcal{X}},\mathcal{C}_\textrm{e-e}\bm{\mathcal{X}})^\textrm{intra} = \frac{1}{2}\cdot\frac{1}{4}\beta^2\int_{\bm{p}}\int_{\bm{k}}\int_{\bm{p'}}\int_{\bm{k'}} |M_{LL}^{LL}(\bm{p}, \bm{k}; \bm{p}', \bm{k}')|^2f_0(p)f_0(k)[1-f_0(p')][1-f_0(k')] \\
    &\label{starting_collision_integral}
    (2\pi)^4\delta(p+k-p'-k')\delta^{(3)}(\bm{p}+\bm{k}-\bm{p}'-\bm{k}')|\Delta \mathcal{X}|^2,
\end{eqnarray}
\end{widetext}
\twocolumngrid
%
%
%
\twocolumngrid

\begin{figure}
    \includegraphics[width = 0.48\textwidth]{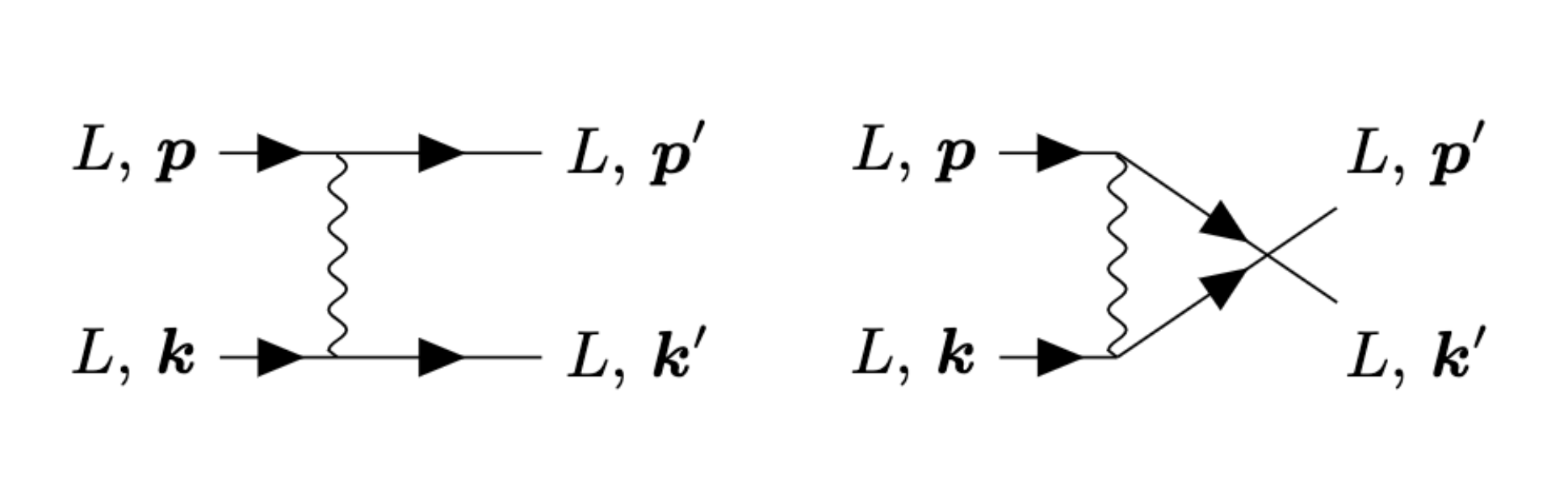}
    \caption{Intranode Feynman diagrams, with the $t$-channel corresponding to $A_t$ (left) and the $u$-channel corresponding to $A_u$ (right).}
    \label{fig:LLLL_diagrams}
\end{figure}
where $|\Delta \mathcal{X}|^2 \equiv [\mathcal{X}(p)\hat{\bm{p}} + \mathcal{X}(k)\hat{\bm{k}}  - \mathcal{X}(p')\hat{\bm{p}}'  - \mathcal{X}(k')\hat{\bm{k}}' ]^2$ and $|M_{LL}^{LL}(\bm{p}, \bm{k}; \bm{p}', \bm{k}')|^2$ is the square of the scattering amplitude for two  left-handed fermions with momenta $\bm{p}$ and $\bm{k}$ to scatter to two final electron states with momenta $\bm{p}'$ and $\bm{k}'$ of the same chirality. To see how the action functional collision integral (\ref{starting_collision_integral}) is derived from the collision integral $\mathcal{I}_\textrm{e-e}[f_\lambda]$ appearing in the Boltzmann equation (\ref{collision_integral_contributions}), see section B of \cite{SupplementaryMaterial}. The scattering can occur on two different channels, depending on how the initial states are connected to the final states, as represented via Feynman diagrams in Fig.(\ref{fig:LLLL_diagrams}). We call the channel where $\bm{p}$ is connected to $\bm{p}'$ the $t$-channel, and the one where $\bm{p}$ is connected to $\bm{k}'$ the $u$-channel, following the convention in Quantum Electrodynamics. The scattering amplitude for the $t$-channel is given by
\begin{equation}\label{scattering_amplitude_ee}
A_t = -C\frac{q^2}{\omega^2 - \omega_{\bm{q}}^2 +i\omega_{\bm{q}}\Gamma_{\bm{q}}}\braket{+\bm{p}'|+\bm{p}}\braket{+\bm{k}'|+\bm{k}}. 
\end{equation}
Note that since we are considering left chiral states, technically $\ket{+\bm{p}}$ represents $\ket{+,\bm{-b} + \delta\bm{p}}$. Here $\bm{q} \equiv \bm{p}' - \bm{p} = \bm{k} - \bm{k}'$ is the virtual phonon momentum and $\omega \equiv p' - p = k - k'$ is its energy. The $u$-channel amplitude is obtained by interchanging $\bm{p}' \leftrightarrow \bm{k}'$. The total scattering amplitude is then given by $M_{LL}^{LL} \equiv A_t - A_u$, where the minus sign is due to the fermion exchange from the $t$ to the $u$ diagrams. After some manipulations, one finds that
\begin{widetext}
$$
(\bm{\mathcal{X}},\mathcal{C}_\textrm{e-e}\bm{\mathcal{X}})^\textrm{intra} = \frac{\beta^2}{4(2\pi)^6}\int_0^{q_D} dq \int_{-q}^q d\omega \int_{\frac{q-\omega}{2}}^\infty dp\ pp'\int_{\frac{q+\omega}{2}}^\infty dk\ kk'\int_0^{2\pi}d\varphi_{q,pk}
$$
\begin{equation}\label{intermediate_collision_integral}
    |M_{LL}^{LL}|^2(\beta\omega)^2b_0(\omega)[1+b_0(\omega)]f_0(p)[1-f_0(p)]f_0(k)[1-f_0(k)]|\Delta \mathcal{X}|^2,
\end{equation}
\end{widetext}
where now $p'$ and $k'$ should be understood as shorthands for $p' \equiv p + \omega$ and $k' \equiv k - \omega$. A Debye cutoff $q_D$ to the phonon momentum has been introduced, although it will not have any effect due to the assumption that $T_\textrm{BG} < T_\textrm{D}$, which effectively restricts $q$ to values under $2p_f < q_D$. To proceed further analytically, it is necessary to study this integral in asymptotic high and low temperature regimes, $T\gg T_\textrm{BG}$ and $T\ll T_\textrm{BG}$ respectively.

\subsection{High Temperatures}\label{high_T_intraband_section}
The full square of the scattering amplitude equates to $|M_{LL}^{LL}|^2 = |A_t|^2 + |A_u|^2 - 2\textrm{Re}\left[A_t^*A_u\right]$. The $|A_t|^2$ and $|A_u|^2$ give identical contributions to the collision integral as can be seen by relabeling $\bm{p}' \leftrightarrow \bm{k}'$ in the original integral (\ref{starting_collision_integral}). These induce term  of the form
\begin{equation}\label{omegaintegral}
    I_\omega \equiv \int_{-1}^1 \frac{f(x)}{[(x - c)^2 + \frac{\Gamma^2_q}{4q^2}][(x + c)^2 + \frac{\Gamma^2_q}{4q^2}]}dx,
\end{equation}
inside the $q$ integral, where $x \equiv \omega/q$ is a dimensionless form of the virtual phonon energy and $f(x)$ is some function. The residue theorem can be used to perform the integral, with the enclosing curve being far enough from the poles appearing in the denominator of (\ref{omegaintegral}). If $f(x)$ contains no further poles within this contour, then expanding $f(x)$ in a Taylor series around $x = 0$ and performing the integral evaluates to
\begin{equation}\label{omegaintegral_result}
    I_\omega = \frac{\pi q}{\Gamma_q}\left[\frac{1}{c^2}f(0)+\frac{1}{2}f''(0)\right] + O(1),
\end{equation}
where the $O(1)$ refers to terms that are proportional to a power of $0$ or more of the small parameters $c$ and $\Gamma_q/(2q)$. A similar procedure can be carried out for the interference term that depends on $2\textrm{Re}[A_t^*A_u]$ (see section B1a of \cite{SupplementaryMaterial}). However, in that case, the results are of a higher-order in the expansion parameters than the non-interference contribution (\ref{omegaintegral_result}), and can thus be neglected.

The next-to leading order term proportional to $f''(0)$ is considered because the contribution from the leading term proportional to $f(0)$ is actually vanishing at the leading order in the Sommerfeld expansion, where $p = k = p' = k' = p_f$ and the $|\Delta\mathcal{X}|^2$ term in (\ref{starting_collision_integral}) is easily seen to vanish because of momentum conservation. This is not the case for the $f''(0)$ term, which is non-vanishing at the leading order in the Sommerfeld expansion. This implies that the ratio between the contributions of $f(0)$ and $f''(0)$ is of order $ (T/cp_f)^2\sim (T/T_\textrm{BG})^2$. Therefore, for temperatures above the Bloch-Gr\"uneisen temperature, the $f(0)$ term dominates over the one proportional to $f''(0)$ so, in practice, only the $f(0)$ term is relevant in this subsection.

The assumption that $f(x)$ has no other poles inside the contour only applies for $T \gg T_\textrm{BG}$ as the Fermi-Dirac distribution functions contained in $f(x)$ also have imaginary poles at the Matsubara frequencies, being the lowest of order $\beta^{-1} = T$. Therefore, when $ T \gg T_\textrm{BG}$, we can integrate along contours not containing Matsubara poles. This is not the case if $T \lesssim T_\textrm{BG}$, to be considered in the next subsection. The relevance of the Bloch-Grüneisen temperature will become clear shortly.

Assuming that the dominant contribution to the phonon linewidth comes from electron-phonon interactions, that is, Eq.(\ref{phonon_linewidth_bound}) is an (approximate) equality, one finds that
\begin{equation}\label{ee_intraband_high_T_result}
    (\bm{\mathcal{X}},\mathcal{C}_\textrm{e-e}\bm{\mathcal{X}})^\textrm{intra} =  \frac{4c_2(Up_f^2)\beta^{-2}}{3(2\pi)^3c^2}\left[X'(p_F)p_f\right]^2.
\end{equation}
where $X(p) \equiv \mathcal{X}(p)/p$ and $c_2$ is a numeric constant presented previously in (\ref{e-ph_relaxation_low_T}) and defined in (\ref{definition c_2}). Expanding $X(p)$ as a linear combination of the basis functions (\ref{basis}) to extract the matrix components of $\bm{C}_\textrm{e-e}$, only the $C^{11}_\textrm{e-e}$ component is non-zero, as expected from the fact that that $F^0(p)$ is a zero-mode. For this reason, this collision integral does not contribute to the electric conductivity, again as expected from Eq.(\ref{electric_conductivity_limit}). On the other hand, it contributes to the thermal conductivity as much as the interaction with real phonons (\ref{e-ph_relaxation_high_T}),
\begin{equation}\label{electron_electron_rt_high_T}
    \tau_\textrm{$\kappa$, e-e}^{-1} = \tau_\textrm{$\kappa$, e-ph}^{-1}.
\end{equation}
In physical terms, this can be understood from the fact that at $T>T_\textrm{BG}$, the dominant contribution to the electron-electron collision integral is from virtual phonons satisfying the on-shell condition $\omega = c q$ stemming from the poles in Eq.(\ref{omegaintegral}). It is natural to expect that the processes in fig. (\ref{fig:LLLL_diagrams}) share the same time scale as the emission and absorption of real phonons, $\tau_{ \textrm{e-e}}^{-1} = \tau_\textrm{emission}^{-1} + \tau_\textrm{absorption}^{-1}\equiv \tau_\textrm{e-ph}^{-1}$. 
\\
\\
If impurity scattering is negligible (which is natural to expect at sufficiently high temperatures), and ignoring chirality-violating e-e processes, then one sees that
\begin{equation}\label{lorenz_ratio_ht_no_chirality_breaking}
    L = \frac{L_0}{2}     ,
\end{equation}
as the thermal conductivity would now have half the relaxation time of the electric conductivity, oblivious to electron-electron interactions. This equality is only valid when the inequality in Eq.(\ref{phonon_linewidth_bound}) saturates.
For any larger phonon linewidth, electron-electron scattering mediated by phonons would be suppressed at $T>T_\textrm{BG}$. Physically this corresponds to the scenario where the on-shell virtual phonons, acting like real phonons, have decay channels stronger than those of emission and absorption by electrons, thus suppressing the emission and absorption-dependent processes depicted in Fig.(\ref{fig:LLLL_diagrams}). In this limit, $L\rightarrow L_0$, independently on whether chirality-violating processes are considered or not.

\subsection{Low Temperatures}
At temperatures below $T_{\textrm{BG}}$ the Matsubara poles start to play a role and the approximation in Eq.(\ref{omegaintegral_result}) is no longer valid. Instead of attempting to calculate the contribution of these new poles, we make the observation that, in this regime, the Bose-Einstein distribution in Eq.(\ref{intermediate_collision_integral}) can no longer be approximated by its $\beta\omega \rightarrow 0$ limit. We expect then that, for $T \ll T_{\textrm{BG}}$, the equilibrium Bose-Einstein distribution will be the dominating contribution, fixing $\omega \sim \beta^{-1}$ instead of the on-shell condition $\omega \sim cq$. This observation, together with the defining condition for this temperature regime, implies that $\omega \ll \omega_{\bm{q}}$ and hence the $\omega$ dependence of the propagator in Eq.(\ref{phonon_propagator}) can be dropped. Consequently the scattering amplitude for the $t$-channel becomes
\begin{gather}\nonumber
A_t = U\frac{q^2}{\omega_{\bm{q}}^2 }\braket{+\bm{p}'|+\bm{p}}\braket{+\bm{k}'|+\bm{k}}\\
\label{scattering_amplitude_ee_low_T}
\simeq\frac{U}{c^2 }\braket{+\bm{p}'|+\bm{p}}\braket{+\bm{k}'|+\bm{k}},
\end{gather}
where we have dropped the phonon linewidth $\Gamma \ll \omega_{\bm{q}}$ as it is no longer necessary to make the integral convergent.

In the $T\gg T_\textrm{BG}$ regime we argued that the interference term in the square of the scattering amplitude could be dropped as it was subleading in the $c$ and $\Gamma/(2q)$ expansion stemming from Eq.(\ref{omegaintegral_result}). That statement is no longer valid, as the low temperature approximation for the amplitude Eq.(\ref{scattering_amplitude_ee_low_T}) obviously has no poles, and we need to consider the interference contribution from $-2\textrm{Re}[A_t^*A_u]$. The physical intuition for this is that for temperatures above $T_\textrm{BG}$, phonons behave classically and quantum interference effect is negligible. For temperatures below $T_\textrm{BG}$, phonons  regain their quantum behavior and interference terms are no longer subleading. Nevertheless, the simple form of the amplitudes in Eq.(\ref{scattering_amplitude_ee_low_T}) makes calculations simple. For instance, one can show that, in the low temperature limit,
\begin{equation}\label{scattering_amplitude_squared_low_temperature}
    |M_{LL}^{LL}|^2 = \frac{U^2}{4c^4}\left(1 - \cos\theta_{pk}\right)^2,
\end{equation}
where $\theta_{pk}$ is the angle between $\bm{p}$ and $\bm{k}$. It is important to note that the equivalent intraband square of the scattering amplitude would vanish in a model diagonal in the electron basis of the Hamiltonian, as then the wavefunction overlaps would be trivial and $A_t = A_u$ in Eq.(\ref{scattering_amplitude_ee_low_T}). This would mean that higher order terms in $\omega$ would be required for the scattering amplitude, which would suppress the contribution of the collision integral by a power of the small parameter $T/p_f$ and hence the contribution of electron-electron scattering mediated by phonons to transport effects.

After some intermediate calculations, the collision integral reduces to
\begin{equation}\label{ee_intraband_low_T_result}
   (\bm{\mathcal{X}},\mathcal{C}_\textrm{e-e}\bm{\mathcal{X}})^\textrm{intra} = \frac{2^4U^2p_f^7\beta^{-3}}{35(2\pi)^5c^4} \left[4c_2^2 + 2b_4\right]X'(p_F)^2,
\end{equation}
where $c_2$ and $b_4$ are O(1) numeric constants defined in (\ref{definition c_2}) and (\ref{definition_b4}). The inverse relaxation time $\tau_{\kappa, \textrm{e-e}}^{-1}$ will now scale with temperature as $T^2$, just like in the case of Coulomb scattering, and with the electron-phonon coupling strength as $U^2$. This is to be expected as, in the case of Quantum Electrodynamics,  virtual photons are always in the $T \ll cq$ regime where the $\omega^2$ in the denominator of the photon propagator is negligible, as in the case Coulomb interaction. The prefactors however are obviously different in the case of Coulomb and virtual phonons.

\subsection{Chirality-violating processes}
Until now we have studied processes within an single Weyl node, valid when interactions do not induce changes in the chirality. In Weyl systems, nodal points come in pairs, as is the case for our starting model Eq.(\ref{two_band_Hamiltonian}), one for left $(L)$ and another for right-handed $(R)$ Weyl fermions. Therefore, besides the intraband $(L,L)\rightarrow (L,L)$ scattering processes we had been considering up until now, we should consider $(L,L)\rightarrow (R,R)$, $(L,R) \rightarrow (L,L)$, $(L,R)\rightarrow (L,R)$ etc. Out of all these possibilities, only the process $(L,R)\rightarrow (L,R)$ conserves chirality. While the associated collision integral is more complicated due to the anisotropic wavefunction overlaps in the square of the scattering amplitude (Eq. B.48 in the Supplementary Material), the integral is essentially solved in similar fashion to the chirality-conserving situation.

Chirality-violating processes however, are a different story. In the reference frame of one of the nodes and for processes that violate chirality once, such as $(L,R) \rightarrow (L,L)$, Dirac deltas associated to momentum conservation in the collision integrals take the form
\begin{equation}\label{chirality_violating_delta}
    \delta^{(3)}(\bm{p}+\bm{k} - \bm{p}' -\bm{k}' \pm 2\bm{b}).
\end{equation}
 For processes that violate chirality twice, such as $(L,L)\rightarrow (R,R)$ we have $\delta^{(3)}(\bm{p}+\bm{k} - \bm{p}' -\bm{k}' \pm 4\bm{b})$ . These Dirac deltas are similar to the ones appearing in umklapp processes with a lattice vector $2\bm{b}$ separating two copies of the same Fermi surface. Note however that momentum here is strictly conserved, as can be seen if the momenta are put back into the original reference frame of Eq.(\ref{two_band_Hamiltonian}).

At large temperatures, $T\gg T_\textrm{BG,W}$, the only relevant poles in the $\omega$ integral are those provided by the phonon propagator in Eq.(\ref{phonon_propagator}) so that the expansion in Eq.(\ref{omegaintegral_result}) still holds, making the phonons effectively on-shell. In the limit  where the inequality Eq.(\ref{phonon_linewidth_bound}) is saturated, one recovers the result in Eq.(\ref{electron_electron_rt_high_T}) and an analogously similar result,
\begin{equation}\label{ht_electric_conductivity_chiral_collision_integral}
     (\bm{\mathcal{X}}_0, \mathcal{C}_\textrm{e-e}\bm{\mathcal{X}}_0)_\textrm{h.t.} = \frac{\beta^4(4b)^2}{2}\int_{\bm{q}}\frac{\Gamma^\textrm{inter}_\textrm{e-ph}(\bm{q})}{(\beta\omega_{\bm{q}})^2},
\end{equation}
where $\Gamma^\textrm{inter}_\textrm{e-ph}$ is the contribution to the phonon linewidth from a $L-R$ fermion bubble. This shows how $\tau_{\sigma, \textrm{e-e}}^{-1}$ is directly related to the linewidth associated to interband processes, which are the only electron scattering processes that can relax the electric current in the system. The contribution to the electric conductivity (\ref{ht_electric_conductivity_chiral_collision_integral}), quadratically dependent on the internodal distance, leads to a Lorenz ratio of the form,
\begin{equation}\label{lorenz_ratio_chirality_breaking}
    L = \frac{1}{2}\left(1 + \frac{16b^2}{10p_f^2}\right)L_0.
\end{equation}
For temperatures below $T_\textrm{BG,W}$ but still above $T_\textrm{BG}$, electron-electron scattering processes that violate chirality no longer contribute and the Lorenz ratio satisfies the equality in Eq.(\ref{lorenz_ratio_ht_no_chirality_breaking}). As in the case of intravalley scattering, this is only valid where the dominant phonon decay channel is associated to scattering with electrons. Otherwise, $L\rightarrow L_0$.

At low temperatures, virtual phonons cannot be considered on-shell. Scattering amplitudes for processes that include more than one chirality involve more complicated wavefunction overlaps but otherwise share the relatively similar simple form of Eq.(\ref{scattering_amplitude_squared_low_temperature}) (see section \ref{appendix_hamiltonian_and_scattering_amplitudes} of the appendix). The integral for the chirality conserving process $(L,R) \rightarrow (L,R)$ is analogous to the one performed for $(L,L) \rightarrow (L,L)$ in Eq.(\ref{ee_intraband_low_T_result}), whereas the one for chirality-violating processes now includes the momentum conservation delta shown in Eq.(\ref{chirality_violating_delta}). This delta, along with the restriction that all electron momenta lie on the Fermi sphere, restricts the available phase space as is the case with umklapp scattering \cite{PhysRevB.23.2718,Ziman:100360}. For $2b \ge 4p_f$, these processes are not kinematically allowed. The only interesting parameter space for us to consider is then the relation $2p_f < 2b < 4p_f$, as our model relies on the assumption that $2p_f < b$, where chiralities are well defined. While these integrals can be performed analytically, the end result are somewhat complicated expressions of $b$, and are written in section B4 of the Supplementary Material \cite{SupplementaryMaterial} for reference. In the end one finds that they also lead to a $U^2T^2$ scaling for the inverse relaxation times just like the chirality-conserving processes (\ref{ee_intraband_low_T_result}). Conversely to that result however, they also contibute to the relaxation of the electric current because of the Umklapp-like nature of (\ref{chirality_violating_delta}). 

\begin{figure*}[ht]
    \begin{center}
    \setlength\tabcolsep{-4mm}
    \begin{tabular}{cc}
        \small a) & \small b) \vspace{-5mm}\\
         \includegraphics[width = 0.5\textwidth]{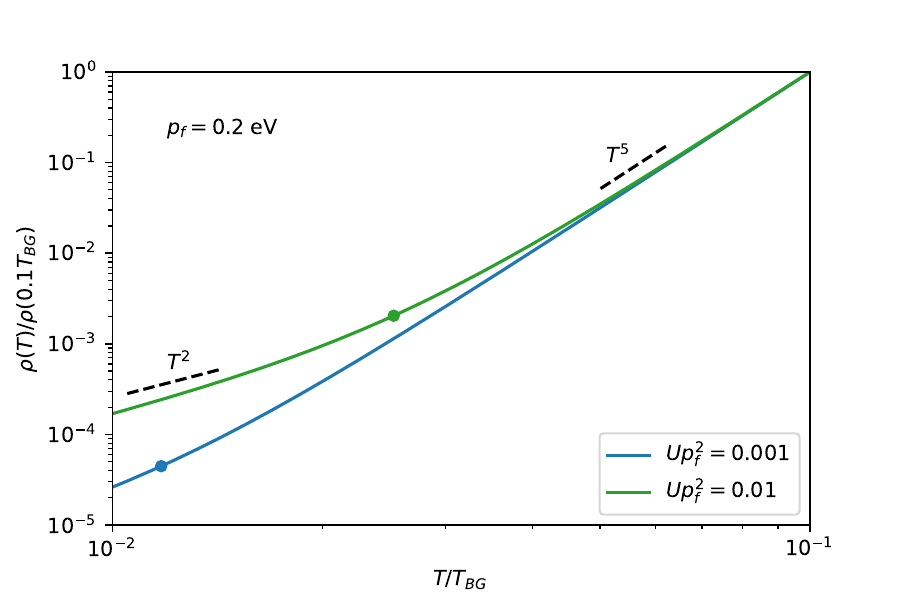} &  \includegraphics[width = 0.5\textwidth]{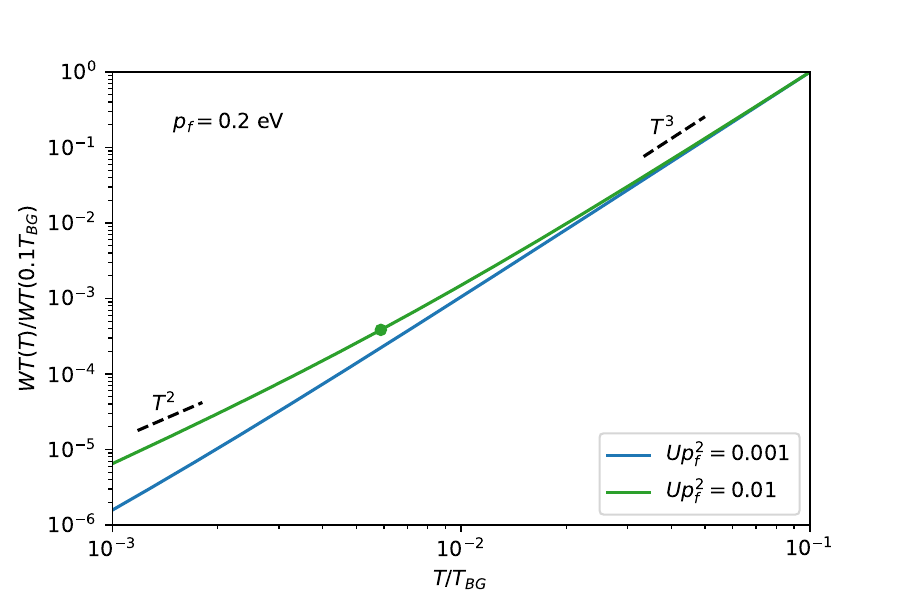} \vspace{-1mm}\\
         \small c) & \small d) \vspace{-5mm}\\
         \includegraphics[width = 0.5\textwidth]{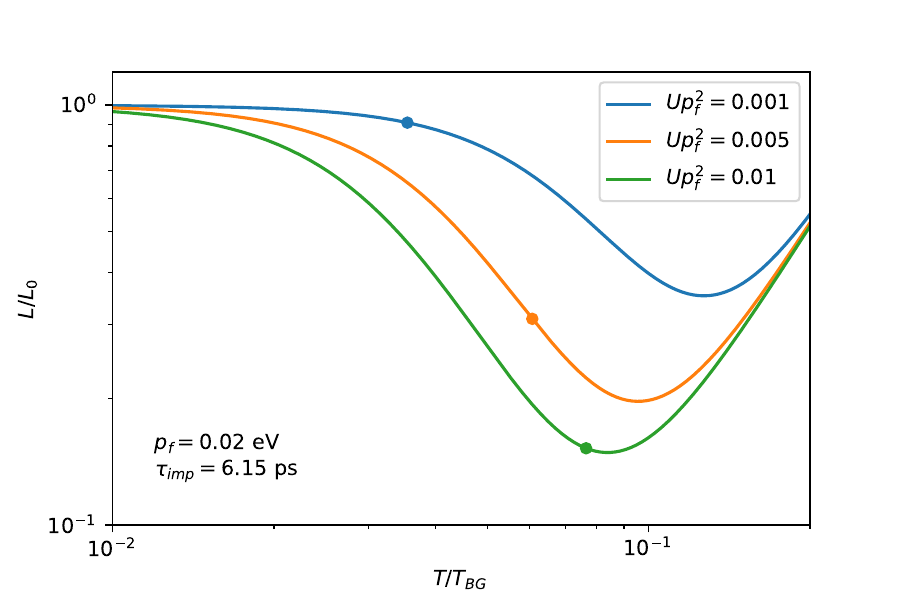} &
         \includegraphics[width = 0.5\textwidth]{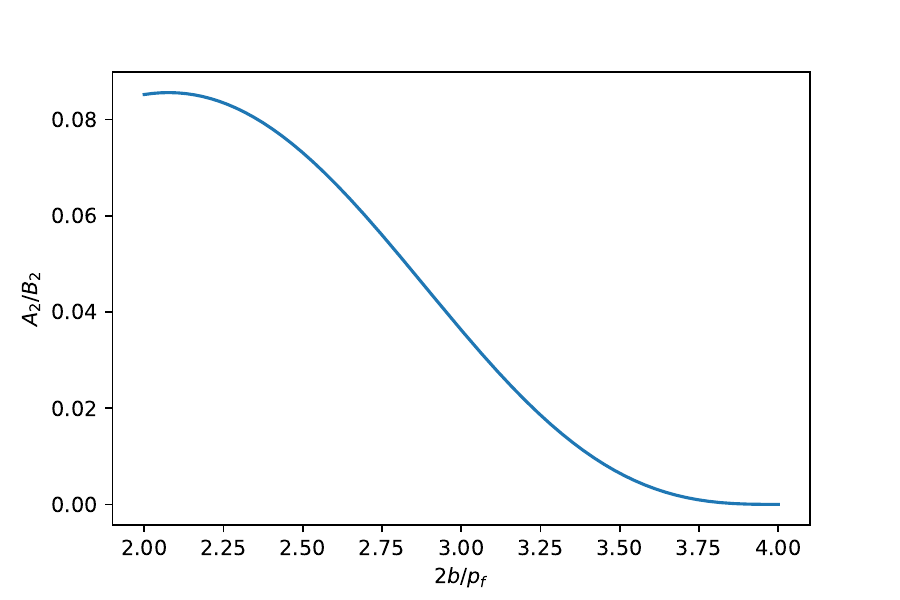}\vspace{-4mm}
    \end{tabular}
    \end{center}    
    \caption{\textbf{a} and \textbf{b}: Electric (a) and thermal (b) conductivities for $T \ll T_\textrm{BG}$ for different electron-phonon interaction strengths $Cp_f^2$. Points indicate temperature below which electron-electron scattering is stronger than electron-phonon scattering. In the case of the thermal conductivity (b) for the smallest electron-phonon coupling, the point is below the represented temperature scale because the $T^3$ scaling from absorption/emission of phonons does not decrease as fast as the $T^5$ scaling from the same source for the electric resistivity. The constant contributions from impurity scattering, i.e. $A_0$ and $B_0$ terms in (\ref{resistivity_parametrization}, \ref{t_resistivity_parametrization}) have been ommitted to show the onset of the $T^2$ scaling behavior.\textbf{c}: Lorenz ratio at small temperatures $T \ll T_\textrm{BG}$ for different coupling strengths. Points indicate where current-conserving interactions start dominating over current-relaxing ones (see fig. \ref{fig:hydro}) for temperatures below the point. \textbf{d}: Quotient between the $A_2$ and $B_2$ terms in  (\ref{resistivity_parametrization}, \ref{t_resistivity_parametrization}) which corresponds to the Lorenz ratio of the electron-electron contributions to the resistivities. Plots a-c are drawn for $p_f = 0.2 \textrm{ eV}$ and $c = 0.02 v_f$ with an internodal separation is set to $2b = 3p_f$. Plot d is independent of these parameters.}
    \label{fig:results}
\end{figure*}

\section{Discussion}

We have shown how the non-trivial structure of the Weyl Hamiltonian in Eq.(\ref{two_band_Hamiltonian}) had the important consequence of allowing phonon-mediated $T^2$ inverse relaxation times at low temperatures due to the nontrivial wavefunction overlaps $\braket{+\bm{k}'|+\bm{k}}$. It was the key point also in allowing relaxation of electric current through chirality-violating intervalley scattering, having important properties both at $T>T_\textrm{BG}$ with a non-trivial Lorenz ratio (Eq.(\ref{lorenz_ratio_chirality_breaking})) as well as low temperatures, as we will discuss further below. Nevertheless, more unique topological properties of the material fail to make an impact, at least in the absence of a quantum anomaly-inducing magnetic field. Its inclusion is left for future work.

Having discussed already the high-temperature behavior in the previous section, we focus here on the properties at temperatures $T \ll T_\textrm{BG}$, relevant for experimentally observing hydrodynamic signatures. When studying thermoelectric transport in experiments, it is useful to consider the electric and thermal resistivities, which can be conveniently parametrized respectively as 
\begin{gather}
    \label{resistivity_parametrization}
   \rho \equiv \sigma^{-1} = A_0 + A_2T^2 + A_5T^5, \\
   \label{t_resistivity_parametrization}
   WT \equiv TL_0/\kappa = B_0 + B_2T^2 + B_3T^3,
\end{gather}
 following the scaling of the resistivities given in Ref.\cite{jaoui2018departure} for WP$_2$. While WP$_2$ is a semimetal with a Fermi surface more complex than the one studied here, similar features in the thermoelectric transport are seen to arise. The resistivities are plotted in Figs.(\ref{fig:results}a-b) in the absence of impurity scattering. The full Lorenz ratio is plotted in Fig.(\ref{fig:results}c). 

\begin{figure*}[ht]
    \begin{center}
    \includegraphics[width = 1\textwidth]{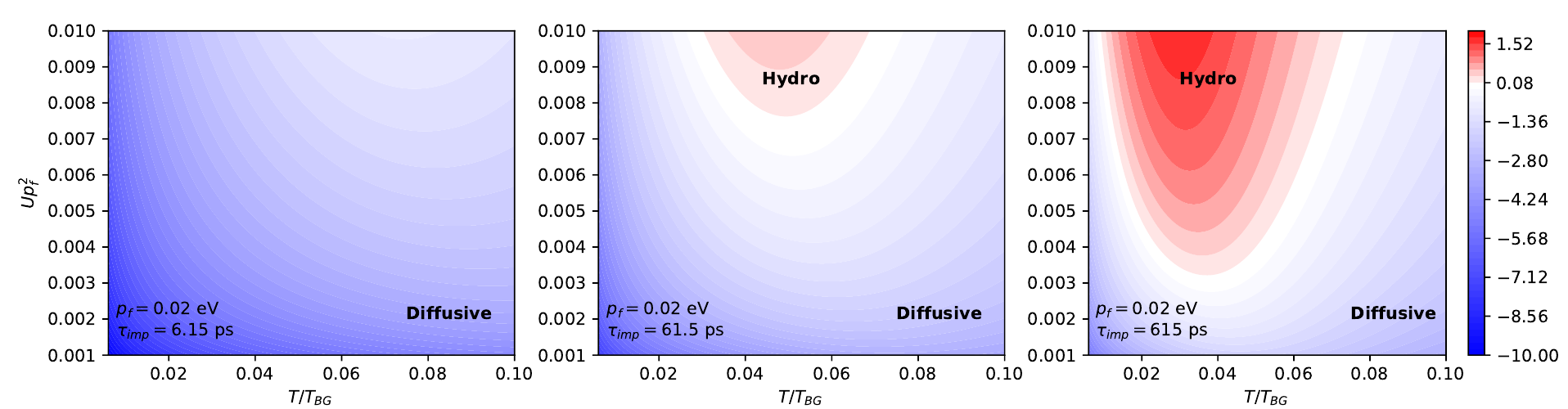}
    \end{center}    
    \caption{Plot of $\log\left(\frac{\tau_{\kappa,e-e}^{-1}}{\tau_{\sigma,\textrm{e-e}}^{-1} + \tau_{\sigma,\textrm{e-ph}}^{-1} + \tau^{-1}_\textrm{imp}}\right)$ as a function of temperature and electron-phonon coupling strength, plotted from left to right for three different impurity relaxation times, $\tau_{\textrm{imp},0} = 6.15$ ps, $10\tau_{\textrm{imp},0}$ and $100\tau_{\textrm{imp},0}$. The red region indicates  where electric current-conserving interactions dominate, whereas the blue region indicates where current-relaxing interactions with virtual and real phonons, as well as impurities dominate. The latter is labeled "Diffusive", as transport in this region can never be dominated by electric current-conserving collisions, whereas the former is labeled "Hydro" as this is where these collisions might dominate, depending on impurity scattering.} 
    \label{fig:hydro}
\end{figure*}

As is clear from our model at low temperatures, $A_0$ and $B_0$ are the coefficients associated with impurity scattering, $A_2$ and $B_2$ with electron scattering (\ref{ee_intraband_low_T_result}), and $A_5$ and $B_3$ with thermal phonon emission/absorption by electrons (\ref{e-ph_relaxation_low_T}). In a system that is at some point dominated by electron-electron interactions, one then expects the Lorenz ratio to level out at the constant value $L/L_0 = \rho/WT = \tau_{\sigma, \textrm{e-e}}^{-1}/\tau_{\kappa, \textrm{e-e}}^{-1} = A_2/B_2$. For our model, $A_2$ is non-vanishing only for $2b < 4p_f$ as otherwise electric current-relaxing intervalley electron-electron interactions are absent. This leads to the curve plotted in Fig.(\ref{fig:results}d), where the ratio $A_2/B_2$ is plotted as a function of $2b/p_f$. It is important to note that this quantity is completely independent of the electron-phonon coupling $U$ and the phonon velocity $c$, and therefore only depends on the geometric factor $2b/p_f$. The ratio is expectedly small as intervalley electron-electron scattering contributes much less than the intravalley counterpart present in thermal conduction.
However, one would expect that such a ratio could increase for systems with different geometries or carriers \cite{Kumar22,Stern22}. For example, in a compensated (inversion-breaking) Weyl semimetal with an electron band and a hole band (instead of the two electron bands), chirality-conserving $(L,R) \rightarrow (L,R)$ processes would be able to relax the electron current owing to the fact that the fermions of each valley have group velocities with the opposite sign. Following the procedure of Ref.\cite{PhysRevB.98.245134} one can deduce that the ratio $A_2/B_2$ in such a system takes the form,
\begin{equation}
    \frac{A_2}{B_2} = \frac{10\alpha^2\left(4305 + 441 \alpha^2 - 1980 \alpha^4 + 700 \alpha^6\right)}{3\left(27090 + 3885\alpha^2 + 63\alpha^4 + 900\alpha^6\right)},
\end{equation}
where $\alpha = \textrm{min}\left[T_\textrm{D}/T_\textrm{BG}, 1\right]$. In a system with a sufficiently small Fermi surface so that $T_\textrm{BG} < T_\textrm{D}$, this would lead to a value of $A_2/B_2 \sim 0.36$, which is of the same order of the observed values in several materials, see Ref.\cite{jaoui2018departure}. In the opposite limit, where $T_\textrm{BG} > T_\textrm{D}$, then one has $A_2/B_2 \sim 0.53 \alpha^2$, i.e. the Lorenz ratio can become arbitrarily small. This is a similar conclusion to one reached in Ref.\cite{PhysRevB.98.245134}, although a screened Coulomb potential is behind that result, whereas Coulomb interactions are expected to be subdominant with respect to electron-phonon interactions in realistic semimetals \cite{Coulter18, osterhoudt2021evidence, vool2021imaging}. Here, by contrast, this result is reached for a constant potential (modulo structure factors) provided by the low temperature phonon-mediated electron-electron processes (\ref{scattering_amplitude_squared_low_temperature}). As in the original model, mixed-chirality processes are essential for a $T^2$ electric conductivity.


To achieve electron hydrodynamic behavior, electric current-conserving interactions need to dominate over those that relax it. The latter correspond to the the relaxation times pertaining to the electric conductivity, i.e. impurities $\tau_\textrm{imp}$, phonon absorption/emission $\tau_{\sigma,  \textrm{e-ph}}$ and intervalley (Baber) scattering $\tau_{\sigma, \textrm{ee}}$. The former correspond to the electron-electron collisions that conserve flow, i.e. intravalley scattering (\ref{starting_collision_integral}), that dominate the electron-electron component of the thermal conductivity, $\tau_{\kappa, \textrm{ee}}$, as the collision operator for thermal conduction does not filter out these contributions. This can easily be seen from Fig.(\ref{fig:results}d); if intervalley scattering dominated the electron-electron contribution to $\kappa$ as it does for $\sigma$, then one would have $A_2/B_2 \sim 1$. Fig (\ref{fig:results}d) clearly indicates this is not the case. Hence, to search for a window for electron hydrodynamic behavior, it is sensible to compare $\tau_{\kappa, e-e}$ with the electric current-relaxing timescales \cite{jaoui2018departure}. The relaxation time associated to the viscosity and stemming from electron-electron interactions should be of this same order, as flow-conserving collisions are not associated to any zero-mode of the collision integral (\ref{boltzmann}) for perturbations from thermal equilibrium proportional to spatial gradients of the flow velocity. With this in mind, the logarithm of the ratio between $\tau_{\kappa,e-e}^{-1}$ and the electric current inverse relaxation time, $ \tau_{\sigma,\textrm{e-e}}^{-1} + \tau_{\sigma,\textrm{e-ph}}^{-1} + \tau_\textrm{imp}^{-1}$ is plotted in Fig. (\ref{fig:hydro}). Hydrodynamic behavior is favored for larger electron-phonon coupling strengths and for a limited temperature window where impurities do not yet dominate.  For a typical value of $\tau_\textrm{imp,0} = 6.15 $ ps (corresponding to a mean scattering time of $1.9$ $\mu$m \cite{vool2021imaging} and a Fermi velocity of $3.09\cdot 10^5$ ms$^{-1}$ as in WTe$_2$ \cite{li2017evidence}), it is shown in the leftmost plot of Fig.(\ref{fig:hydro}) that a hydrodynamic regime is basically inaccessible for the represented parameter ranges. Considering impurity scattering times 10 and 100 times stronger, such a window opens for $Up_f^2 \gtrsim 0.007$ and $Up_f^2 \gtrsim 0.003$, respectively, as is shown in the center and right plots of Fig. (\ref{fig:hydro}). Realistic electron phonon-coupling constants in the range $\lambda \sim 0.1-1$ translate to a range $Up_f^2 \sim 0.0005-0.005$ \ref{appendix_section_eph_coupling}, which clearly shows that the impurity scattering time needs to be of one or two orders of magnitude larger than the value of $\tau_{\textrm{imp},0}^{-1}$ considered for momentum-conserving scattering to be relevant. For example, for the Weyl SM TaAs, the coupling constant for acoustic phonons is $\lambda \sim 0.34$ \cite{han2022effects}, which would need impurity relaxation times two orders of magnitude larger than $\tau_{\textrm{imp},0}^{-1}$. It is worth noting that while impurity scattering is a fundamental aspect in predicting electron hydrodynamics, its nature (i.e. short vs long range Coulomb impurities) is not relevant as the possible differences are subleading in the regime $T/p_f \ll 1$ considered here \cite{PhysRevB.91.035201}.

As a conclusion, we have described how the Bloch-Grüneisen temperature $T_\textrm{BG}$ divides the behavior of phonon-mediated electron-electron collisions into two regions as it does for absorption and emission of real phonons. In the case where the phonon interactions are stronger than anharmonic effects, we have seen how at $T> T_\textrm{BG}$ the Lorenz ratio is modified to a multiple of the Lorenz number (\ref{lorenz_ratio_ht_no_chirality_breaking}, \ref{ht_electric_conductivity_chiral_collision_integral}). If phonon decay is driven by other interactions, then the Lorenz ratio in this regime is unaltered. At small temperatures, we have obtained that the electric and thermal resistivities scale as (\ref{resistivity_parametrization}) and (\ref{t_resistivity_parametrization}). We concluded that in the simple type-I Weyl SM model studied here with isotropic fermion pockets and electron-phonon interactions a hydrodynamic regime would necessitate highly-pure samples. We speculate then that the anisotropy of the Fermi surface may magnify the effect of electron-electron interactions by an order of magnitude so as to lead to the observable hydrodynamic behavior in real materials, as has been suggested elsewhere \cite{Coulter18}. Nevertheless, our analytical results seem to align qualitatively well with the first principles estimations of the relaxation times. In those, it is seen that for high temperatures electron-electron collisions mediated by phonons contribute as much as absorption/emission of phonons, whereas at low temperatures the electron-electron relaxation times decay similarly to $T^2$ \cite{Coulter18, vool2021imaging, garcia2021anisotropic}.

\section{Acknowledgements}
J.B. is supported by FPU grant FPU20/01087. A.C. acknowledges financial support from the Ministerio de Ciencia e Innovaci\'on through the grant PID2021-127240NB-I00 and the Ram\'on y Cajal program through the grant No. RYC2018- 023938-I.

%

\appendix
\section{Hamiltonian and Scattering Amplitudes}\label{appendix_hamiltonian_and_scattering_amplitudes}
As mentioned in the main text, a two-band Hamiltonian matrix of the form
\begin{equation}
    \mathcal{H}_0(\bm{p}) = \bm{d}(\bm{p})\cdot\bm{\sigma},
\end{equation}
where $\bm{d} = (d_1,d_2,d_3)^\textrm{T}$ has positive and negative energy $\epsilon_{\pm, \bm{p}} = \pm d(\bm{p}) \equiv \pm \sqrt{\bm{d}^2(\bm{p})}$ eigenstates, which are respectively given by ($c\leftrightarrow+$, $v\leftrightarrow-$)
\begin{gather}
    \ket{+\bm{p}} \equiv \frac{1}{\sqrt{2d(d+d_3)}}
    \begin{pmatrix}
   d + d_3\\
    d_1 + i d_2
    \end{pmatrix},\\
    \quad \ket{-\bm{p}} \equiv \frac{1}{\sqrt{2d(d+d_3)}}
    \begin{pmatrix}
    d_1 - id_2 \\
    -d - d_3
    \end{pmatrix},
\end{gather}
where $c$ stands for conduction band ($+d(\bm{p})$) and $v$ for valence band ($-d(\bm{p})$). The particle fermion operators are related to the original fermion operators by the unitary transformation $\psi_{\bm{p}} = U_{\bm{p}}c_{\bm{p}}$, where $U_{\bm{p}} \equiv (\ket{+\bm{p}},\ket{-\bm{p}})$. In the particle-hole basis, the electron-phonon interaction given by the deformation potential (\ref{deformation_potential}) becomes
\begin{equation}
    H_\textrm{e-ph} = \int_{\bm{p}',\bm{p},\bm{q}}g_{\bm{p}',\bm{p}}^{\bm{q}}\left(a_{\bm{q}}  + a^{\dagger}_{\bm{q}}\right)c^{\dagger}_{\bm{p'}}U^{\dagger}_{\bm{p}'}U_{\bm{p}}c_{\bm{p}}\ \delta^{(3)}(\bm{p}'-\bm{p} - \bm{q}),
\end{equation}
For $\mu \gg T$, the effects of the negative energy band are exponentially suppressed, so we ignore its effects from now on. For the fermion-phonon interaction, it implies that we only consider care about the term proportional to $c^\dagger_{+,\bm{p}}c_{+,\bm{p}}$, hence the effective fermion coupling to the relevant fermion sector is given by the expression given in (\ref{effective_electron_phonon_coupling}), i.e.
\begin{equation}
    g_{\bm{p}',\bm{p}}^{\textrm{eff}\ \bm{q}} \equiv g_{\bm{p}',\bm{p}}^{\bm{q}}\braket{+\bm{p}|+\bm{p}'} ,
\end{equation}
where for simplicity of notation we have used the label $u_{\bm{p}} \equiv u_{\bm{+,\bm{p}}}$, and will continue to do so henceforth. The scattering amplitude for fermion collisions with incoming momenta $\bm{p},\bm{k}$ and outgoing momenta $\bm{p}',\bm{k}'$ mediated by virtual phonons is given by the interference of the $t$ and $u$ channels, $M = A_t - A_u$
\begin{equation}
    A_t = -g_{\bm{p}',\bm{p}}^{\bm{q}}g_{\bm{k},\bm{k}'}^{\bm{q}} \mathcal{D}(\bm{q},\omega)\braket{+\bm{p}'|+\bm{p}}\braket{-\bm{k}'|-\bm{k}},
\end{equation}
where $\bm{q} = \bm{p}'-\bm{p}$ and $\omega = d(\bm{p}') - d(\bm{p})$. The amplitude for the $u$ channel $A_u$ is obtained by exchanging $\bm{p}'\leftrightarrow \bm{k}'$ in the expression for $A_t$. The square of the amplitude relevant for the Boltzmann equation can be shown then to be
\begin{widetext}
\begin{gather}
    |M|^2(\bm{p},\bm{k},\bm{p}',\bm{k}') = |A_t|^2 + |A_u|^2 - 2\textrm{Re}[A_t^*A_u] =\nonumber \\ \left\{|\mathcal{D}|^2(\bm{q}_t,\omega_t)|g_{\bm{p}',\bm{p}}^{\bm{q}_t}g_{\bm{k},\bm{k}'}^{\bm{q}_t}|^2 -\textrm{Re}\left[\mathcal{D}^*(\bm{q}_t,\omega_t)\mathcal{D}(\bm{q}_u,\omega_u)\right]g_{\bm{p}',\bm{p}}^{\bm{q}_t}g_{\bm{k},\bm{k}'}^{\bm{q}_t}g_{\bm{k}',\bm{p}}^{\bm{q}_u}g_{\bm{k},\bm{p}'}^{\bm{q}_u}\right\}\left[1 +\hat{\bm{d}}(\bm{p})\cdot\hat{\bm{d}}(\bm{p}')\right]\left[1 +\hat{\bm{d}}(\bm{k})\cdot\hat{\bm{d}}(\bm{k}')\right] \nonumber\\
    + \left\{|\mathcal{D}|^2(\bm{q}_u,\omega_u)|g_{\bm{k}',\bm{p}}^{\bm{q}_u}g_{\bm{k},\bm{p}'}^{\bm{q}_u}|^2-\textrm{Re}\left[\mathcal{D}^*(\bm{q}_t,\omega_t)\mathcal{D}(\bm{q}_u,\omega_u)\right]g_{\bm{p}',\bm{p}}^{\bm{q}_t}g_{\bm{k},\bm{k}'}^{\bm{q}_t}g_{\bm{k}',\bm{p}}^{\bm{q}_u}g_{\bm{k},\bm{p}'}^{\bm{q}_u}\right\} \left[1 +\hat{\bm{d}}(\bm{p})\cdot\hat{\bm{d}}(\bm{k}')\right]\left[1 +\hat{\bm{d}}(\bm{k})\cdot\hat{\bm{d}}(\bm{p}')\right] \nonumber\\ \label{full_scattering}
    +\textrm{Re}\left[\mathcal{D}^*(\bm{q}_t,\omega_t)\mathcal{D}(\bm{q}_u,\omega_u)\right]g_{\bm{p}',\bm{p}}^{\bm{q}_t}g_{\bm{k},\bm{k}'}^{\bm{q}_t}g_{\bm{k}',\bm{p}}^{\bm{q}_u}g_{\bm{k},\bm{p}'}^{\bm{q}_u}\left[1 -\hat{\bm{d}}(\bm{p})\cdot\hat{\bm{d}}(\bm{k})\right]\left[1 -\hat{\bm{d}}(\bm{p}')\cdot\hat{\bm{d}}(\bm{k}')\right].
\end{gather}
\end{widetext}
Following the discussion of the main text, a distinction between the behavior at high and low temperatures can be made so as to simplify (\ref{full_scattering}). At high temperatures $T \gg T_\textrm{BG}$, the interference terms can be dropped. In addition, since the collision integral is symmetric with the exchange $\bm{p}'\leftrightarrow \bm{k}'$, inside the collision integral one effectively has that $|A_t|^2 = |A_u|^2$. 

At low temperatures $T \ll T_\textrm{BG}$, the $\omega$ dependence of the phonon propagators can be dropped at the lowest order in $\omega/\omega_{\bm{q}}$ expansion. The small linewidth can be dropped as well, leading to the cancellation of the first two terms in (\ref{full_scattering}). All in all, these considerations lead to the square scattering amplitudes
\begin{widetext}
\begin{gather}\label{generic_high_temperature_amplitude}
    |M|^2_{\textrm{h.t.}}(\bm{p},\bm{k},\bm{p}',\bm{k}') = \frac{2U^2}{4}\left\lvert\frac{q^2}{\omega^2 - \omega_{\bm{q}}^2 +i\omega_{\bm{q}}\Gamma_{\bm{q}}}\right\rvert^2\left[1 +\hat{\bm{d}}(\bm{p})\cdot\hat{\bm{d}}(\bm{p}')\right]\left[1 +\hat{\bm{d}}(\bm{k})\cdot\hat{\bm{d}}(\bm{k}')\right], \\
    \label{generic_low_temperature_amplitude}
    |M|^2_{\textrm{l.t.}}(\bm{p},\bm{k},\bm{p}',\bm{k}') = \frac{U^2}{4c^4}\left[1 -\hat{\bm{d}}(\bm{p})\cdot\hat{\bm{d}}(\bm{k})\right]\left[1 -\hat{\bm{d}}(\bm{p}')\cdot\hat{\bm{d}}(\bm{k}')\right],
\end{gather}
\end{widetext}
where the labels "h.t." and "l.t." stand are for high temperatures and low temperatures, respectively. The explicit forms of the electron-phonon coupling (\ref{electron_phonon_coupling}) and the phonon propagator (\ref{phonon_propagator}) have been used.

\begin{widetext}
\section{Phonon Linewidth}
Considering only intravalley interactions, the phonon linewidth is calculated from the first diagram in Fig. \ref{fig:linewidth_diagrams} of the main text. To do so, one calculates the self-energy,
\begin{gather}
    \Sigma(\bm{q}, i\omega_n) = \beta^{-1}\int_{\bm{p}}\sum_{i\nu_m}\left(ig_{\bm{p}',\bm{p}}^{\textrm{eff}\ \bm{q}}\right)\left(ig_{\bm{p}',\bm{p}}^{\textrm{eff}\ \bm{q}*}\right)\mathcal{G}_0(\bm{k},i\nu_m)\mathcal{G}_0(\bm{k} + \bm{q},i\nu_m + i\omega_n) \\
    \label{self_energy_1}
    = -\beta^{-1}\int_{\bm{p}}\sum_{i\nu_m}(g_{\bm{p}',\bm{p}}^{\bm{q}})^2 \frac{ |\braket{+\bm{p}|+\bm{p}'}|^2 }{\left[i\nu_m+p_f - p\right]\left[i\nu_m + i\omega_n + p_f - p'\right]} =  \int_{\bm{p}}(g_{\bm{p}',\bm{p}}^{\bm{q}})^2|\braket{+\bm{p}|+\bm{p}'}|^2 \frac{f_0(p)-f_0(p')}{i\omega_n + p-p'}.
\end{gather}
where $\bm{p}' = \bm{p} + \bm{q}$ and the Matsubara sum has been performed in the last equality. After analytically continuing to real frequencies $i\omega_n\rightarrow \omega - i\delta$, the phonon linewidth can be obtained
\begin{gather}\label{appendix_phonon_linewidth}
    \Gamma^{\textrm{e-ph}}_{\bm{q}} \equiv 2\textrm{Im}\left\{\Sigma(\bm{q}, \omega_{\bm{q}} - i\delta)\right\} = (2\pi)\int_{\bm{p}}(g_{\bm{p}',\bm{p}}^{\bm{q}})^2|\braket{+\bm{p}|+\bm{p}'}|^2 \left[f_0(p)-f_0(p')\right]\delta(p'-p-\omega_{\bm{q}}),
\end{gather}
where we have used the Cauchy-Dirac relation $\frac{1}{x-a-i\delta} = P\frac{1}{x-a} + i\pi\delta(x-a)$. Note that the structure factor is given by
\begin{equation}\label{appendix_structure_factor}
    |\braket{+\bm{p}|+\bm{p}'}|^2 = \frac{1 +\hat{\bm{d}}(\bm{p})\cdot\hat{\bm{d}}(\bm{p}')}{2}.
\end{equation}
We will also use the approximation where
\begin{equation}
    f_0(p)-f_0(p') \approx \omega_{\bm{q}}[-f_0'(p)] \approx \omega_{\bm{q}}\delta(p-p_f),
\end{equation}
where the first equality comes from $p,p' \sim p_f \gg cq$ and the latter is the lowest order in the Sommerfeld approximation.
The intranode contribution is given by the loop with fermions of the same chirality. After a simple calculation, it can be shown that
\begin{equation}\label{appendix_phonon_linewidth_intra}
    \Gamma_\textrm{e-ph}^\textrm{intra}(\bm{q}) = 
    \begin{cases}
    2\cdot \frac{U}{4\pi}qp_f^2\left(1 - \frac{q^2}{4p_f^2}\right), & \textrm{if $q \le 2p_f$,} \\
        0, & \textrm{if $q > 2p_f$}.
    \end{cases}
\end{equation}
The leading factor of $2$ comes from considering both $L$ intraband and $R$ intraband contributions, which are identical. The interband contribution on the other hand can be shown to be
\begin{equation}\label{appendix_phonon_linewidth_inter}
    \Gamma_\textrm{e-ph}^\textrm{inter}(\bm{q}) = 
    \begin{cases}
    \frac{U}{4\pi}\frac{q^2}{|\bm{q}-2\bm{b}|}p_f^2\left[1 - \frac{|\bm{q} - 2\bm{b}|^2}{4p_f^2} + \frac{|\bm{q} - 2\bm{b}|^2}{2p_f^2}\cos^2\theta_{b,q-2b} - \left(1-\frac{|\bm{q} - 2\bm{b}|^2}{4p_f^2}\right)\frac{\sin^2\theta_{b,q-2b}}{2}\right], & \textrm{if $|\bm{q}-2\bm{b}| \le 2p_f$,} \\
        0, & \textrm{if $|\bm{q}-2\bm{b}| > 2p_f$}.
    \end{cases}
\end{equation}
The calculation of these linewdiths is similar to the calculation of the collision matrix terms for phonon absorption and emission by fermions provided in the next section, so we refer the reader to it for more detail. The full linewidth is simply given by 
\begin{equation}\label{sum_of_linewidths}
    \Gamma^{\textrm{e-ph}}_{\bm{q}} = \Gamma_\textrm{e-ph}^\textrm{intra}(\bm{q}) + \Gamma_\textrm{e-ph}^\textrm{inter}(\bm{q}).
\end{equation}
\end{widetext}

\section{Electron-Phonon Coupling} \label{appendix_section_eph_coupling}
The electron-phonon coupling strength is usually parameterized \cite{coleman2015introduction} in terms of the dimensionless
\begin{equation}
    \lambda = \frac{1}{\pi \int_{\bm{p}}\delta(\varepsilon_{\bm{p}}-\mu)}\int_{\bm{q}}\frac{\Gamma_{\bm{q}}^\textrm{e-ph}}{\omega_{\bm{q}}^2}.
\end{equation}
The integral can be evaluated using the phonon linewidths for intraband (\ref{appendix_phonon_linewidth_intra}) and interband scattering (\ref{appendix_phonon_linewidth_inter}) using (\ref{sum_of_linewidths}). This in turn gives
\begin{gather}
    \lambda = \frac{5Up_f^2}{6\pi^2c^2}
    \Rightarrow Up_f^2 = \frac{6\pi^2}{5}\lambda c^2 \sim 12\lambda c^2.
\end{gather}

\section{Numeric Integrals}\label{appendix_numeric_integrals}
Several numeric integrals appear often in the calculations and are related to the integrals of momenta that include the Fermi-Dirac distribution and the Bose Einstein distribution. We list them here for convenience:
\begin{gather}
    \label{definition c_2}
    c_2 = \int_{-\infty}^{\infty} x^2 \frac{e^x}{\left(1 + e^{x}\right)^2} = \frac{1}{2}\int_{-\infty}^{\infty} x^2 \frac{e^x}{\left(1 - e^{x}\right)^2}  = \frac{\pi^2}{3}, \\
    \label{definition_b4}
    b_4 = \int_{-\infty}^{\infty} x^4 \frac{e^x}{\left(1 - e^{x}\right)^2} = \frac{4\pi^4}{15}, \\
    \label{definition_gn}
    g_n = \int_{-\infty}^\infty dx \frac{e^x}{(e^x+1)^2}\left[\zeta(n) - \textrm{Li}_n\left(-e^{-x}\right)\right],
\end{gather}
where $g_4 \approx 4.3$, $g_5 \approx 5.2$ and $g_6 \approx 6.1$.

\end{document}


\title{Supplementary Material:\\
Bounds on Phonon-mediated Hydrodynamic Transport in a Type-I Weyl Semimetal}
\author{Joan Bernabeu}%
\affiliation{Departamento de F\'isica de la Materia Condensada, Universidad Aut\'onoma de Madrid, Cantoblanco, E-28049 Madrid, Spain}
\email{joan.bernabeu@uam.es}

\author{Alberto Cortijo}
\affiliation{Departamento de F\'isica de la Materia Condensada, Universidad Aut\'onoma de Madrid, Cantoblanco, E-28049 Madrid, Spain}
\affiliation{Condensed Matter Physics Center (IFIMAC), Cantoblanco, E-28049 Madrid, Spain}
\email{alberto.cortijo@uam.es}

\maketitle
\onecolumngrid
\section{Phonon Absorption/Emission}
The collision integral pertaining to phonon absorption or emission by fermionic excitations is given by
\begin{gather}
    \mathcal{I}_\textrm{e-ph}[f_\lambda](\bm{p}) = \sum_{\lambda'}\int_{\bm{q,p'}}|M^{\lambda\lambda'}_\textrm{abs}|^2(\bm{p},\bm{q},\bm{p}') (2\pi)^4\delta^{(4)}(p+q-p')\left\{f_\lambda(\bm{p})b(\bm{q})[1-f_{\lambda'}(\bm{p'})] - f_{\lambda'}(\bm{p'}))[1-f_\lambda(\bm{p})][1+b(\bm{q})]  \right\} \nonumber \\
    +\int_{\bm{q,p'}}|M^{\lambda\lambda'}_\textrm{emi}|^2(\bm{p},\bm{q},\bm{p}') (2\pi)^4\delta^{(4)}(p-q-p')\left\{f_\lambda(\bm{p})[1+b(\bm{q})][1-f_{\lambda'}(\bm{p'})] - b(\bm{q})f_{\lambda'}(\bm{p'}))[1-f_\lambda(\bm{p})] \right\},
\end{gather}
where $\lambda$ and $\lambda'$ refer to the chirality, $L$ or $R$. After linearizing, removing the external field dependence and setting $f_L = f_R$ as in the main text, the corresponding operator $\mathcal{C}_\textrm{e-ph}\bm{\mathcal{X}}$ appearing in the operator $\mathcal{C}\bm{\mathcal{X}}$ of equation (\ref{linearized_boltzmann_equation}) can be shown to be
\begin{gather}
    f_0(\bm{p})[1-f_0(\bm{p})]\mathcal{C}_\textrm{e-ph}\bm{\mathcal{X}}(\bm{p}) = \beta\sum_{\lambda'}\int_{\bm{q,p'}}|M^{L\lambda'}_\textrm{abs}|^2(\bm{p},\bm{q},\bm{p}') (2\pi)^4\delta^{(4)}(p+q-p')f_0(p)b_0(cq)[1-f_0(p')][\bm{\mathcal{X}}(\bm{p})-\bm{\mathcal{X}}(\bm{p}')] \nonumber \\
    +\int_{\bm{q,p'}}|M^{L\lambda'}_\textrm{emi}|^2(\bm{p},\bm{q},\bm{p}') (2\pi)^4\delta^{(4)}(p-q-p')f_0(p)[1+b_0(cq)][1-f_0(p')][\bm{\mathcal{X}}(\bm{p})-\bm{\mathcal{X}}(\bm{p}')].
\end{gather}
Introducing the inner product (\ref{inner_product}), the corresponding collision term in the variational functional (\ref{action_functional}) can be expressed as
\begin{equation}\label{appendix_eph_collision_term}
    (\bm{\mathcal{X}}, \mathcal{C}_\textrm{e-ph}\bm{\mathcal{X}}) = \frac{1}{2}\cdot2\beta^2\sum_{\lambda'}\int_{\bm{p,q,p'}}|M^{L\lambda'}_\textrm{abs}|^2(\bm{p},\bm{q},\bm{p}') (2\pi)^4\delta^{(4)}(p+q-p')f_0(p)b_0(cq)[1-f_0(p')][\bm{\mathcal{X}}(\bm{p})-\bm{\mathcal{X}}(\bm{p}')]^2.
\end{equation}
The factor of $1/2$ comes from the multiplicity of the function $\bm{\mathcal{X}}(\bm{p})$ inside the integral and the factor of 2 is due to the emission term being the same as the absorption one under the exchange of the integration variables $\bm{p} \leftrightarrow \bm{p}'$. We now turn to considering each individual component of $|M^{L\lambda'}_\textrm{abs}|^2(\bm{p},\bm{q},\bm{p}')$.
\subsection{Intranode Absorption/Emission}

In this case, the square of the scattering amplitude has the form
\begin{gather}
    \label{absorption_LL}
    |M^{LL}_\textrm{abs}|^2(\bm{p},\bm{p}',\bm{q}) = |M^{LL}_\textrm{emi}|^2(\bm{p},\bm{p}',\bm{q}) = \frac{Uq^2}{2\omega_{\bm{q}}}\cdot\frac{1+\cos\theta_{pp'}}{2}.
\end{gather}
 Note that intranodal $RR$ emission and absorption scattering amplitudes are the same as (\ref{absorption_LL}). To solve the integral, we undertake the following steps:
\begin{itemize}
    \item The $\bm{p}'$ integration is removed through the momentum delta.
    \item The energy conservation delta is rewritten as 
    \begin{equation}\label{absorption_LL_energy_conservation_delta}
        \delta(p+cq-p') = \frac{p'}{pq}\delta(\cos\theta_{pq} - \cos\theta_0) \quad \textrm{where} \ \ \cos\theta_0 \equiv \frac{q}{2p}\left(c^2 -1\right) + c.
    \end{equation}
    Since $c \ll 1$, this effectively restricts the $q$ integration range to values below $2p$.
    \item Using conservation of momentum, it can be shown that
    \begin{equation}
        \cos\theta_{pp'} = \frac{p}{p'}+\frac{q}{p'}\cos\theta_{pq} = 1 + \frac{q^2}{2pp'}(c^2-1) ,
    \end{equation}
    where the restriction from (\ref{absorption_LL_energy_conservation_delta}) has been used in the second equality.
    \item The equilibrium distribution functions are rearranged:
    \begin{equation}
        b_0(cq) [1-f_0(p')] = [1-f_0(p)][b_0(cq) + f_0(p')].
    \end{equation}
    
    \item Exploting that $cq \ll p,p' \sim p_f$, the scattering amplitude for the process is approximated to
    \begin{equation}
        |M^{LL}_\textrm{abs}|^2(\bm{p},\bm{k},\bm{p}',\bm{k}') \approx \frac{Uq}{2c}\left(1-\frac{q^2}{4p^2}\right).
    \end{equation}
    
    \item Since in the end we use the finite basis (\ref{basis}) to solve the Boltzmann equation, we will need to calculate the components $(\bm{\mathcal{X}}_i, \mathcal{C}_\textrm{e-ph}\bm{\mathcal{X}}_j)$ with $i,j = 0,1$. To do so, we need $|\Delta \bm{\mathcal{X}}|^2_{ij} \equiv [\bm{\mathcal{X}}_i(\bm{p})-\bm{\mathcal{X}}_i(\bm{p}')][\bm{\mathcal{X}}_j(\bm{p})-\bm{\mathcal{X}}_j(\bm{p}')]$ for each component:
    \begin{gather}
        |\Delta\bm{\mathcal{X}}|^2_{00} = \beta^2(\bm{p} - \bm{p}')^2 = \beta^2q^2,\\
        |\Delta\bm{\mathcal{X}}|^2_{11} =  2[1-\cos\theta_{pp'}] = 
        \frac{q^2}{pp'}\left[1-c^2 \right], \\
        |\Delta\bm{\mathcal{X}}|^2_{01} =|\Delta\bm{\mathcal{X}}|^2_{10} = \beta(p+p')[1-\cos\theta_{pp'}] = \beta(p+p')\frac{q^2}{2pp'}\left[1-c^2 \right] .
    \end{gather}
\end{itemize}
Hence, the collision terms are
\begin{equation}
    (\bm{\mathcal{X}}_i, \mathcal{C}_\textrm{e-ph}\bm{\mathcal{X}}_j) = \frac{U}{2c}2(2\pi)^2 \frac{1}{(2\pi)^5} \beta^2\int_0^\infty p^2dpf_0(p)[1-f_0(p)]\int_0^{2p}q^2dq\cdot q\left(1-\frac{q^2}{4p^2}\right) \frac{p'}{pq}[b_0(cq) + f_0(p')]|\Delta\bm{\mathcal{X}}|^2_{ij}.
\end{equation}
The result fundamentally depends on the $q$ integral, which can be solved in two different asymptotic regimes, depending on $\beta cq \sim T_\textrm{BG}/T$.

\subsubsection{High Temperatures}
In this case $\beta T_\textrm{BG}\ll 1$, which implies that for most of the integration region, $\beta cq \ll 1$. Therefore the Bose-Einstein distribution function can be approximated as $b_0(cq) \approx \frac{1}{\beta cq}$ and $f_0(p')$ can be disregarded as it is nonsigular in the small $cq$ limit. One then can show that
\begin{equation}
    \int_0^{2p}q^4dq\cdot q\left(1-\frac{q^2}{4p^2}\right) \frac{p'}{pq}[b_0(cq) + f_0(p')] \approx \beta^{-1}c^{-1}\frac{(2p)^4}{12},
\end{equation}
where we have taken $p' \approx p$. Therefore
\begin{gather}
    (\bm{\mathcal{X}}_i, \mathcal{C}_\textrm{e-ph}\bm{\mathcal{X}}_j)_\textrm{h.t.}^{\textrm{intra}}  = \frac{2^4}{12}\frac{U}{2c^2}2(2\pi)^2 \frac{1}{(2\pi)^5}\beta\int_0^\infty p^2dpf_0(p)[1-f_0(p)]\beta^{2-i-j}p^{4-i-j} \nonumber\\
    \label{appendix_e-ph_collision_integral_intranode_result_ht}
    = \frac{4U}{3(2\pi)^3c^2}\beta^{2-i-j}p_f^{6-i-j} \left[1 + r^{-2}c_2\binom{6-i-j}{2} + O\left(r^{-4}\right)\right],
\end{gather}
where we have expanded up to next-to leading order in the Sommerfeld expansion using the parameter $r \equiv \beta p_f$.

\subsubsection{Low temperatures}
In this case In this case $\beta T_\textrm{BG}\gg 1$, which now implies that the upper limit of integration for the $q$ integral can be taken to infinity. It also means that the $\frac{q^2}{4p^2}$ term from the square of the scattering amplitude can be neglected. In this case, one cannot simply take the lowest order term in $c$, since now terms of the form $p'-p = cq$ can dominate over appearances of the scale $T$. Nevertheless, this can only happen in sums/subtractions of $p$ and $p'$, so terms like $p/p'$ can still be take to be $\approx 1$.The integration of $q$ in this case is then seen to be
\begin{equation}
   \int_0^{2p}q^4dq\cdot q\left(1-\frac{q^2}{4p^2}\right) \frac{p'}{pq}[b_0(cq) + f_0(p')] \approx \beta^{-5}c^{-5}\cdot24\left[\zeta(5)-\textrm{Li}_5\left(-e^{-\beta(p-p_f)}\right)\right].
\end{equation}
Now one has that
\begin{gather}
    (\bm{\mathcal{X}}_i, \mathcal{C}_\textrm{e-ph}\bm{\mathcal{X}}_j)_\textrm{l.t.}^{\textrm{intra}} = 24\frac{U}{2c^6}2(2\pi)^2 \frac{1}{(2\pi)^5}\beta^{-3}\int_0^\infty p^2dpf_0(p)[1-f_0(p)]\left[\zeta(5)-\textrm{Li}_5\left(-e^{-\beta(p-p_f)}\right)\right][1-c^2]^{ij}\beta^{2-i-j}p^{-i-j} \nonumber\\
    \label{appendix_e-ph_collision_integral_intranode_result_lt}
    = \frac{24Ug_5}{(2\pi)^3c^6}[1-c^2]^{i\cdot j}\beta^{-2-i-j}p_f^{2-i-j},
    \end{gather}
where $g_5 \approx 5.2$ is defined in (\ref{definition_gn}).

\subsection{Internode Absorption/Emission}
Now the momentum conservation delta is given by $\delta^{(3)}(\bm{p} + \bm{q} - \bm{p}' - 2\bm{b})$ and the square of the scattering amplitude is
\begin{equation}
    \label{appendix_LR_amplitude_squared}
    |M^{LR}_\textrm{abs}|^2(\bm{p},\bm{p}',\bm{q}) = |M^{LR}_\textrm{emi}|^2(\bm{p},\bm{p}',\bm{q}) = \frac{Uq^2}{2\omega_{\bm{q}}}\cdot\frac{1+\cos\theta_{pp'} -2 \cos\theta_{pb}\cos\theta_{p'b}}{2}.
\end{equation}
We take similar steps as the intranode case:
\begin{itemize}
    \item The $\bm{p}'$ integration is removed through the momentum delta.
    \item The energy conservation delta is rewritten as 
    \begin{equation}\label{absorption_LR_energy_conservation_delta}
        \delta(p+cq-p') = \frac{p'}{pQ}\delta(\cos\theta_{pQ} - \cos\theta_0) \quad \textrm{where} \ \ \cos\theta_0 \equiv \frac{c^2q^2-2cpq-Q^2}{2pQ},
    \end{equation}
    where $\bm{Q} \equiv \bm{q} - 2\bm{b}$.
    Similarly to before, this delta imposes the restriction $Q \le 2p + cq \approx 2p$. However, the restriction 
    \begin{equation}
        cq \le Q,
    \end{equation}
    must also be satisfied, which for large internodal separations is effectively $4cb \ll 2p_f$. For extremely large internodal separations this could pose an issue but we disregard such scenarios, effectively expecting that $c \ll p_f/b$.
    \item We perform the convenient change of variable $\cos\theta_{qb} \rightarrow Q$ by exploiting
    \begin{equation}
        \cos\theta_{qb} = \frac{q^2+(2b)^2-2q(2b)}{2q(2b)}, \quad d\cos\theta_{qb} = -\frac{Q}{q(2b)}.
    \end{equation}
    The lower limit of integration for $Q$ is thus $|q-2b|$ whereas the upper limit is given by $\textrm{min}[q+2b, 2p]$. Clearly the upper limit depends on the value of $q$, as $p\sim p_f$, hence it is convenient to translate this upper limit to the $q$ integration. It translates into two regions:
    \begin{enumerate}
        \item $0 \le q \le \textrm{max}[2p - 2b,0]$ and $|q-2b| \le Q \le q + 2b$.
        \item $|2p-2b| \le q \le 2p + 2b$ and $|q-2b| \le Q \le 2p$.
    \end{enumerate}
    
    \item Using conservation of momentum, it can be shown that
    \begin{equation}
        \cos\theta_{pp'} = \frac{p}{p'}+\frac{Q}{p'}\cos\theta_{pQ} = 1 + \frac{1}{2pp'}(c^2q^2-Q^2) ,
    \end{equation}
    where the restriction from (\ref{absorption_LR_energy_conservation_delta}) has been used in the second equality.
    \item The equilibrium distribution functions are rearranged:
    \begin{equation}
        b_0(cq) [1-f_0(p')] = [1-f_0(p)][b_0(cq) + f_0(p')].
    \end{equation}
    
    \item The cosines with respect to the internodal distance can be reexpressed as
    \begin{gather}
        \cos\theta_{pb}\cos\theta_{p'b} =
   \frac{Q}{p'}\cos\theta_{pb}\cos\theta_{Qb} + \frac{p}{p'}\cos^2\theta_{pb}.
    \end{gather}
    Then, using that
    \begin{gather}
        \cos\theta_{pb} = \cos\theta_{pQ}\cos\theta_{Qb} + \sin\theta_{pQ}\sin\theta_{Qb}\cos\varphi_{Q,pb}\ \textrm{  and  }\ \cos\theta_{Qb} = \frac{q^2-Q^2-(2b)^2}{2Q(2b)},
    \end{gather}
    along with the restriction on $\cos\theta_{pQ}$ (\ref{absorption_LR_energy_conservation_delta}) one can calculate the relevant azimuthal integral of the angular dependence in the square of the scattering amplitude (\ref{appendix_LR_amplitude_squared}),
    \begin{gather}
        \int_{0}^{2\pi}d\varphi_{Q,pb}(1 + \cos\theta_{pp'} -2\cos\theta_{pb}\cos\theta_{p'b}) = 4\pi - \pi\sin^2\theta_{Qb}\left(1+ \frac{3Q^2}{4p_f^2}\right),
    \end{gather}
    where we have taken $p=p'=p_f$, which is effectively imposed by  $f_0(p)[1-f_0(p)]$ at leading order in the Sommerfeld expansion.
    
    \item Since in the end we use the finite basis (\ref{basis}) to solve the Boltzmann equation, we will need to calculate the components $(\bm{\mathcal{X}}_i, \mathcal{C}_\textrm{e-ph}\bm{\mathcal{X}}_j)$ with $i,j = 0,1$. To do so, we need to calculate $|\Delta \bm{\mathcal{X}}|^2_{ij} \equiv [\bm{\mathcal{X}}_i(\bm{p})-\bm{\mathcal{X}}_i(\bm{p}')][\bm{\mathcal{X}}_j(\bm{p})-\bm{\mathcal{X}}_j(\bm{p}')]$ for each component:
    \begin{gather}
        |\Delta\bm{\mathcal{X}}|^2_{00} = \beta^2(\bm{p} - \bm{p}')^2 = \beta^2Q^2,\\
        |\Delta\bm{\mathcal{X}}|^2_{11} =  2[1-\cos\theta_{pp'}] = 
        \frac{1}{pp'}\left[Q^2- c^2q^2 \right], \\
        |\Delta\bm{\mathcal{X}}|^2_{01} =|\Delta\bm{\mathcal{X}}|^2_{10} = \beta(p+p')[1-\cos\theta_{pp'}] = \beta(p+p')\frac{1}{2pp'}\left[Q^2- c^2q^2 \right] .
    \end{gather}
    \item Assuming the internodal separation is much larger than the Fermi radius $b \gg p_f$ implies that for temperatures below $T_\textrm{BG,W} = 2cb$, the integral is suppressed since phonons do not have enough energy to exchange fermions between the two nodes. Hence, only the high-temperature exchange is relevant, which as before, implies that we only need to keep the leading order in $c$, i.e. $p \approx p'$ . Again, $f_0(p+cq)$ can be dropped since it is small in comparison to $b_0(cq) \approx (\beta cq)^{-1} \gg 1$.
\end{itemize}
The collision terms are then
\begin{gather}
    (\bm{\mathcal{X}}_i, \mathcal{C}_\textrm{e-ph}\bm{\mathcal{X}}_j)^{\textrm{inter}}  = \frac{1}{2b}\frac{U}{4c} \frac{1}{(2\pi)^4}\beta^2\int_0^\infty p^2dpf_0(p)[1-f_0(p)]\int_0^{2p-2b}q^2dq[b_0(cq)+f_0(p+cq)]\int_{|q-2b|}^{q+2b}dQ \nonumber \\
    \left[4\pi - \pi\sin^2\theta_{Qb}\left(1+ \frac{3Q^2}{4p_f^2}\right)\right]\frac{p'}{p}|\Delta\bm{\mathcal{X}}|_{ij}^2 \nonumber \\
    + \frac{1}{2b}\frac{U}{4c} \frac{1}{(2\pi)^4}\beta^2\int_0^\infty p^2dpf_0(p)[1-f_0(p)]\int_{|2p-2b|}^{2p+2b}q^2dq[b_0(cq)+f_0(p+cq)]\int_{|q-2b|}^{2p}dQ  \nonumber \\
    \label{appendix_LR_collision_integral_after_manipulations}
    \left[4\pi - \pi\sin^2\theta_{Qb}\left(1+ \frac{3Q^2}{4p_f^2}\right)\right]\frac{p'}{p}|\Delta\bm{\mathcal{X}}|_{ij}^2.
\end{gather}
Notice how for $2b = 0$, only the first term in 
(\ref{appendix_LR_collision_integral_after_manipulations}) is non-zero, whereas for $2b > 2p$, only the second term is. Since effectively $p = p_f$, it makes sense thus to study the regimes $2b > 2p_f$ and $2b \le 2p_f$ separately.
\subsubsection{$2b > 2p_f$}
The high temperature regime is defined by $\beta cq \ll 1$. This occurs for temperatures that satisfy $T \gg T_\textrm{BG,W} \sim c2b$. One can take $b_0(cq) \approx (\beta cq)^{-1}$ dominates over the $f_0(p + cq)$ term. In addition, all terms depending on $cq$ in $|\Delta \bm{\mathcal{X}}|^2_{ij}$ as they are smaller than all other energy scales, $p\sim p'\sim p_f$ and $p'-p\sim T$. Taking the leading order in the Sommerfeld expansion, the collision integral is 
\begin{gather}
    (\bm{\mathcal{X}}_i, \mathcal{C}_\textrm{e-ph}\bm{\mathcal{X}}_j)^{\textrm{inter}}_\textrm{h.t.}  = \frac{1}{2b}\frac{U}{4c^2}\frac{1}{(2\pi)^4}(\beta p_f)^{2-i-j}\int_{2b-2p_f}^{2p_f+2b}qdq\int_{|q-2b|}^{2p_f}Q^2dQ \left[4\pi - \pi\sin^2\theta_{Qb}\left(1+ \frac{3Q^2}{4p_f^2}\right)\right] \nonumber \\
    \label{appendix_LR_collision_integral_result_2b}
    = \frac{1}{2b}\frac{U}{4c^2}\frac{1}{(2\pi)^3}(\beta p_f)^{2-i-j}8(2b)p_f^4 = \frac{2Cp_f^4}{c^2(2\pi)^3}(\beta p_f)^{2-i-j}.
\end{gather}
Notice how the result is indepent on the value of $b$.

On the other hand, for $\beta cq \ll 1$, which occurs when $T \ll T_\textrm{BG,W}$ temperatures, the $q$ integral of the relevant term in (\ref{appendix_LR_collision_integral_after_manipulations}) is generically exponentially suppressed unless $2b$ is fairly close to $2p_f$. Hence, we take the contribution at $2b \ge 2p_f$ of this integral to be zero at low temperatures,
\begin{gather}\label{appendix_LR_collision_integral_result_2b_lt}
    (\bm{\mathcal{X}}_i, \mathcal{C}_\textrm{e-ph}\bm{\mathcal{X}}_j)^{\textrm{inter}}_\textrm{l.t.} \approx 0.
\end{gather}
\subsubsection{$2b \le 2p_f$}
Now we need to consider both integrals in (\ref{appendix_LR_collision_integral_after_manipulations}). Since now $2p_f \ge 2b$, the transition temperature is now given by $T_\textrm{BG}$. Manipulating them in the same way as was done in the high temperature $2p_f < 2b$ case, one finds the same result,
\begin{gather}
    (\bm{\mathcal{X}}_i, \mathcal{C}_\textrm{e-ph}\bm{\mathcal{X}}_j)^{\textrm{inter}}_\textrm{h.t.}  = \frac{1}{2b}\frac{U}{4c^2}\frac{1}{(2\pi)^4}(\beta p_f)^{2-i-j}\int_{0}^{2p_f-2b}qdq\int_{|q-2b|}^{q+2b}Q^2dQ \left[4\pi - \pi\sin^2\theta_{Qb}\left(1+ \frac{3Q^2}{4p_f^2}\right)\right]\nonumber \\
    \label{appendix_LR_collision_integral_result_2p}
    +\frac{1}{2b}\frac{U}{4c^2}\frac{1}{(2\pi)^4}(\beta p_f)^{2-i-j}\int_{2p_f-2b}^{2p_f+2b}qdq\int_{|q-2b|}^{2p_f}Q^2dQ \left[4\pi - \pi\sin^2\theta_{Qb}\left(1+ \frac{3Q^2}{4p_f^2}\right)\right]
    = \frac{2Up_f^4}{c^2(2\pi)^3}(\beta p_f)^{2-i-j}.
\end{gather}
For low temperatures $T\ll T_\textrm{BG}$ we find that, as in the $b > 2p_f$ case, the the second integral in (\ref{appendix_LR_collision_integral_after_manipulations}) is exponentially suppressed unless $2b \rightarrow 2p_f$, a case we again do not consider. Hence we keep only the first integral. To make the problem tractable analytically, we note that in this regime the Bose-Einstein distribution forces $cq \sim T$. Since in this regime we expect $b,p_f \gg T/c \sim q$, we can keep the lowest order in $q$. Due to the limits of the $Q$ integral, this implies that $Q\sim 2b$. This implies that $\cos\theta_{bQ} \approx 1$, meaning of course that $\sin\theta_{bQ} \approx 0$. Therefore, the collision integral in this regime is
\begin{gather}
 (\bm{\mathcal{X}}_i, \mathcal{C}_\textrm{e-ph}\bm{\mathcal{X}}_j)^{\textrm{inter}}_\textrm{l.t.}  \approx 4\pi \frac{1}{2b}\frac{U}{4c} \frac{1}{(2\pi)^4}\beta^2\int_0^\infty p^2dpf_0(p)[1-f_0(p)]\int_0^{2p-2b}q^2dq[b_0(cq)+f_0(p+cq)]\int_{2b-q}^{q+2b}dQ|\Delta\bm{\mathcal{X}}|_{ij}^2 \nonumber \\
\approx 4\pi \frac{1}{2b}\frac{U}{4c} \frac{1}{(2\pi)^4}\beta^2\int_0^\infty p^2dpf_0(p)[1-f_0(p)]\int_0^{2p-2b}q^2dq[b_0(cq)+f_0(p+cq)]2q|\Delta\bm{\mathcal{X}}|_{ij}^2(Q = 2b) \nonumber \\
 = \pi \frac{1}{2b}\frac{2U}{c^5} \frac{1}{(2\pi)^4}\beta^{-2-i-j}\int_0^\infty p^{2-i-j}dpf_0(p)[1-f_0(p)] \nonumber \\\left\{(2b)^26\left[\zeta(4)-\textrm{Li}_4\left(-e^{\beta(p-p_f)}\right)\right] -\delta^{i+j>0}\beta^{-2}120\left[\zeta(6)-\textrm{Li}_6\left(-e^{-\beta(p-p_f)}\right)\right]\right\} \nonumber \\
 \label{appendix_LR_collision_integral_result_2p_lt}
 = \pi \frac{1}{2b}\frac{2U}{c^5} \frac{1}{(2\pi)^4}\beta^{-2-i-j}p_f^{2-i-j}\left[6g_4(2b)^2 - 120g_6\beta^{-2}\delta^{i+j>0}\right],
\end{gather}
where $\delta^{i+j>0}$ is vanishing for $i=j=0$ and is equal to 1 otherwise, and $g_n$ is a numeric constant defined in $\ref{definition_gn})$.
Note that (\ref{appendix_LR_collision_integral_result_2b}) is suppressed by the small factor of $c$ with respect to its intranode version (\ref{appendix_e-ph_collision_integral_intranode_result_lt}), so it can effectively be ignored.

\section{Electron-Electron Collisions}\label{appendix_electron-electron}
The collision integral describing 2-body fermion collisions is given by
\begin{gather}
   \mathcal{I}_\textrm{e-e}[f_a](\bm{p}) = \frac{1}{2}\sum_{b,c,d}\int_{\bm{k,p',k'}}|M^{ab}_{cd}|^2(\bm{p},\bm{k},\bm{p}',\bm{k}') (2\pi)^4\delta^{(4)}(p+k-p'-k')\times \nonumber \\
   \label{starting_ee_collision_integral}
    \left\{f_a(\bm{p})f_b(\bm{k})[1-f_c(\bm{p'})][1-f_d(\bm{k'})] - f_c(\bm{p'})f_d(\bm{k'})[1-f_a(\bm{p})][1-f_b(\bm{k})]  \right\},
\end{gather}
where $a,b,c,d = L,R$ and the leading one half factor cancels the multiplicity of interband scattering from the sum and provides the correct symmetry factor for intraband scattering. Following the same procedure as in the collision integral for phonon emission/absorption of linearizing the distribution functions and introducing the inner product, the corresponding collision term in the variational functional (\ref{action_functional}) is 
\begin{gather}\label{appendix_ee_collision_integral}
    (\bm{\mathcal{X}}, \mathcal{C}_\textrm{e-e}\bm{\mathcal{X}}) =\frac{\beta^2}{8}\sum_{a,b,c,d}\int_{\bm{p,k,p',k'}}|M^{ab}_{cd}|^2(\bm{p},\bm{k},\bm{p}',\bm{k}') (2\pi)^4\delta^{(4)}(p+k-p'-k')f(p)f(k)[1-f(p')][1-f(k')]|\Delta\mathcal{X}|^2,
\end{gather}
where $|\Delta\mathcal{X}|^2 \equiv |\bm{\mathcal{X}}(\bm{p})+\bm{\mathcal{X}}(\bm{k})-\bm{\mathcal{X}}(\bm{p}')-\bm{\mathcal{X}}(\bm{k}')|^2$ and the extra factor of $1/4$ with respect to (\ref{starting_ee_collision_integral}) takes into account the multiplicity of the perturbation. As in the case of phonon emission/absorption, it is useful to consider intranode and internode scattering separately.

\subsection{Intranode Scattering}
Here we consider the case where the scattering is between left-handed fermions. For now we do not assume any particular behavior in any temperature regime, so the square of the scattering amplitude is $|M^{LL}_{LL}|^2 = |A^{LL}_{LL,t}|^2+|A^{LL}_{LL,u}|^2-2\textrm{Re}\left\{A^{LL*}_{LL,t}A^{LL}_{LL,u}\right\}$. Here $A^{LL*}_{LL,t}$ is the same as $A_t$ given in the main text \label{scattering_amplitude_ee}. To deal with the collision integral, we perform the following manipulations:
\begin{itemize}
    \item The $\bm{k}'$ integration is removed through the momentum conservation delta.
    \item The integrals of the non-interference terms $|A^{LL}_{LL,t}|^2$ and $|A^{LL}_{LL,u}|^2$  are the same, as can be seen by performing a relabelling $\bm{p}' \leftrightarrow \bm{k}'$. Furthermore, we now perform a change of integration variables from the remaining outgoing electron momentum to the virtual phonon momentum, i.e. $\bm{p}' \rightarrow \bm{q} = \bm{p}' - \bm{p}$.
    \item Introduce the virtual phonon energy $\omega$ through
    \begin{gather}
        \delta(p+k-p'-k') = \int_{-\infty}^\infty d\omega \ \delta(\omega - p'+p)\delta(\omega - k+ k') \nonumber \\
        \label{appendix_ee_intraband_energy_conservation}
        = \int_{-q}^q d\omega\ \frac{p'}{pq}\delta\left( \frac{\omega}{q} + \frac{\omega^2-q^2}{2pq} -\cos\theta_{pq} \right)\frac{k'}{kq}\delta\left(\frac{\omega}{q} - \frac{\omega^2-q^2}{2kq} -\cos\theta_{kq} \right)\Theta\left(p - \frac{q-\omega}{2}\right)\Theta\left(k - \frac{q+\omega}{2}\right),
    \end{gather}
    where the integration limits on $\omega$ and the theta functions are needed to ensure the cosines have zeros for $|\cos\theta|\le 1$.
    \item Employ the relation
    \begin{gather}
        f_0(p)[1-f_0(p+\omega)]f_0(k)[1-f_0(k-\omega)]= b_0(\omega)[1+b_0(\omega)][f_0(p)-f_0(p+\omega)][-f_0(k)+ f_0(k-\omega)] \nonumber \\
        = (\beta\omega)^2f_0(p)[1-f_0(p)]f_0(k)[1-f_0(k)]
    \end{gather}
    To make the problem more tractable analytically, we have taken the limit where $\omega \ll p_f$ in the last equality, which as we shall shortly see, is well justified.
    
\end{itemize}
These manipulations result in
\begin{gather}
    (\bm{\mathcal{X}}, \mathcal{C}_\textrm{e-e}\bm{\mathcal{X}})^\textrm{intra} =\frac{2}{(2\pi)^6}\frac{\beta^2}{8}\int_0^{k_\textrm{D}} dq\int_{-q}^qd\omega\ 
\int_0^{2\pi}d\varphi_{q,pk} \int_{\frac{q-\omega}{2}}^\infty p dp \int_{\frac{q+\omega}{2}}^\infty k dk \nonumber \\
\label{appendix_intermediate_ee_collision_integral}
|M_{LL}^{LL}|^2(\beta\omega)^2b_0(\omega)[1+b_0(\omega)]f_0(p)[1-f_0(p)]f_0(k)[1-f_0(k)]|\Delta \mathcal{X}|^2,
\end{gather}
which is the same expression as the one given in (\ref{intermediate_collision_integral}). One can see that the non-interference terms will contain the integral
\begin{equation}\label{appendix_omegaintegral}
    \int_{-q}^qd\omega\left\lvert\frac{1}{\omega^2 - \omega_{\bm{q}}^2 +i\omega_{\bm{q}}\Gamma_{\bm{q}}}\right\rvert^2 b_0(\omega)[1+b_0(\omega)]\cdot F(\omega) = \int_{-q}^qd\omega\frac{ b_0(\omega)[1+b_0(\omega)]}{[(\omega - cq)^2 + \frac{\Gamma^2}{4}][(\omega + cq)^2 + \frac{\Gamma^2}{4}]} \cdot F(\omega),
\end{equation}
where $F(\omega)$ is some function of $\omega$ with no singularities on the complex plane. The other functions however, do contain sigularities. The propagators provide four poles
\begin{equation}
    \omega_{+,\pm} = \pm cq + i\frac{\Gamma}{4}, \quad \omega_{-,\pm} = \pm cq - i\frac{\Gamma}{4},
\end{equation}
whereas the Bose-Einstein distribution functions have poles at the Matsubara frequencies, i.e.
\begin{equation}\label{matsubara_frequencies}
    \omega_n = i(2n\pi)\beta^{-1}, \quad \forall n \in \mathbb{Z}.
\end{equation}
A similar situation is found for the interference terms.

\subsubsection{High Temperatures} \label{appendix_ee_intra_high_T}
For now we keep focusing on the non-interference terms, whose exclusion we justify afterwards. This amounts to considering a square scattering amplitude of the form,
\begin{gather}
    |M^{LL}_{LL}|^2_{\textrm{h.t.}}(\bm{p},\bm{k},\bm{p}',\bm{k}') = |M^{RR}_{RR}|^2_{\textrm{h.t.}}(\bm{p},\bm{k},\bm{p}',\bm{k}') = \frac{2U^2}{4}\left\lvert\frac{q^2}{\omega^2 - \omega_{\bm{q}}^2 +i\omega_{\bm{q}}\Gamma_{\bm{q}}}\right\rvert^2\left[1 +\cos\theta_{pp'}\right]\left[1 +\cos\theta_{kk'}\right].
\end{gather}
At temperatures $T\gg c\langle q\rangle = T_\textrm{BG}$, the poles will satisfy the hierarchy $\omega_n \gg \omega_{\pm,\pm}$ for all values of $n$. Therefore, the integral (\ref{appendix_omegaintegral}) can be performed by using a complex contour on the upper half of the complex plane that only encloses the propagator poles $\omega_{+,\pm}$. To do this, first perform a change of variables $\omega \rightarrow \omega/q$, to recover
\begin{equation}\label{appendixomegaintegral2}
    I_\omega = \int_{-1}^1 \frac{f(x)}{[(x - c)^2 + \frac{\Gamma^2}{4q^2}][(x + c)^2 + \frac{\Gamma^2}{4q^2}]}dx,
\end{equation}
given in (\ref{omegaintegral}). Note that $f(x)$ is the function that has absorbed all the non-propagator factors appearing in (\ref{appendix_omegaintegral}) and is not to be confused with the Fermi-Dirac distribution. A contour can now be chosen such that
\begin{equation}
    I_\omega  = 2\pi i \sum_{z= z_{+,\pm}} \textrm{Res}\left[\frac{f(z)}{[(z - c)^2 + \frac{\Gamma^2}{4q^2}][(z + c)^2 + \frac{\Gamma^2}{4q^2}]},z\right] - I_C,
\end{equation}
where $I_C$ is the integral over the path that closes the contour and $x_{+,\pm} \equiv \omega_{+,\pm}/q$. Since $c\ll1$ and $\frac{\Gamma}{4q} \ll 1$, the closing path can be chosen such that the variable of integration $z \gg z_{+,\pm}$. Therefore, $I_C = O\left(x_{+,\pm}^0\right)$. To calculate the residues, one formally expands the function $f(x)$ around 0, i.e. $f(z) = f(0) + f'(0)z+\frac{1}{2}f''(0)z^2 + O(z^3)$. Keeping only the next-to-leading order in the small parameters $z_{+,\pm}$ implies that one needs the expansion of $f(z)$ up to the quadratic term, whose integral is of order $O\left(x_{+,\pm}^{-1}\right)$. Thus, $I_C$ is subleading and can be ignored, so that finally the expression in (\ref{omegaintegral_result}) is recovered, i.e.
\begin{equation}\label{appendix_omegaintegral_result}
     I_\omega = \frac{\pi q}{\Gamma}\left[\frac{1}{c^2}f(0)+\frac{1}{2}f''(0)\right].
\end{equation}

The condition $x = \omega/q = 0$ imposed by the lowest order term in $I_\omega$ implies that $p = p'$ and $k = k'$. In this limit, one can show that
\begin{gather}
    \bm{\mathcal{X}}(\bm{p}) - \bm{\mathcal{X}}(\bm{p}') = 
    \mathcal{X}(p)\hat{\bm{p}} - \mathcal{X}(p')\hat{\bm{p}}' = -\frac{\mathcal{X}(p)}{p}\bm{q}, \nonumber \\
    \bm{\mathcal{X}}(\bm{k}) - \bm{\mathcal{X}}(\bm{k}') = 
    \mathcal{X}(k)\hat{\bm{k}} - \mathcal{X}(p')\hat{\bm{k}}' = \frac{\mathcal{X}(k)}{k}\bm{q} ,
\end{gather}
so the perturbation-dependent term becomes
\begin{gather}
    |\Delta \mathcal{X}|^2\biggr|_{\substack{\omega = 0}} = q^2\left[X(p)-X(k)\right]^2 ,
\end{gather}
where $X(p) = \mathcal{X}(p)/p$. It is simple to see that at the lowest order in the Sommerfeld expansion, where $p=k=p_f$, this expression is vanishing. To calculate the leading order contribution, it is convenient expand around the Fermi momentum as is mentioned in the main text,
\begin{equation}
    X(p) = X(p_f) + \delta_pX'(p_f) + O(\delta_p^2),
\end{equation}
where $\delta_p \equiv p - p_f$. By taking this expansion we are implicitly assuming that $X'(p_f) \ne 0$. For this to happen, the perturbation should satisfy $\mathcal{X}(p) \propto (p-p_f)^n$ for $n>1$. We do not expect this to occur since the source term in the Boltzmann equation is at most $\propto (p-p_f)^1$, in the case of a thermal gradient (\ref{linearized_vector_boltzmann_equation}).
Finally, is simple to show that under this expansion
\begin{equation}
    |\Delta \mathcal{X}|^2\biggr|_{\substack{\omega = 0}} = X'(p_f)^2\left[\delta_p^2 + \delta_k^2 - 2\delta_p\delta_k\right] + O(\delta^3).
\end{equation}
The fact that the leading order term in the $c$ expansion in $I_\omega$ is vanishing in the Sommerfeld expansion explains why we retain the next-to-leading order term in the $c$ expansion. It can be shown that this term, proportional to $f''(0)$ leads to a non-vanishing term at leading order in the Sommerfeld expansion, as the second derivative of the perturbation term in this case is
\begin{gather}
\frac{q^2}{2}\frac{d^2|\Delta \mathcal{X}|^2}{d\omega^2}\biggr|_{\substack{p=k=p_f \\\omega = 0}} = 2q^2p_f^2\left[\chi'(p_f) - \frac{\chi(p_F)}{p_f}\right]^2\left[1 + \frac{q^2}{4p_f^2} - \left(1 - \frac{q^2}{4p_f^2}\right)\cos\varphi_{q,pk}\right] \nonumber \\
= 2q^2p_f^2\left[\chi'(p_f) - \frac{\chi(p_f)}{p_f}\right]^2\left[1 + \frac{q^2}{4p_f^2}\right] = 2q^2p_f^2\left[1 + \frac{q^2}{4p_f^2}\right]X'(p_F)^2,
\end{gather}
where values of $\cos\theta_{pq}$ and $\cos\theta_{kq}$ fixed by energy conservation (\ref{appendix_ee_intraband_energy_conservation}) have been used to evaluate 
\begin{equation}\label{appendix_cospk}
    \cos\theta_{pk} = \cos\theta_{pq}\cos\theta_{kq}+\sin\theta_{pq}\sin\theta_{kq}\cos\varphi_{q,pk}.
\end{equation} The second equality is obtained by noticing that the rest of the integral has no terms depending on $\cos\varphi_{q,pk}$, so its integration over $\varphi_{q,pk}$ is vanishing. No derivatives over $\omega$ need to be taken elsewhere to evaluate $f''(0)$, as $\Delta |\mathcal{X}|'$ is also vanishing in the $p=k=p'=k'=p_f$ limit.Hence, the rest of the integrand, including the square of the scattering amplitude, can be directly evaluated at $\omega = 0$. Then, at the end of the day one finds
\begin{gather}
    (\bm{\mathcal{X}}, \mathcal{C}_\textrm{e-e}\bm{\mathcal{X}})^\textrm{intra}_\textrm{h.t.} = 2U^2\cdot\frac{\pi}{c^2}\frac{2}{(2\pi)^5}\frac{\beta^2}{8}\int_0^{2p_f}dq\frac{q^4}{\Gamma(q)}\nonumber \\
\times\int_{0}^\infty  dp\ f_0(p)[1-f_0(p)]p^2\left(1- \frac{q^2}{4p^2}\right)\int_{0}^\infty  dk\ f_0(k)[1-f_0(k)]k^2\left(1- \frac{q^2}{4k^2}\right) \nonumber \\
    \times X'(p_f)^2\left[\left(\delta_p^2 + \delta_k^2 -2\delta_p\delta_k\right) + 2c^2p_f^2\left(1 + \frac{q^2}{4p_f^2}\right)\right],
\end{gather}
where we have used that $\omega = 0$, momentum conservation and the fixed cosines from energy conservation (\ref{appendix_ee_intraband_energy_conservation}) to show that
\begin{equation}
    (1+\cos\theta_{pp'}) = 2\left(1-\frac{q^2}{4p^2}\right), \quad (1+\cos\theta_{kk'}) = 2\left(1-\frac{q^2}{4k^2}\right).
\end{equation}
Now, evaluating the $p$ and $k$ integrals, the $\delta_p\delta_k$ as the rest of the integrand is even in $\delta_p$ and $\delta_k$ at leading order in the Sommerfeld expansion. The resulting integral is
\begin{gather}\label{appendix_ee_intra_q_integral}
    (\bm{\mathcal{X}}, \mathcal{C}_\textrm{e-e}\bm{\mathcal{X}})^\textrm{intra}_\textrm{h.t.} =  X'(p_f)^2(Up_f)^2\cdot\frac{\pi}{c^2}\frac{1}{(2\pi)^5}\int_0^{2p_f}dq\ \frac{q^4}{\Gamma(q)}\left(1- \frac{q^2}{4p_f^2}\right)^2\left[c_2\beta^{-2}+ c^2p_f^2\left(1 + \frac{q^2}{4p_f^2}\right)\right].
\end{gather}
Notice that choosing $2p_f \ll k_\textrm{D}$ implies that the second term in the square brackets is suppressed in the high temperature regime $\beta T_\textrm{BG} \sim \beta cp_f \ll 1$, which is in any case the regime where this expression is valid. Had we chosen $k_\textrm{D}\ll 2p_f$, then there would be a regime of temperatures $T_\textrm{D} \ll T \ll T_\textrm{BG}$ where the second term in (\ref{appendix_ee_intra_q_integral}) would dominate. We only consider the $2p_f < k_\textrm{D}$, and we will hence drop this term.

To proceed now, an explicit value for $\Gamma(q)$ needs to be given. We choose
\begin{equation}
    \Gamma(q) = \Gamma_\textrm{e-ph}(q) + \gamma,
\end{equation}
where $\gamma$ is some linewidth, representing some other channel through which phonons can decay, e.g. impurities, and the phonon linewidth $\Gamma_\textrm{e-ph}(q)$ from interactions with electrons is given by the intraband contribution (\ref{appendix_phonon_linewidth_intra}), although omitting for now the extra factor of two from considering both left and right contributions. Then,
\begin{equation}
     (\bm{\mathcal{X}}, \mathcal{C}_\textrm{e-e}\bm{\mathcal{X}})^\textrm{intra}_\textrm{h.t.} =  X'(p_f)^2(Up_f^2)\cdot\frac{1}{c^2}\frac{(2p_f)^4}{(2\pi)^3}A(\tilde{\gamma})c_2\beta^{-2},
\end{equation}
where
\begin{equation}
\int_0^{2p_f}dq\frac{q^4\left(1- \frac{q^2}{4p_f^2}\right)^2}{\frac{Cp_f^2}{4\pi}q\left(1-\frac{q^2}{4p_f^2}\right) + \gamma} = \frac{4\pi}{(Cp_f^2)}\left(2p_f\right)^4\int_0^1dx\ \frac{x^4(1-x^2)^2}{x(1-x^2) + \tilde{\gamma}} \equiv \frac{4\pi}{(Cp_f^2)}\left(2p_f\right)^4 A(\tilde{\gamma}),    
\end{equation}
and $\tilde{\gamma} = \frac{2\pi\gamma}{p_f(Up_f^2)}$. In the case where the dominant phonon decay channel is given by the electron-phonon interaction, then $\tilde{\gamma} = 0$ and $A(0) = \frac{1}{12}$, hence
\begin{equation}\label{appendix_ee_intraband_high_T_result}
    (\bm{\mathcal{X}}, \mathcal{C}_\textrm{e-e}\bm{\mathcal{X}})^\textrm{intra}_\textrm{h.t.} =  \left[X'(p_f)p_f^2\right]^2(Up_f^2)\frac{4c_2\beta^{-2}}{3(2\pi)^3c^2}
\end{equation}
i.e. the same expression as that of (\ref{ee_intraband_high_T_result}). \\

To finalize the discussion of intraband fermion scattering we now prove that the interference term in $|M^{LL}_{LL}|^2$ is subleading with respect to the non-interference terms. Fundamentally, the interference term $2\textrm{Re}\left\{A_t^*A_u\right\}$ includes the product of the $t$ and $u$ channel phonon propagators,
\begin{equation}
|M^{LL}_{LL}|^2_\textrm{interference} \sim \left[ \frac{q_t^2}{\omega_t^2 - \omega_{{\bm{q}}_t}^2 + i\eta_t}\frac{q_u^2}{\omega_u^2 - \omega_{{\bm{q}}_u}^2 - i\eta_u} + c.c.\right] ,
\end{equation}
where $\omega_t = p' - p$, $\omega_u = k' - p$, $\bm{q}_t = \bm{p}'-\bm{p}$ and $\bm{q}_u = \bm{k}' - \bm{p}$ and the $\eta$ include the phonon linewidth. To perform the corresponding collision integral (\ref{appendix_ee_collision_integral}) of this term, we proceed with the following manipulations:
\begin{itemize}
    \item Remove the $\bm{k}$ integration through the momentum conservation delta. Then two changes of integration variables are performed, $\bm{p}' \rightarrow \bm{q}_t$ and $\bm{k}' \rightarrow \bm{q}_u$
    \item The energy conservation delta is now dealt with by introducing two dummy variables $\omega_t$ and $\omega_u$:
    \begin{gather}
        \delta(p+k-p'-k') = \int_{-\infty}^\infty d\omega_t\int_{-\infty}^\infty d\omega_u \ \delta(\omega_t - p' + p)\delta(\omega_u - k' + p)\delta(k - \omega_t -\omega_u - p) \nonumber \\
        \label{appendix_ee_intraband_energy_conservation}
        = \int_{-q_t}^{q_t} d\omega_t \int_{-q_u}^{q_u} d\omega_u\ \frac{p'}{pq_t}\delta\left( \frac{\omega}{q_t} + \frac{\omega_t^2-q_t^2}{2pq_t} -\cos\theta_{pq_t} \right)\frac{k'}{pq_u}\delta\left(\frac{\omega_u}{q_u} + \frac{\omega_u^2-q_u^2}{2pq_u} -\cos\theta_{pq_u} \right)\frac{k}{q_tq_u}\delta\left(\cos\theta_{q_tq_u} - \frac{\omega_t\omega_u}{q_tq_u}\right) \nonumber \\
        \Theta\left(p - \frac{q_t-\omega_t}{2}\right)\Theta\left(p - \frac{q_u-\omega_u}{2}\right).
    \end{gather}
    \item The relevant angular integrations will be over $\cos\theta_{pq_t}$, $\cos\theta_{pq_u}$ and $\varphi_{p,q_tq_u}$. The remaining cosine appearing in the deltas above can be expressed as a combination of the former variables, $\cos\theta_{q_tq_u} = \cos\theta_{pq_u}\cos\theta_{pq_t} + \sin\theta_{pq_u}\sin\theta_{pq_t}\cos\varphi_{p,q_tq_u}$. Hence, the delta for $\cos\theta_{q_tq_u}$ is implicitly a delta for $\varphi_{p,q_tq_u}$.
    \item The relevant poles will come from the integrals
    \begin{equation}
        \int_{-q_t}^{q_t} d\omega_t \int_{-q_u}^{q_u} d\omega_u \left[ \frac{q_t^2}{\omega_t^2 - \omega_{\bm{q}_t}^2 + i\eta_t}\frac{q_u^2}{\omega_u^2 - \omega_{\bm{q}_u}^2 - i\eta_u} + c.c.\right]f(\omega_t,\omega_u,...) .
    \end{equation}
    Assuming the function $f(\omega_t,\omega_u,...)$ has no further poles near the small poles of the propagator, which is the case in the present regime $T\gg T_\textrm{BG}$, we can use the same contour procedure as in the case of the non-interference terms (\ref{appendix_omegaintegral}). Expanding $f$ as a power series of $\omega_t$ and $\omega_u$ around $\omega_t = \omega_u = 0$, one can show that this integral is proportional to a power of $-2$ of the small parameters $c$ and $\eta$. Such a contribution is subleading with respect to the non-interference terms, where the small parameters appear as a power of $-3$ (\ref{appendix_omegaintegral_result}), so it can be neglected. This in turn justifies the form (\ref{generic_high_temperature_amplitude}), where only the square of the scattering amplitudes $|A_t|^2$ and $|A_u|^2$ are taken into account, ignoring the interference term.
\end{itemize}

\subsubsection{Low temperatures}
When $T\lesssim T_\textrm{BG}$, the Matsubara poles from the Bose-Einstein distribution function can no longer be left out of the contour of integration used to calculate (\ref{appendix_omegaintegral}). As the temperature decreases, more and more Matsubara poles enter the contour, and in fact become smaller than the propagator poles $\sim cq \sim T_\textrm{BG}$. Therefore, we expect that these poles will dominate the integral, which will translate to $\omega \sim T$ in the denominator of the phonon propagator. By the definition of this regime, $\omega^2 \sim T^2 \ll \omega_{\bm{q}}^2 \sim T_\textrm{BG}^2$, so we effectively take the phonon propagator (\ref{phonon_propagator}) to be
\begin{equation}\label{appendix_low_T_phonon_propagator}
    \mathcal{D}(\bm{q}) \approx  \frac{2\omega_{\bm{q}}}{- \omega_{\bm{q}}^2} + O\left(\frac{\omega}{\omega_{\bm{q}}}\right),
\end{equation}
where we have also omitted the phonon linewidth, since it is no longer needed to find a finite solution. The form of (\ref{appendix_low_T_phonon_propagator}), along with the $q$ factor coming from the phonon coupling (\ref{electron_phonon_coupling}) will make the scattering amplitudes $A_t$ and $A_u$ independent of the phonon momentum, $\bm{q}_t$ and $\bm{q}_u$ respectively. This facilitates the calculation of the square of the scattering amplitudes in this temperature regime
\begin{equation}
    \label{appendix_low_T_intraband_amplitude_squared}
    |M^{LL}_{LL}|^2_{\textrm{l.t.}}(\bm{p},\bm{k},\bm{p}',\bm{k}') = |M^{RR}_{RR}|^2_{\textrm{l.t.}}(\bm{p},\bm{k},\bm{p}',\bm{k}') = \frac{U^2}{4c^4}\left[1 -\cos\theta_{pk}\right]\left[1 -\cos\theta_{p'k'}\right],
\end{equation}
 and the computation of the collision integral itself (\ref{appendix_intermediate_ee_collision_integral}). Substituting (\ref{appendix_low_T_intraband_amplitude_squared}) into it, one obtains
\begin{gather}
     (\bm{\mathcal{X}}, \mathcal{C}_\textrm{e-e}\bm{\mathcal{X}})^\textrm{intra}_\textrm{l.t.} = \frac{U^2}{4c^4}\frac{2}{(2\pi)^6}\frac{\beta^2}{8}\int_0^{k_\textrm{D}}  dq\int_{-q}^qd\omega (\beta\omega)^2b_0(\omega)[1+b_0(\omega)]
   \int_0^{2\pi}d\varphi_{q,pk} \nonumber \\
   \label{appendix_itermediate_low_T_ee_collision_integral}
    \times\int_{\frac{q-\omega}{2}}^\infty  dp\ p^2\int_{\frac{q+\omega}{2}}^\infty  dk\ k^2\ \left(1 - \cos\theta_{pk}\right)^2 f_0(p)f_0(k)[1-f_0(p)][1-f_0(k)]|\Delta \mathcal{X}|^2.
\end{gather}
Since $\omega \sim T \ll p,k \sim p_f$, we again expect that effectively $\omega = 0$ everywhere. However, as in the high temperature case, the perturbation term $|\Delta \mathcal{X}|^2$ vanishes at the lowest order in the Sommerfeld expansion, so we again need to consider the next-to leading order term in $\omega$:
\begin{equation}\label{appendix_ee_lowT_perturbation}
    |\Delta \mathcal{X}|^2 = |\Delta \mathcal{X}|^2\biggr|_{\substack{\omega = 0}} + \frac{\omega^2}{2}\frac{d^2|\Delta \mathcal{X}|^2}{d\omega^2}\biggr|_{\substack{p=k=p_f \\\omega = 0}} + O(\omega^3).
\end{equation}
As was mentioned in the high temperature calculation, the first derivative of $|\Delta \mathcal{X}|^2$ is vanishing when $p =k= p' = k' = p_f$, so the second derivative is needed. This results in
\begin{equation}
    |\Delta \mathcal{X}|^2 = X'(p_f)^2\left[q^2(\delta_p^2 + \delta_k^2 -2\delta_p\delta_k) + 2\omega^2p_f^2\left(1 - \cos\theta_{pk}\right)\right].
\end{equation}
Plugging it back into (\ref{appendix_itermediate_low_T_ee_collision_integral}), one can now simply carry out the $\omega$, $p$ and $k$ integrations. For the $\varphi_{q,pk}$ integration, one can use (\ref{appendix_cospk}) and the restrictions on $\cos\theta_{pq}$ and $\cos\theta_{kq}$ (\ref{appendix_ee_intraband_energy_conservation}) evaluated at $p=k=p_f$ and $\omega = 0$, i.e.
\begin{equation}\label{appendix_cospk_with_restrictions}
    \cos \theta_{pk} = - \frac{q^2}{4p_f^2} + \left(1-\frac{q^2}{4p_f^2}\right)\cos\varphi_{q,pk},
\end{equation}
to calculate the integrals
\begin{gather}
    \int_0^{2\pi}d\varphi_{q,pk}\ \left(1-\cos\theta_{pk}\right)^2 = 2\pi\left(1+\frac{q^2}{4p_f^2}\right)^2 + \pi\left(1-\frac{q^2}{4p_f^2}\right)^2 ,\\
    \int_0^{2\pi}d\varphi_{q,pk}\ \left(1-\cos\theta_{pk}\right)^3 = 2\pi\left(1+\frac{q^2}{4p_f^2}\right)^3 + 3\pi\left(1+\frac{q^2}{4p_f^2}\right)\left(1-\frac{q^2}{4p_f^2}\right)^2.
\end{gather}
One then obtains    
\begin{gather}
     (\bm{\mathcal{X}}, \mathcal{C}_\textrm{e-e}\bm{\mathcal{X}})^\textrm{intra}_\textrm{l.t.} = \left[X'(p_f)^2 p_f^2\right]^2\frac{U^2}{4c^4}\frac{2}{(2\pi)^6}\frac{\beta^{-3}}{8}\int_0^{2p_f}  dq \nonumber \\
     \label{appendix_ee_intra_lowT_intermediate}
    \left\{4c_2^2q^2\left[2\pi\left(1+\frac{q^2}{4p_f^2}\right)^2 + \pi\left(1-\frac{q^2}{4p_f^2}\right)^2\right] + 2b_4p_f^2\left[2\pi\left(1+\frac{q^2}{4p_f^2}\right)^3 + 3\pi\left(1+\frac{q^2}{4p_f^2}\right)\left(1-\frac{q^2}{4p_f^2}\right)^2\right]\right\}.
\end{gather}
Finally, we use the following integrals
\begin{gather}
  \int_0^{2p_f}dq \ q^2\left[2\pi\left(1+\frac{q^2}{4p_f^2}\right)^2 + \pi\left(1-\frac{q^2}{4p_f^2}\right)^2\right] = \frac{2^6\pi}{35}(2p_f)^3 ,\\
\int_0^{2p_f}dq \ p_f^2\left[2\pi\left(1+\frac{q^2}{4p_f^2}\right)^3 + 3\pi\left(1+\frac{q^2}{4p_f^2}\right)\left(1-\frac{q^2}{4p_f^2}\right)^2\right] = \frac{2^6\pi}{35}(2p_f)^3  ,
\end{gather}
to obtain
\begin{gather}
     (\bm{\mathcal{X}}, \mathcal{C}_\textrm{e-e}\bm{\mathcal{X}})^\textrm{intra}_\textrm{l.t.} = \left[X'(p_f)^2 p_f^2\right]^2\frac{U^2}{4c^4}\frac{2}{(2\pi)^6}\frac{\beta^{-3}}{8}\frac{2^6\pi}{35}(2p_f)^3  \left[4c_2^2 + 2b_4\right] = \left[X'(p_f)^2 p_f^2\right]^2\frac{2^4U^2p_f^3\beta^{-3}}{35(2\pi)^5c^4}\left[4c_2^2 + 2b_4\right],
\end{gather}
which is the same expression as the one given in the main text (\ref{ee_intraband_low_T_result}).

\subsection{Generic High Temperature Scattering}\label{appendix_generic_high_temperature_scattering}
Instead of performing the calculation for high-temperature $T \gg T_\textrm{BG}$ specifically for electron-electron scattering between different Weyl nodes, we will provide a derivation for a generic system and show that a similar result to (\ref{appendix_ee_intraband_high_T_result}) is found.  In that discussion it was shown that interference terms are negligible in the regime where $c\bm{q} \ll T$. We can therefore write the generic collision term for a generic system with arbitrary band structure ignoring such interference terms,
\begin{gather}
 (\bm{\mathcal{X}}, \mathcal{C}_\textrm{e-e}\bm{\mathcal{X}})_\textrm{h.t.} = 2\sum_{abcd}\frac{1}{8}\cdot(2\pi)\cdot \beta^2\int_{\bm{q},\bm{p},\bm{k}}\int_{-\infty}^{\infty}d\omega\ \left\lvert\frac{2\omega_{\bm{q}}g_{-\bm{q}}^{b\bm{k},d\bm{k}'*}g_{\bm{q}}^{a\bm{p},c\bm{p}'}}{\omega^2-\omega_{\bm{q}}^2 + i\omega_{\bm{q}}\Gamma_{\bm{q}}}\right\rvert^2 \nonumber \\ \label{before_omega_integration}
    \times \delta(\omega - \varepsilon_{c\bm{p}'} + \varepsilon_{a\bm{p}} )\delta(\omega - \varepsilon_{b\bm{k}} + \varepsilon_{d\bm{k}'})f_0(\varepsilon_{a\bm{p}})f_0(\varepsilon_{b\bm{k}})[1-f_0(\varepsilon_{c\bm{p}'})][1-f_0(\varepsilon_{d\bm{k}'})]|\Delta\mathcal{X}|^2,
\end{gather}
where $a,b,c,d$ are the band indices and the leading factor of $2$ takes into account the contributions of $t$ and $u$ channels. Using the $\omega$-integral result (\ref{appendix_omegaintegral_result}) in this regime, one obtains
\begin{gather}
    (\bm{\mathcal{X}}, \mathcal{C}_\textrm{e-e}\bm{\mathcal{X}})_\textrm{h.t.} = 2\sum_{abcd}\frac{1}{8}\cdot(2\pi)\cdot \beta^2\int_{\bm{q}}\frac{\pi4\omega_{\bm{q}}^2}{\Gamma_{\bm{q}}\omega_{\bm{q}}^2}\int_{\bm{k}} |g_{-\bm{q}}^{b\bm{k},d\bm{k}'}|^2f_0(\varepsilon_{b\bm{k})} [1-f_0(\varepsilon_{b\bm{k}})] \delta(\varepsilon_{d\bm{k}'}-\varepsilon_{b\bm{k}}) \nonumber\\
    \int_{\bm{p}}|g_{\bm{q}}^{a\bm{p},c\bm{p}'}|^2f_0(\varepsilon_{a\bm{p}})[1-f_0(\varepsilon_{a\bm{p}})]\delta(\varepsilon_{c\bm{p}'}-\varepsilon_{a\bm{p}})|\Delta\mathcal{X}|^2 .
\end{gather}

\subsubsection{Thermal Conductivity}
The perturbation term for $\omega = 0$ is of the form
\begin{gather}
    |\Delta\mathcal{X}|^2 = |\bm{\mathcal{X}}_a(\varepsilon_{a\bm{p}},\hat{\bm{p}})-\bm{\mathcal{X}}_c(\varepsilon_{a\bm{p}},\hat{\bm{p}}')|^2+|\bm{\mathcal{X}}_b(\varepsilon_{b\bm{k}},\hat{\bm{k}})-\bm{\mathcal{X}}_d(\varepsilon_{b\bm{k}},\hat{\bm{k}}')|^2 \nonumber \\
    +2 \left(\bm{\mathcal{X}}_a(\varepsilon_{a\bm{p}},\hat{\bm{p}})-\bm{\mathcal{X}}_c(\varepsilon_{a\bm{p}},\hat{\bm{p}}')\right)\cdot(\bm{\mathcal{X}}_b(\varepsilon_{b\bm{k}},\hat{\bm{k}})-\bm{\mathcal{X}}_d(\varepsilon_{b\bm{k}},\hat{\bm{k}}')).
\end{gather}
For the thermal conductivity $\bm{\mathcal{X}}_a(\varepsilon_{a\bm{p}},\hat{\bm{p}}) \propto (\varepsilon_{a\bm{p}}-\mu)$. This implies that the squared terms are proportional to $(\varepsilon_{a\bm{p}}-\mu)^2$ and $(\varepsilon_{b\bm{k}}-\mu)^2$, whereas the cross-term is proportional to $(\varepsilon_{a\bm{p}}-\mu)(\varepsilon_{b\bm{k}}-\mu)$. Since the latter is clearly odd in both the $(\varepsilon_{a\bm{p}}-\mu)$ and $(\varepsilon_{b\bm{k}}-\mu)$ whereas the rest of the integrand is even at the lowest order in those variables, making the integral over $\varepsilon_{a\bm{p}}-\mu$ and $\varepsilon_{b\bm{p}}-\mu$  (obtained from a change of variables with the $p$ and $k$ variables as usual) vanish. Therefore, effectively one has
\begin{gather}
    (\bm{\mathcal{X}}, \mathcal{C}_\textrm{e-e}\bm{\mathcal{X}})_\textrm{h.t.} = 4\sum_{abcd}\frac{1}{8}\cdot(2\pi)\cdot \beta^2\int_{\bm{q}}\frac{\pi4\omega_{\bm{q}}^2}{\Gamma_{\bm{q}}\omega_{\bm{q}}^2}\int_{\bm{k}} |g_{-\bm{q}}^{b\bm{k},d\bm{k}'}|^2f_0(\varepsilon_{b\bm{k})} [1-f_0(\varepsilon_{b\bm{k}})] \delta(\varepsilon_{d\bm{k}'}-\varepsilon_{b\bm{k}})|\bm{\mathcal{X}}_b(\varepsilon_{b\bm{k}},\hat{\bm{k}})-\bm{\mathcal{X}}_d(\varepsilon_{b\bm{k}},\hat{\bm{k}}')|^2 \nonumber\\
    \int_{\bm{p}}|g_{\bm{q}}^{a\bm{p},c\bm{p}'}|^2f_0(\varepsilon_{a\bm{p}})[1-f_0(\varepsilon_{a\bm{p}})]\delta(\varepsilon_{c\bm{p}'}-\varepsilon_{a\bm{p}}) \nonumber \\
    \label{appendix_generic_ee_ht_result}
    = \sum_{bd}(2\pi)\cdot \beta^2\int_{\bm{q}}\frac{1}{\beta\omega_{\bm{q}}}\int_{\bm{k}} |g_{-\bm{q}}^{b\bm{k},d\bm{k}'}|^2f_0(\varepsilon_{b\bm{k})} [1-f_0(\varepsilon_{b\bm{k}})] \delta(\varepsilon_{d\bm{k}'}-\varepsilon_{b\bm{k}})|\bm{\mathcal{X}}_b(\varepsilon_{b\bm{k}},\hat{\bm{k}})-\bm{\mathcal{X}}_d(\varepsilon_{b\bm{k}},\hat{\bm{k}}')|^2 = (\bm{\mathcal{X}}, \mathcal{C}_\textrm{e-ph}\bm{\mathcal{X}})_\textrm{h.t.},
\end{gather}
where in the latter two equalities we have employed
\begin{equation}
    \Gamma_{\bm{q}} = \sum_{ab} (2\pi)\beta\omega_{\bm{q}}\int_{\bm{p}}|g_{\bm{q}}^{a\bm{p},b\bm{p}'}|^2f_0(\varepsilon_{a\bm{p}})[1-f_0(\varepsilon_{a\bm{p}})]\delta(\varepsilon_{b\bm{p}'}-\varepsilon_{a\bm{p}}),
\end{equation}
\begin{equation}
    (\bm{\mathcal{X}}, \mathcal{C}_\textrm{e-ph}\bm{\mathcal{X}})_\textrm{h.t.} = \sum_{ab}(2\pi)\cdot \beta^2\int_{\bm{q},\bm{p}}\frac{1}{\beta\omega_{\bm{q}}} |g_{\bm{q}}^{a\bm{p},b\bm{p}'}|^2f_0(\varepsilon_{a\bm{p})} [1-f_0(\varepsilon_{a\bm{p}})] \delta(\varepsilon_{b\bm{p}'}-\varepsilon_{b\bm{p}})|\bm{\mathcal{X}}_a(\varepsilon_{a\bm{p}},\hat{\bm{p}})-\bm{\mathcal{X}}_b(\varepsilon_{a\bm{p}},\hat{\bm{p}}')|^2 .
\end{equation}
These expressions are just the generalizations of (\ref{appendix_phonon_linewidth}) and (\ref{appendix_eph_collision_term}). Note that at high temperatures $b_0(\omega_{\bm{q}}) \approx \frac{1}{\beta\omega_{\bm{q}}}$. \\

As mentioned before, this derivation fundamentally requires that $c\bm{q} \ll T$. If we return to our particular model with two Weyl nodes (\ref{weyl_semimetal_d}), this means that the temperature must be larger than the largest possible phonon energy, which in the case where $b \gg p_f$ is equivalent to $T > T_\textrm{BG,W} \approx 2cb$. In the intermediate regime where $2cp_f \ll T \ll 2cb$, emission and absorption of phonons between different nodes is exponentially suppressed. In the case of two-body scattering of the type $LR \rightarrow LR$,  the $t$-channel, where different chiralities do not mix, remains unaffected, as $q_t \sim p_f$, whereas  the $u$-channel, where $q_u \sim b$, recovers it's low-temperature form, i.e. where the propagator is effectively independent of the phonon frequency and the virtual phonon is "off-shell". One can show that the terms in $|M^{LR}_{LR}|^2$ depending on the $u$-channel are suppressed in comparison with the square of the $t$-channel amplitude $|A_t|^2$, so that effectively in this regime $|M^{LR}_{LR}|^2 \approx |A_t|^2$. 

\subsubsection{Electric Conductivity}
For the electric conductivity, we take the relevant mode (see main text) $\mathcal{X} \propto F^0(p) = \beta p$. For processes that do not violate chirality, one then has that $|\Delta\mathcal{X}|^2 =0$. For processes that do violate it once, e.g. $LR \rightarrow LL$, then $|\Delta\mathcal{X}|^2 =\beta^2 (2b)^2$. Processes that violate chirality twice, e.g. $LL \rightarrow RR$ will have $|\Delta\mathcal{X}|^2 =\beta^2 (4b)^2$. Noting that $\Gamma^{LR}(\bm{q}) = \Gamma^{RL}(\bm{q}) = \Gamma^\textrm{inter}_\textrm{e-ph}(\bm{q})$ and $\Gamma^{LL}(\bm{q}) = \Gamma^{RR}(\bm{q})= \Gamma^\textrm{intra}_\textrm{e-ph}(\bm{q})$, one can show that
\begin{gather}
    (\bm{\mathcal{X}}, \mathcal{C}_\textrm{e-e}\bm{\mathcal{X}})_\textrm{h.t., $\sigma$} = 2\sum_{abcd}\frac{1}{8}\cdot(2\pi)\cdot \beta^2\int_{\bm{q}}\frac{\pi4\omega_{\bm{q}}^2}{\Gamma_{\bm{q}}\omega_{\bm{q}}^2}\int_{\bm{k}} |g_{-\bm{q}}^{b\bm{k},d\bm{k}'}|^2f_0(\varepsilon_{b\bm{k})} [1-f_0(\varepsilon_{b\bm{k}})] \delta(\varepsilon_{d\bm{k}'}-\varepsilon_{b\bm{k}}) \nonumber\\
    \cdot \beta^2\left[(2b)^2\delta_{a+c,b+d \pm 1} + (4b)^2 \delta_{a+c, b+d \pm 2}\right] \int_{\bm{p}}|g_{\bm{q}}^{a\bm{p},c\bm{p}'}|^2f_0(\varepsilon_{a\bm{p}})[1-f_0(\varepsilon_{a\bm{p}})]\delta(\varepsilon_{c\bm{p}'}-\varepsilon_{a\bm{p}}) \nonumber \\
    \label{appendix_generic_ee_ht_result_electric_conductivity}
    = \frac{\beta^4}{2}\int_{\bm{q}}\frac{1}{(\beta\omega_{\bm{q}})^2}\frac{8(2b)^2\Gamma^\textrm{inter}_\textrm{e-ph}(\bm{q})\Gamma^\textrm{intra}_\textrm{e-ph}(\bm{q}) + 2(4b)^2\Gamma^\textrm{inter}_\textrm{e-ph}(\bm{q})\Gamma^\textrm{inter}_\textrm{e-ph}(\bm{q})}{2\Gamma^\textrm{intra}_\textrm{e-ph}(\bm{q})  + 2\Gamma^\textrm{inter}_\textrm{e-ph}(\bm{q})} = \frac{\beta^4(4b)^2}{2}\int_{\bm{q}}\frac{\Gamma^\textrm{inter}_\textrm{e-ph}(\bm{q})}{(\beta\omega_{\bm{q}})^2}
    = \beta^2(4b)^2\frac{2Up_f^4}{3(2\pi)^3c^2}.
\end{gather}
This result rests on the assumption that $cq \ll T$ for all relevant $q$ in these processes. This is the case for $T \gg 2T_{BG,W}$, as then all chirality-violating processes satisfy this requirement. For processes with $T \ll 2T_\textrm{BG,W}$, then the result (\ref{appendix_omegaintegral_result}) would no longer be valid for processes that violate chirality twice, e.g. $L,L\rightarrow R,R$. The integral for these processes would take the low-temperature form, which is $\propto (\beta p_f)^{-2}$ and is therefore suppressed with respect to the remaining high-temperature terms $\propto (\beta p_f)^{-1}$. The regime $T_\textrm{BG,W} \ll T \ll 2T_\textrm{BG,W}$ where the only processes that violate chirality once are allowed.

\subsection{Chirality-conserving Internode Scattering at Low Temperatures}
For temperatures below $T_\textrm{BG}$, intranode phonon absorption/emission also declines, although not exponentially, and all phonon propagators are effectively independent of $\omega$ so that 
\begin{equation}
    \label{appendix_LRLR_low_T_amplitude_squared}
|M^{LR}_{LR}|^2_{\textrm{l.t.}}(\bm{p},\bm{k},\bm{p}',\bm{k}') = 
    \frac{U^2}{4c^4}\left[1 -\cos\theta_{pk} +2 \cos\theta_{pb}\cos\theta_{kb}\right]\left[1 -\cos\theta_{p'k'}+2 \cos\theta_{p'b}\cos\theta_{k'b}\right],
\end{equation}
 is recovered for chirality conserving internode processes $LR \rightarrow LR$.
The relevant collision integral is
\begin{gather}\label{appendix_ee_collision_integral_LRLR}
    (\bm{\mathcal{X}}, \mathcal{C}_\textrm{e-e}\bm{\mathcal{X}}) =\frac{\beta^2}{4}\int_{\bm{p,k,p',k'}}|M^{LR}_{LR}|^2(\bm{p},\bm{k},\bm{p}',\bm{k}') (2\pi)^4\delta^{(4)}(p+k-p'-k')f(p)f(k)[1-f(p')][1-f(k')]|\Delta\mathcal{X}|^2,
\end{gather}
where the scattering amplitude is now given by (\ref{appendix_LRLR_low_T_amplitude_squared}). Note that while the momenta $\bm{p}$ and $\bm{p}'$ are in the reference frame of the $L$ node and $\bm{k}$ and $\bm{k}'$ are in the reference frame of the $R$ node, the internodal distance  $2\bm{b}$ vector cancels out of the momentum conservation delta, so that the actual distance $2b$ makes no appearence in the integral. Only the direction is relevant, as it in the square of the scattering amplitude $|M^{LR}_{LR}|^2$. As in the previous integrals, integration is performed with respect to the vectors $\bm{p}$, $\bm{k}$ and $\bm{q} \equiv \bm{p}' - \bm{p} = \bm{k} - \bm{k}'$. The relevant angular variables are $\cos\theta_{bq}$, $\cos\theta_{pq}$, $\cos\theta_{kq}$, $\varphi_{q,pb}$, and $\varphi_{q,kb}$. In terms of these variables, the anisotropic quantities appearing in can be expressed as
\begin{gather}
(\hat{\bm{p}}\cdot\hat{\bm{b}})(\hat{\bm{k}}\cdot\hat{\bm{b}}) = \left[-\frac{q}{2p_f}\cos\theta_{bq} + \sqrt{1-\frac{q^2}{4p_f^2}}\sin\theta_{bq}\cos\varphi_{q,pb}\right]\left[\frac{q}{2p_f}\cos\theta_{bq} + \sqrt{1-\frac{q^2}{4p_f^2}}\sin\theta_{bq}\cos\varphi_{q,kb}\right], \\
(\hat{\bm{p}}'\cdot\hat{\bm{b}})(\hat{\bm{k}}'\cdot\hat{\bm{b}}) = \left[q\cos\theta_{bq} + \hat{\bm{p}}\cdot\hat{\bm{b}}\right]\left[-q\cos\theta_{bq} + \bm{k}\cdot\hat{\bm{b}}\right] \nonumber \\
= \left[\frac{q}{2p_f}\cos\theta_{bq} + \sqrt{1-\frac{q^2}{4p_f^2}}\sin\theta_{bq}\cos\varphi_{q,pb}\right]\left[-\frac{q}{2p_f}\cos\theta_{bq} + \sqrt{1-\frac{q^2}{4p_f^2}}\sin\theta_{bq}\cos\varphi_{q,kb}\right],
\end{gather}
where we have replaced $\cos\theta_{pq}$ and $\cos\theta_{kq}$ by the values imposed by energy conservation (\ref{appendix_ee_intraband_energy_conservation}) at $p = k = p' = k' = p_f$. It is also convenient to calculate the following terms
\begin{gather}
(1-\cos\theta_{pk})(\hat{\bm{p}}'\cdot\hat{\bm{b}})(\hat{\bm{k}}'\cdot\hat{\bm{b}}) = -\frac{q^2}{4p_f^2}\left(1+\frac{q^2}{4p_f^2}\right)\cos^2\theta_{bq} - \left(1-\frac{q^2}{4p_f^2}\right)^2\sin^2\theta_{bq}\cos^2\varphi_{q,pb}\cos^2\varphi_{q,kb} ,\\
(1-\cos^2\theta_{pk})(\hat{\bm{p}}'\cdot\hat{\bm{b}})(\hat{\bm{k}}'\cdot\hat{\bm{b}}) = -\frac{q^2}{4p_f^2}\left[\left(1 + \frac{q^2}{4p_f^2}\right)^2 + \left(1-\frac{q^2}{4p_f^2}\right)^2\left(\cos^2\varphi_{q,pb}\cos^2\varphi_{q,kb} + \sin^2\varphi_{q,pb}\sin^2\varphi_{q,kb}\right)\right]\cos^2\theta_{bq} \nonumber \\
\label{appendix_chirality_conserving_funky_term}
-2\left(1+\frac{q^2}{4p_f^2}\right)\left(1-\frac{q^2}{4p_f^2}\right)^2\cos^2\varphi_{q,pb}\cos^2\varphi_{q,kb} \sin^2\theta_{bq},
\end{gather}
where we have ommited terms that vanish under the integration of the azimuthal angles $\varphi_{q,pb}$ and $\varphi_{q,kb}$. We have also made use of $\varphi_{q,pk} = \varphi_{q,pb}-\varphi_{q,kb}$, needed to expand $\cos\theta_{pk}$ (\ref{appendix_cospk_with_restrictions}). The second term (\ref{appendix_chirality_conserving_funky_term}) arises from the fact that the perturbation has the form
\begin{equation}
    |\Delta\mathcal{X}|^2 = |\Delta\mathcal{X}|^2\biggr|_{\substack{\omega = 0}} + \frac{\omega^2}{2}\frac{d^2|\Delta\mathcal{X}|^2}{d\omega^2}\biggr|_{\substack{p=k=p_f \\\omega = 0}}= X'(p_f)^2\left[q^2(\delta_p^2 + \delta_k^2 -2\delta_p\delta_k) + 2\omega^2p_f^2\left(1 - \cos\theta_{pk}\right)\right],
\end{equation}
which includes a term dependent on $(1-\cos\theta_{pk})$. Substituting in the value of $|M^{LR}_{LR}|^2$ given in (\ref{appendix_LRLR_low_T_amplitude_squared}) into the collision integral term, one can see that the the angular integrals needed are
\begin{gather}
    \mathcal{I}_0(q) = \int_{-1}^1d\cos\theta_{qb}\int_0^{2\pi}d\varphi_{q,pb}\int_0^{2\pi}d\varphi_{q,kb} \nonumber \\
    q^2\left[\left(1-\cos\theta_{pk}\right)^2 + 2\left(1-\cos\theta_{pk}\right)(\hat{\bm{p}}'\cdot\hat{\bm{b}})(\hat{\bm{k}}'\cdot\hat{\bm{b}})+2\left(1-\cos\theta_{pk}\right)(\hat{\bm{p}}\cdot\hat{\bm{b}})(\hat{\bm{k}}\cdot\hat{\bm{b}}) + 4(\hat{\bm{p}}\cdot\hat{\bm{b}})(\hat{\bm{k}}\cdot\hat{\bm{b}})(\hat{\bm{p}}'\cdot\hat{\bm{b}})(\hat{\bm{k}}'\cdot\hat{\bm{b}})\right], \\
    \mathcal{I}_1(q) = p_f^2\int_{-1}^1d\cos\theta_{qb}\int_0^{2\pi}d\varphi_{q,pb}\int_0^{2\pi}d\varphi_{q,kb} \nonumber \\
    \label{appendix_LRLL_scattering_amplitude_squared}
\left(1-\cos\theta_{pk}\right)\left[\left(1-\cos\theta_{pk}\right)^2 + 2\left(1-\cos\theta_{pk}\right)(\hat{\bm{p}}'\cdot\hat{\bm{b}})(\hat{\bm{k}}'\cdot\hat{\bm{b}})+2\left(1-\cos\theta_{pk}\right)(\hat{\bm{p}}\cdot\hat{\bm{b}})(\hat{\bm{k}}\cdot\hat{\bm{b}}) + 4(\hat{\bm{p}}\cdot\hat{\bm{b}})(\hat{\bm{k}}\cdot\hat{\bm{b}})(\hat{\bm{p}}'\cdot\hat{\bm{b}})(\hat{\bm{k}}'\cdot\hat{\bm{b}})\right].
\end{gather}
After performing the integrals over $p,k$ and $\omega$, one recovers an analogous expression to that of (\ref{appendix_ee_intra_lowT_intermediate}),
\begin{gather}
     (\bm{\mathcal{X}}, \mathcal{C}_\textrm{e-e}\bm{\mathcal{X}})_\textrm{l.t.} = \left[X'(p_f)^2 p_f^2\right]^2\frac{U^2}{4c^4}\frac{1}{(2\pi)^7}\frac{\beta^{-3}}{8}\int_0^{2p_f}  dq
    \left[4c_2^2\mathcal{I}_0 + 2b_4\mathcal{I}_1\right].
\end{gather}
The integrals in $q$ can be performed, resulting in
\begin{equation}
    \int_0^{2p_f}dq\ \mathcal{I}_0(q) =  \int_0^{2p_f}dq\ \mathcal{I}_1(q) = \frac{2^8\cdot11\cdot19\pi^2}{3^2\cdot5^2\cdot7}p_f^3. 
\end{equation}
Therefore,
\begin{equation}\label{appendix_chirality_conserving_low_T_result}
    (\bm{\mathcal{X}}, \mathcal{C}_\textrm{e-e}\bm{\mathcal{X}})_\textrm{l.t.} =\left[X'(p_f) p_f^2\right]^2\frac{2\cdot11\cdot19U^2p_f^3\beta^{-3}}{3^2\cdot5^2\cdot7(2\pi)^5c^4} \left[4c_2^2+ 2b_4\right].
\end{equation}

\subsection{Chirality-violating Internode Scattering}\label{appendix_chirality_violating_internode_scattering}
\subsubsection{Low Temperatures : Electric Conductivity}
Unlike the 2-body processes considered up until now, processes where chirality is violated do expectedly depend on the magnitude of the internodal distance vector $2\bm{b}$. For instance, $LR \rightarrow LL$ processes are governed by the collison integral term
\begin{gather}\label{appendix_ee_collision_integral_LRLL}
    (\bm{\mathcal{X}}, \mathcal{C}_\textrm{e-e}\bm{\mathcal{X}})^{LR,LL}_\textrm{l.t.} =4\frac{\beta^2}{8}\int_{\bm{p,k,p',k'}}|M^{LR}_{LL}|^2(\bm{p},\bm{k},\bm{p}',\bm{k}') (2\pi)^4\delta(p+k-p'-k')\delta^{(3)}(\bm{p}+\bm{k}-\bm{p}'-\bm{k}' + 2\bm{b}) \nonumber \\
    f(p)f(k)[1-f(p')][1-f(k')]|\Delta\mathcal{X}|^2.
\end{gather}
The leading factor of $4$ counts the multiplicity of the process. This integral resembles the one for Umklapp scattering with a crystal lattice vector $2\bm{b}$. Note however that this is purely formal, as in reality momentum is conserved, and the $2\bm{b}$ term is purely a consequence of the different reference frame for the incoming $R$ Weyl fermion. The scattering amplitude is given by
\begin{gather}
    |M^{LR}_{LL}|^2_\textrm{l.t.} = \frac{U^2}{4c^4}\left[1 - \hat{\bm{d}}(-\bm{b}+\bm{p}) \hat{\bm{d}}(\bm{b}+\bm{k}) \right]\left[1  - \hat{\bm{d}}(-\bm{b}+\bm{p}') \hat{\bm{d}}(-\bm{b}+\bm{k}') \right] \nonumber \\
    = \frac{U^2}{4c^4}\left(1- \cos\theta_{pk}+2(\hat{\bm{p}}\cdot\hat{\bm{b}})(\hat{\bm{k}}\cdot\hat{\bm{b}})\right)\left(1-\cos\theta_{p'k'}\right).
\end{gather}
Notice that since $\bm{\mathcal{X}}(\bm{p}) \equiv \mathcal{X}(p)\bm{v}_{\bm{p}}$, one can easily see that $\mathcal{X}(p) = F^0(p) \equiv \beta p$ is no longer a zero-mode of this integral due to the momentum conservation delta being Umklapp-like. As long as $\mathcal{X}(p)$ is non-vanishing at $p = p_f$, which is the case for the electrical conductivity, then the function $\mathcal{X}(p)$ chosen to represent the perturbation is not important, as they are all equivalent under the inner product (\ref{inner_product}). Since it is the relevant mode for the electric conductivity, we choose $\mathcal{X}(p) = F^0(p) =\beta p$. 

To solve the collision integral (\ref{appendix_ee_collision_integral_LRLL}), we follow a similar approach to the one given in \cite{PhysRevB.23.2718}:
\begin{itemize}
\item As was done in the previous collision integrals, the $\bm{k}'$ integration is removed through the momentum conservation delta introduce the dummy variable $\omega$ to separate the energy conservation delta, and perform the integral
\begin{gather}
    \int_{-\infty}^\infty d\omega \ (\beta\omega)^2b_0(\omega)[1+b_0(\omega)] f_0(p)[1-f_0(p)]f_0(k)[1-f_0(k)]\delta(p'-p-\omega)\delta(k-k'-\omega) \nonumber \\
    \label{restriction_to_surface}
    \approx 2c_2\beta^{-3}\delta(p-p_f)\delta(k-p_f)\delta(p'-p_f)\delta(k'-p_f).
\end{gather}
This result assumes $\omega$ is vanishing elsewhere in the integrand, which is a sound approximation since $\omega \sim T \ll p,k,p',k' \sim p_f$.
\item For the chosen $\mathcal{X}(p) = F^0(p) = \beta p$, the perturbation term is easily seen to be, using the restrictions $p=k=p'=k'$ from (\ref{restriction_to_surface}),
\begin{equation}\label{appendix_umklapp_perturbation}
    |\Delta \mathcal{X}|^2 = \beta^2\left(\bm{p}+\bm{k}-\bm{p}'-\bm{k}'\right)^2= \beta^2 4b^2,
\end{equation}
because of momentum conservation.
\item The energy conservation delta corresponding to $k'$ in (\ref{restriction_to_surface}) can be dealt with by integrating $\hat{\bm{p}}'$ with respect to its angle with the vector
\begin{equation}\label{appendix_K}
    \bm{k} \equiv \bm{p} + \bm{k} + 2\bm{b},
\end{equation}
so that
\begin{equation}\label{appendix_umklapp_kinematic_restriction}
    \int_{\hat{p}'}\delta(|\bm{p} + \bm{k} - \bm{p}' - \bm{g}| - p_f) = 2\pi\int_{-1}^1 d\cos\theta_{p'K}\ \frac{1}{K}\delta\left(\cos\theta_{p'K} - \frac{K}{2p_f}\right) = 2\pi\Theta(2p_f -K)\frac{1}{K}.
\end{equation}
where the step function enforces that the cosine is between -1 and 1.
\item Integrating over the remaining energy deltas in (\ref{restriction_to_surface}), one finds
\begin{equation}\label{appendix_ee_chirality_breaking_intermediate1}
    (\bm{\mathcal{X}}, \mathcal{C}_\textrm{e-e}\bm{\mathcal{X}})^{LR,LL}_\textrm{l.t.,$\sigma$} =p_f^6\frac{2c_2}{(2\pi)^7}\beta^24b^2\frac{\beta^{-1}}{2}\int_{\hat{\bm{p}},\hat{\bm{k}}}|M^{LR}_{LR}|^2(\bm{p},\bm{k},\bm{p}',\bm{k}')\Theta(2p_f -K)\frac{1}{K}.
\end{equation}
\item One can now perform a change of integration variables $cos\theta_{pb},\cos\theta_{kP}\rightarrow P,K$, where $\bm{k}$ is defined as above and
\begin{equation}\label{appendix_P}
    \bm{p} \equiv \bm{k} - \bm{k} = \bm{p} + 2\bm{b}.
\end{equation}
\end{itemize}
In doing so, the collision integral (\ref{appendix_ee_chirality_breaking_intermediate1}) becomes
\begin{equation}\label{appendix_ee_chirality_breaking_intermediate2}
    (\bm{\mathcal{X}}, \mathcal{C}_\textrm{e-e}\bm{\mathcal{X}})^{LR,LL}_\textrm{l.t.,$\sigma$} = \frac{p_f^4}{2b}\frac{2c_2}{(2\pi)^6}\beta^24b^2\frac{\beta^{-1}}{2}\int_{|2b-p_f|}^{P_\textrm{max}}dP\int_{|P-p_f|}^{K_\textrm{max}}dK \int_0^{2\pi} d\varphi_{P,kb}\ |M^{LR}_{LR}|^2(\bm{p},\bm{k},\bm{p}',\bm{k}'),
\end{equation}
where the upper integration limits, imposed by $\cos\theta_{pb} \le 1$ and $\cos\theta_{kP} \le 1$, as well as the kinematic restriction that $K \le 2p_f$ (which also implies $P = |\bm{k} - \bm{k}| \le 3p_f$), are 
\begin{gather}
P_\textrm{max = }
\begin{cases}
    b + p_f & \textrm{if $b < 2p_f$}, \\
    3p_f & \textrm{if $b \ge 2p_f$}.
\end{cases} \\
K_\textrm{max = }
\begin{cases}
    P + p_f & \textrm{if $ P < p_f$}, \\
     2p_f & \textrm{if $ P \ge p_f$}.
\end{cases}
\end{gather}
All that remains is to write the square of the scattering amplitude (\ref{appendix_LRLL_scattering_amplitude_squared}) in terms of these integration variables. This can be done using
\begin{gather}
  p_f^2\cos\theta_{pk} = \cos\theta_{Pk}[Pp_f - 2bp_f\cos\theta_{Pb}] - 2bp_f\cos\varphi_{P,kb}\sin\theta_{Pk}\sin\theta_{Pb}, \\
\cos\theta_{pb} = \frac{P^2-p_f^2-(2b)^2}{2p_f(2b)}, \quad \cos\theta_{kb} = \cos\theta_{Pk}\cos\theta_{Pb} + \sin\theta_{Pk}\sin\theta_{Pb}\cos\varphi_{P,kb}, \\
\cos\theta_{Pk} = \frac{K^2-P^2-p_f^2}{2Pp_f}, \quad \cos\theta_{Pb} = \frac{(2b)^2+P^2-p_f^2}{2(2b)P}, \quad \cos\theta_{p'k'} = \frac{K^2-2p_f^2}{2p_f^2}. 
\end{gather}
so that the relevant azimuthal angle integrals for solving (\ref{appendix_ee_chirality_breaking_intermediate2}) are
\begin{gather}
\label{appendix_ee_chirality_breaking_azimuthal_integral1}
\int_0^{2\pi} d\varphi_{P,kb} (1-\cos\theta_{pk})(1-\cos\theta_{p'k'} ) = (2\pi)\left(1 - \frac{K^2 -2p_f^2}{2p_f^2}\right)\left[1-\left(\frac{K^2-P^2-p_f^2}{4P^2p_f^2}\right)\left(P^2+p_f^2-(2b)^2\right)\right], \\
2\int_0^{2\pi} d\varphi_{P,kb} (\hat{\bm{p}}\cdot\hat{\bm{b}})(\hat{\bm{k}}\cdot\hat{\bm{b}})(1-\cos\theta_{p'k'} ) \nonumber \\
\label{appendix_ee_chirality_breaking_azimuthal_integral2}
= 2(2\pi)\left(1 - \frac{K^2 -2p_f^2}{2p_f^2}\right)\left(\frac{p_f^2+(2b)^2-P^2}{2p_f(2b)}\right)\left(\frac{K^2-P^2-p_f^2}{2Pp_f}\right)\left(\frac{p_f^2-(2b)^2-P^2}{2(2b)P}\right).
\end{gather}
Finally, carrying out the $K$ and $P$ integrals results in
\begin{equation}\label{appendix_LRLL_result_lt}
     (\bm{\mathcal{X}}, \mathcal{C}_\textrm{e-e}\bm{\mathcal{X}})^{LR,LL}_\textrm{l.t.,$\sigma$} = \frac{U^2}{4c^4}\frac{p_f^4}{2b}\frac{2c_2}{(2\pi)^6}\beta^24b^2\frac{\beta^{-1}}{2} \cdot
     \begin{cases}
     0 & \textrm{if $ 2b > 4p_f$}, \\
     \pi(4p_f-2b)^3\frac{(2b)^3 + 12(2b)^2p_f + 96(2b)p_f^2 + 320p_f^3}{720p_f^4} & \textrm{if $4p_f \ge 2b > 2p_f$}, \\
     \pi (2b)\frac{3(2b)^5 -320 (2b)^2p_f^3 -2880(2b)p_f^4 +9216p_f^5 }{720p_f^4} & \textrm{if $2p_f \ge 2b $} .
     \end{cases}
\end{equation}

Considering now processes that violate chirality twice, such as $LL \rightarrow RR$, one would have a comparatively simpler scattering amplitude,
\begin{gather}\label{appendix_scattering}
    |M^{LL}_{RR}|^2_\textrm{l.t.} = \frac{U^2}{4c^4}\left[1 - \hat{\bm{d}}(-\bm{b}+\bm{p}) \hat{\bm{d}}(-\bm{b}+\bm{k}) \right]\left[1  - \hat{\bm{d}}(\bm{b}+\bm{p}') \hat{\bm{d}}(\bm{b}+\bm{k}') \right] \nonumber \\
    = \frac{U^2}{4c^4}\left(1- \cos\theta_{pk}\right)\left(1-\cos\theta_{p'k'}\right).
\end{gather}
The resulting integrals however are formally the same as (\ref{appendix_ee_collision_integral_LRLL}) and (\ref{appendix_ee_chirality_breaking_intermediate2}) with the replacement $2b \rightarrow 4b$, but the relevant integral with the azimuthal angle is just (\ref{appendix_ee_chirality_breaking_azimuthal_integral1}), as the direction of $\bm{b}$ no longer appears in the square of the scattering amplitude. The result is
\begin{equation}\label{appendix_LLRR_result_lt}
     (\bm{\mathcal{X}}, \mathcal{C}_\textrm{e-e}\bm{\mathcal{X}})^{LL,RR}_\textrm{l.t. ,$\sigma$} = \frac{U^2}{4c^4}\frac{p_f^4}{4b}\frac{2c_2}{(2\pi)^6}\beta^216b^2\frac{\beta^{-1}}{8} \cdot
     \begin{cases}
     0 & \textrm{if $ 2b > 4p_f$} ,\\
     \pi(4p_f-4b)^4\frac{(4b)^2 + 16(4b)p_f + 40p_f^2}{360p_f^4} & \textrm{if $4p_f \ge 4b > 2p_f$}, \\
     \pi (4b)\frac{-3(4b)^5 +360 (4b)^3p_f^2 -640(4b)^2p_f^3 -2880(4b)p_f^4 +6144p_f^5 }{360p_f^4} & \textrm{if $2p_f \ge 4b $}, 
     \end{cases}
\end{equation}
where we have taken into account that the multiplicity is now $1$ instead of $4$ as in the $L,R \rightarrow L,L$ process.

\subsubsection{Low Temperatures: Thermal Conductivity}
In the previous calculation we assumed that $\mathcal{X} \propto F^0(p) = \beta p$, since that is the relevant mode for the electric conductivity, as is shown in the main text. For the thermal conductivity $\kappa$ however, one should consider $\mathcal{X} \propto \beta(p-p_f)$, so the result will be different. In fact, the perturbation now becomes similar to the low temperature perturbation term used for chirality-conserving processes (\ref{appendix_ee_lowT_perturbation}), 
\begin{gather}
    |\Delta \mathcal{X}|^2 = |\Delta \mathcal{X}|^2\biggr|_{\substack{\omega = 0}} + \frac{\omega^2}{2}\frac{d^2|\Delta \mathcal{X}|^2}{d\omega^2}\biggr|_{\substack{p=k=p_f \\\omega = 0}} + O(\omega^3) \nonumber \\
   = X'(p_f)^2\left[\delta_p^2(\bm{p}'-\bm{p})^2 + \delta_k^2(\bm{k}-\bm{k}')^2 + \omega^2\left(\bm{k} - 2\bm{p}'\right)^2\right],
\end{gather}
where we have omitted odd terms in $\delta_p$ and $\delta_k$, as they will vanish at lowest order in the Sommerfeld expansion. To deal with the terms that depend on the momenta, we note that at the lowest order in the Sommerfeld expansion, their moduli will satisfy $p=k=p'=k'=p_f$, so that
\begin{gather}
    (\bm{p}'-\bm{p})^2 = 2p_f^2(1-\cos\theta_{pp'}), \quad (\bm{k}-\bm{k}')^2 = 2p_f^2(1-\cos\theta_{kk'}), \\
    \left(\bm{k} - 2\bm{p}'\right)^2 = K^2 + 4p_f^2 - 4Kp_f\cos\theta_{Kp'} = 4p_f^2 - K^2,
\end{gather}
where in the latter we have used the restriction imposed by (\ref{appendix_umklapp_kinematic_restriction}). One can show that
\begin{gather}
    2-\cos\theta_{pp'}-\cos\theta_{kk'} = 1 + \frac{K^2}{2p_f^2} - \frac{P}{p_f}\cos\theta_{Pk} -2\frac{P}{p_f}\cos\theta_{Pp'} + \frac{2b}{p_f}\cos\theta_{bp'} \nonumber \\
    = 1 + \frac{K^2}{2p_f^2} - 2\frac{P}{p_f}\cos\theta_{Pk} - \frac{P^2}{p_f^2} + \frac{2bK}{2p_f^2}\cos\theta_{bK} + f(\varphi_{K,p'}),
\end{gather}
where $f(\varphi_{K,p'})$ is a function that is linear in cosines of azimuthal angles of $\bm{p}'$ with respect to the pole $\bm{k}$, and will hence vanish under their integration. Therefore one has 
\begin{gather}
    |\Delta \mathcal{X}|^2 = X'(p_f)^2\left[2\delta^2p_f^2\left(1 + \frac{K^2}{2p_f^2} - 2\frac{P}{p_f}\cos\theta_{Pk} - \frac{P^2}{p_f^2} + \frac{2bK}{2p_f^2}\cos\theta_{bK}\right) + \omega^2(4p_f^2-K^2)\right],
\end{gather}
where we have taken $\delta_p^2 = \delta_k^2 = \delta^2$, as once the integration of $p$ and $k$ is carried out, they will provide the same result. Hence, analogously to (\ref{restriction_to_surface}), we find that
\begin{gather}
    \int_0^\infty dp \int_0^\infty dk\int_{-\infty}^\infty d\omega \ (\beta\omega)^2b_0(\omega)[1+b_0(\omega)] f_0(p)[1-f_0(p)]f_0(k)[1-f_0(k)]\delta(p'-p-\omega)\delta(k-k'-\omega) |\Delta \mathcal{X}|^2 \nonumber \\
    \approx \beta^{-5}X'(p_f)^2\left[4c_2^2p_f^2\left(1 + \frac{K^2}{2p_f^2} - 2\frac{P}{p_f}\cos\theta_{Pk} - \frac{P^2}{p_f^2} + \frac{2bK}{2p_f^2}\cos\theta_{bK}\right) + b_4(4p_f^2-K^2)\right] \nonumber \\
    \label{restriction_to_surface_thermal_conductivity}
    = \beta^{-5}X'(p_f)^2\left(2p_f^2-\frac{K^2}{2}\right)\left[4c_2^2 + 2b_4\right] +  \beta^{-5}X'(p_f)^2\frac{2bK}{2}\cos\theta_{bK}4c_2^2.
\end{gather}
Of course, elsewhere in the collision integral we just set $p=k=p'=k'=p_f$. We need to consider the value
\begin{gather}
    (2b)K\cos\theta_{bK} = (2b)^2 + (2b)p_f(\cos\theta_{bp} +\cos\theta_{bk})\nonumber \\
    =  (2b)^2 + (2b)p_f\left(\cos\theta_{bp} + \cos\theta_{Pk}\cos\theta_{Pb} + \sin\theta_{Pk}\sin\theta_{Pb}\cos\varphi_{P,kb}\right).
\end{gather}
The term proportional to $\cos\varphi_{P,kb}$ implies that, in addition to  (\ref{appendix_ee_chirality_breaking_azimuthal_integral1}) and (\ref{appendix_ee_chirality_breaking_azimuthal_integral2}), we also need to consider the following azimuthal integrals
\begin{gather}
\sin\theta_{Pk}\sin\theta_{Pb}\int_0^{2\pi} d\varphi_{P,kb} \cos\varphi_{P,kb}(1-\cos\theta_{pk})(1-\cos\theta_{p'k'} ) = \nonumber \\
\label{appendix_ee_chirality_breaking_azimuthal_integral1_thermal}
\pi\left(1 - \frac{K^2 -2p_f^2}{2p_f^2}\right)\left[1-\left(\frac{K^2-P^2-p_f^2}{2Pp_f}\right)^2\right]\left[1-\left(\frac{(2b)^2 + P^2 -p_f^2}{2P(2b)}\right)^2\right], \\
2\sin\theta_{Pk}\sin\theta_{Pb}\int_0^{2\pi} d\varphi_{P,kb}\cos\varphi_{P,kb} (\hat{\bm{p}}\cdot\hat{\bm{b}})(\hat{\bm{k}}\cdot\hat{\bm{b}})(1-\cos\theta_{p'k'} ) \nonumber \\
\label{appendix_ee_chirality_breaking_azimuthal_integral2_thermal}
= 2\pi\left(1 - \frac{K^2 -2p_f^2}{2p_f^2}\right)\left(\frac{P^2-p_f^2-(2b)^2}{2p_f(2b)}\right)\left[1-\left(\frac{K^2-P^2-p_f^2}{2Pp_f}\right)^2\right]\left[1-\left(\frac{(2b)^2 + P^2 -p_f^2}{2P(2b)}\right)^2\right].
\end{gather}

 Combining (\ref{restriction_to_surface_thermal_conductivity}) with the azimuthal integrals (\ref{appendix_ee_chirality_breaking_azimuthal_integral1}), (\ref{appendix_ee_chirality_breaking_azimuthal_integral2}), (\ref{appendix_ee_chirality_breaking_azimuthal_integral1_thermal}) and (\ref{appendix_ee_chirality_breaking_azimuthal_integral2_thermal}) one can calculate the result of (\ref{appendix_ee_collision_integral_LRLL}) for the thermal conductivity:
\begin{itemize}
    \item if $2b > 4p_f$, then $(\bm{\mathcal{X}}, \mathcal{C}_\textrm{e-e}\bm{\mathcal{X}})^{LR,LL}_\textrm{l.t., $\kappa$} = 0$.
    \item if $4p_f \ge 2b > 2p_f$, then
    \begin{gather}
    \textstyle
        (\bm{\mathcal{X}},\mathcal{C}_\textrm{e-e}\bm{\mathcal{X}})^{LR,LL}_\textrm{l.t., $\kappa$} = \frac{U^2}{4c^4}\frac{p_f^4}{2b}\frac{1}{(2\pi)^6}X'(p_f)^2\frac{\beta^{-3}}{2} \nonumber \\
        \textstyle
         \pi(4p_f-2b)^4\frac{9(2b)^6 + 144(2b)^5p_f+1580(2b)^4p_f^2+5120(2b)^3p_f^3 + 9600(2b)^2p_f^4 + 1024(2b)p_f^5 + 1024p_f^6}{50400(2b)^2p_f^4}\left[4c_2^2 + 2b_4\right]  \nonumber \\
         \textstyle
     + \pi(4p_f-2b)\frac{37(2b)^9+153(2b)^8p_f+1552(2b)^7p_f^2 - 25242(2b)^5p_f^4 + 108262(2b)^4p_f^5-4172(2b)^3p_f^4 -68748(2b)^2p_f^7 + 21928(2b)p_f^8 - 196608p_f^9}{120\cdot840(2b)^2p_f^4}\left[4c_2^2\right]  \nonumber \\
     \label{appendix_LRLL_result_lt_thermal_2b}
     \textstyle
     - \pi \frac{105(2b)(2b-p_f)^3(2b+p_f)^2p_f^4\log\left[\frac{3p_f}{2b-p_f}\right]}{840(2b)^2p_f^4}\left[4c_2^2\right].
    \end{gather}
    \item if $2p_f \ge 2b > 0$, then
    \begin{gather}
    \textstyle
        (\bm{\mathcal{X}},\mathcal{C}_\textrm{e-e}\bm{\mathcal{X}})^{LR,LL}_\textrm{l.t., $\kappa$} = \frac{U^2}{4c^4}\frac{p_f^4}{2b}\frac{1}{(2\pi)^6}X'(p_f)^2\frac{\beta^{-3}}{2} \nonumber \\
        \textstyle
         \pi(2b)\frac{-27(2b)^7 -420(2b)^5p_f^2 + 8640(2b)^4p_f^3-33600(2b)^3p_f^4+28672(2b)^2p_f - 403200(2b)p_f^6 + 1064960p_f^7}{50400p_f^4}\left[4c_2^2 + 2b_4\right] 
         \nonumber \\
        \textstyle
         \pi(2b)\frac{111(2b)^9 + 126(2b)^8p_f + 2835(2b)^7p_f^2 - 26240(2b)^6p_f^3 + 32250(2b)^5p_f^4 + 49224(2b)^4p_f^5 + 81774(2b)^3p_f^6 + 48320(2b)^2p_f^7 +18900(2b)p_f^8+12600p_f^9}{100800(2b)(2b+p_f)p_f^4}\left[4c_2^2 \right]  \nonumber \\
         \label{appendix_LRLL_result_lt_thermal_2p}
         \textstyle
     + \pi\frac{25200p_f^4[(2b)^2-p_f^2]^3\log\left[\frac{p_f}{|p_f-2b|}\right] + 12600p_f^4[(2b)^2-p_f^2]^3\log\left[\frac{p_f}{p_f+2b}\right]}{100800(2b)(2b+p_f)p_f^4}\left[4c_2^2\right]. 
    \end{gather}
\end{itemize}
For $LL \rightarrow RR$, one can proceed analogously and obtain 
\begin{itemize}
    \item if $2b > 4p_f$, then $(\bm{\mathcal{X}}, \mathcal{C}_\textrm{e-e}\bm{\mathcal{X}})^{LL,RR}_\textrm{l.t., $\kappa$} = 0$.
    \item if $4p_f \ge 2b > 2p_f$, then
    \begin{gather}
    \textstyle
        (\bm{\mathcal{X}},\mathcal{C}_\textrm{e-e}\bm{\mathcal{X}})^{LL,RR}_\textrm{l.t., $\kappa$} = \frac{U^2}{4c^4}\frac{p_f^4}{4b}\frac{1}{(2\pi)^6}X'(p_f)^2\frac{\beta^{-3}}{8} \nonumber \\
        \textstyle
         \pi(4p_f-2b)^5\frac{3(2b)^3 + 60(2b)^2p_f + 272(2b)p_f + 448p_f^3}{10080p_f^4}\left[4c_2^2 + 2b_4\right]  \nonumber \\
         \textstyle
     + \pi(4p_f-2b)\frac{-15(2b)^8 - 59(2b)^7p_f + 1632(2b)^6p_f^2-4028(2b)^5p_f^3-7726(2b)^4p_f^4+30766(2b)^3p_f^5-220(2b)^2p_f^6-11292(2b)p_f^7+14216p_f^8}{20160(2b)p_f^4}\left[4c_2^2\right]  \nonumber \\
     \label{appendix_LLRR_result_lt_thermal_2b}
     \textstyle
     - \pi \frac{2520(2b-p_f)^3(2b+p_f)^2p_f^4\log\left[\frac{3p_f}{2b-p_f}\right]}{20160(2b)p_f^4}\left[4c_2^2\right]. 
    \end{gather}
    \item if $2p_f \ge 2b > 0$, then
    \begin{gather}
    \textstyle
        (\bm{\mathcal{X}},\mathcal{C}_\textrm{e-e}\bm{\mathcal{X}})^{LL,RR}_\textrm{l.t., $\kappa$} = \frac{U^2}{4c^4}\frac{p_f^4}{4b}\frac{1}{(2\pi)^6}X'(p_f)^2\frac{\beta^{-3}}{8} \nonumber \\
        \textstyle
         \pi(2b)\frac{9(2b)^7 -1344(2b)^5p_f^2 + 2688(2b)^4p_f^3 + 20160(2b)^3p_f^4-28672(2b)^2p_f^5 - 161280(2b)p_f + 294912p_f^7}{10080p_f^4}\left[4c_2^2 + 2b_4\right].
         \nonumber \\
        \textstyle
         \pi(2b)\frac{-45(2b)^9 -42(2b)^8p_f + 5607(2b)^7p_f^2 - 5328(2b)^6p_f^3 - 38990(2b)^5p_f^4 + 29736(2b)^4p_f^5 + 76566(2b)^3p_f^6 + 9664(2b)^2p_f^7 +3780(2b)p_f^8+2520p_f^9}{20160(2b)(2b+p_f)p_f^4}\left[4c_2^2\right]  \nonumber \\
         \label{appendix_LLRR_result_lt_thermal_2p}
         \textstyle
     + \pi\frac{5040p_f^4[(2b)^2-p_f^2]^3\log\left[\frac{p_f}{|p_f-2b|}\right] + 2520p_f^4[(2b)^2-p_f^2]^3\log\left[\frac{p_f}{p_f+2b}\right]}{20160(2b)(2b+p_f)p_f^4}\left[4c_2^2\right]. 
    \end{gather}
\end{itemize}

\section{Boltzmann Equation in 2-dimensional Basis}\label{appendix_boltzmann_section}
Here we solve the vectorized 2D Boltzmann equation (\ref{functional_boltzmann_equation})
\begin{equation}\label{appendix_functional_boltzmann_equation}
    \bm{S}_\textrm{E,T} = \bm{C}\ \bm{x}_\textrm{E,T},
\end{equation}
in the basis (\ref{basis}) to obtain the electric and thermal conductivities, $\sigma$ and $\kappa$. However, instead of solving a two-dimensional vector equation, we will make use of the results used in the main text (\ref{electric_conductivity_limit}) and (\ref{thermal_conductivity_matrix}), whereby the electric conductivity is accurately described by the momentum mode $F^0(p)$ due to the Sommerfeld expansion, wheras the thermal conductivity lives in the subspace in the subspace perpendicular to this mode:
\begin{equation}\label{appendix_conductivities}
    \sigma = \frac{2e^2\beta}{3}C^{-1,00}(S^0_\textrm{E})^2, \quad \kappa =\frac{2}{3}\left(\frac{P^0}{S^0_\textrm{E}}\right)^2 \bar{\bm{S}}_\textrm{E}^\dagger\bar{\bm{C}}^{-1}\bar{\bm{S}}_\textrm{E}.
\end{equation}
Again, due to the Sommerfeld expansion, only one function in the subspace perpendicular to $F^0(p)$ is neeeded to determine $\kappa$, simplifying calculations. Such function, in terms of our chosen basis (\ref{basis}), is obtained simply by projecting $F^1(p)$ into this subspace,
\begin{equation}\label{appendix_peperndicular_mode}
    \bar{F}^1(p) \equiv (F^0 \hat{\bm{p}}, F^0 \hat{\bm{p}})F^1 - (F^0 \hat{\bm{p}}, F^1 \hat{\bm{p}})F^0 = \frac{2}{(2\pi)^2}\beta^{-2}\left[r^3\beta(p_f-p) + 3c_2r\beta(2p_f - p)\right].
\end{equation}
All quantities appearing in (\ref{appendix_conductivities}) are multiplicative, so there is no danger of generating misleading cancellations if they are calculated to leading order in the Sommerfeld expansion. The quantities dependent on the source terms are
\begin{gather}\label{appendix_boltzmann_source_quantities}
    S^0_\textrm{E} = \frac{2}{(2\pi)^2}\beta^{-2}r^3, \quad P^0 = \frac{2}{(2\pi)^2}\beta^{-2}r^4, \quad \bar{S}^1_\textrm{E} = \left[\frac{2}{(2\pi)^2}\right]^2c_2\beta^{-4}r^4 .
\end{gather}
It is now useful to calculate the collision matrix elements individually. We define the relaxation times for 
\begin{gather}\label{appendix_rt_definitions}
    C^{00}_x = \frac{2}{(2\pi)^2}\beta^{-1}r^4\tau_{\sigma, x}^{-1}, \quad \bar{C}^{11} = \left[\frac{2}{(2\pi)^2}\right]^3c_2\beta^{-5}r^8\tau_{\kappa,x}^{-1}.
\end{gather}
Combining (\ref{appendix_conductivities}), (\ref{appendix_boltzmann_source_quantities}) and (\ref{appendix_rt_definitions}) one can see that the conductivities and the Lorenz ratio are given by
\begin{gather}\label{appendix_conductivities_2}
    \sigma = \frac{4e^2p_f^2}{3(2\pi)^2}\left[\sum_x \tau_{\sigma,x}^{-1}\right]^{-1}, \quad \kappa =\frac{4c_2\beta^{-1}p_f^2}{3(2\pi)^2}\left[\sum_x \tau_{\kappa,x}^{-1}\right]^{-1}, \quad L \equiv \frac{\beta\kappa}{\sigma} = L_0\left[\frac{\sum_x \tau_{\sigma,x}^{-1}}{\sum_x \tau_{\kappa,x}^{-1}}\right].
\end{gather}
where $L_0 \equiv \frac{c_2}{e^2}$.
For impurities (\ref{impurities}), it is simple to find $\tau_\textrm{imp} = \tau_{\sigma, \textrm{imp}} = \tau_{\kappa, \textrm{imp}}$, as expected. For the other collision processes, these quantities depend on the temperature regime.

\subsection{High Temperatures}\label{appendix_be_ht_results}
For temperatures above the Bloch-Gr\"uneisen temperature $T \gg T_\textrm{BG}$, it is simple to show using (\ref{appendix_e-ph_collision_integral_intranode_result_ht}) and (\ref{appendix_LR_collision_integral_result_2b})
\begin{equation}\label{appendix_ht_e-ph_rt}
    \tau_{\sigma,\textrm{e-ph}}^{-1}(T\gg T_\textrm{BG}) = \tau_{\kappa,\textrm{e-ph}}^{-1}(T\gg T_\textrm{BG}) = \beta^{-1} \frac{2(Up_f^2)}{3(2\pi) c^2}\cdot \begin{cases}
        \frac{5}{2} & \textrm{if $T \gg T_\textrm{BG,W}$}, \\
        1 & \textrm{if $T \ll T_\textrm{BG,W}$}.
    \end{cases} 
\end{equation}
The relaxation time for the electric conductivity is extracted simply enough from the sum of the collision integrals (\ref{appendix_e-ph_collision_integral_intranode_result_ht}), (\ref{appendix_LR_collision_integral_result_2b}) and (\ref{appendix_LR_collision_integral_result_2p}) for the $F^0(p)$ mode. The thermal relaxation time can be extracted from these same integrals and the collision integral matrix element for the mode orthogonal mode to $F^0(p)$,
\begin{gather}\label{appendix_e-ph_thermal_collision_element_ht}
    \bar{C}^{11}_\textrm{e-ph, h.t.} =  A^2(\bm{\mathcal{X}}_0, \mathcal{C}_\textrm{e-ph}\bm{\mathcal{X}}_0)_\textrm{h.t.} + B^2(\bm{\mathcal{X}}_1, \mathcal{C}_\textrm{e-ph}\bm{\mathcal{X}}_1)_\textrm{h.t.} - 2AB(\bm{\mathcal{X}}_0, \mathcal{C}_\textrm{e-ph}\bm{\mathcal{X}}_1)_\textrm{h.t.},
\end{gather}
where $A$ and $B$ are simply coefficients defined by through the Gram-Schmidt calculation of the orthogonal mode $\bar{F}^{1} = - AF^0(p) + BF^1(p)$, i.e. $A = (F^0 \hat{\bm{p}}, F^1 \hat{\bm{p}})$ and $B = (F^0 \hat{\bm{p}}, F^0 \hat{\bm{p}})$.
For electron-electron collisions, we showed that at high temperatures these are equivalent to a process of real phonon being emitted by a fermion and absorbed by another for the case of the thermal relaxation time, hence it is given by the expression in (\ref{appendix_ht_e-ph_rt}). In the case of relaxation time associated to the electric conductivity, only processes that violate chirality effectively have non-vanishing relaxation times, and the contribution is given by an expression proportional to (\ref{appendix_generic_ee_ht_result_electric_conductivity})
\begin{equation}
    \tau_{\sigma,\textrm{e-e}}^{-1}(T\gg T_\textrm{BG}) = 
    \begin{cases}
   \beta^{-1}\frac{U(4b)^2}{6(2\pi)c^2} & \textrm{if $T \gg 2T_\textrm{BG,W}$} ,\\
    0 & \textrm{if $T_\textrm{BG,W} \gg T $}.
    \end{cases}
\end{equation}
Note that we have inserted an extra factor of $1/2$ to take into account that we are only considering the relaxation time provided by a single node, whereas the calculation in (\ref{appendix_generic_ee_ht_result_electric_conductivity}) is done considering both nodes.

For temperatures below $T_\textrm{BG,W}$, virtual phonons that break chirality are no longer on-shell, i.e. with energies equal to $\omega_{\bm{q}}$, so the collision integral takes the low temperature form $\propto T^2$ that we discuss in the next section. However, this contribution is suppressed in comparison with the $\tau_{\sigma,\textrm{e-ph}}^{-1}(T\gg T_\textrm{BG})$, so we can effectively take it to be $0$.

\subsection{Low Temperatures}
At low temperatures $T \ll T_\textrm{BG}$, the electric and thermal relaxation times no longer coincide for phonon emission and absorption processes. The electric relaxation time is again extracted from the 00 matrix component given in (\ref{appendix_e-ph_collision_integral_intranode_result_lt}). The thermal relaxation time is extracted from the low-temperature version of (\ref{appendix_e-ph_thermal_collision_element_ht}). Note that now only intraband processes contribute to phonon absorption and emission for large internodal separations $2b > 2p_f$, as internode processes are suppressed because phonons do not have enough energy to connect nodes (\ref{appendix_LR_collision_integral_result_2b_lt}). For small separations $2p_f < 2b$, this is no longer the case and they do provide a nonvanishing contribution (\ref{appendix_LR_collision_integral_result_2p_lt}), but it is suppressed by an extra factor of $c$ with respect to the intranode contribution.  The relaxation times are thus
\begin{gather}
    \tau_{\sigma,\textrm{e-ph}}^{-1}(T\ll T_\textrm{BG}) = \beta^{-5}\frac{12g_5U}{(2\pi)c^6p_f^2},\\
    \tau_{\kappa,\textrm{e-ph}}^{-1}(T\ll T_\textrm{BG}) = \beta^{-3}\frac{12g_5U}{c_2(2\pi)c^4}.
\end{gather}
As for high temperatures, for electron-electron collisions only chirality-violating processes affect electric conduction, whereas both chirality-conserving and violating processes affect thermal conduction. This means that
\begin{gather}
    \tau_{\sigma,\textrm{e-e}}^{-1}(T\ll T_\textrm{BG}) = \tau_{\sigma,\textrm{e-e, $\chi$.v.}}^{-1}(T\ll T_\textrm{BG})  ,\\
    \tau_{\kappa,\textrm{e-e}}^{-1}(T\ll T_\textrm{BG}) = \frac{U^2p_f^3\beta^{-2}}{c_2(2\pi)^3c^4}\frac{1}{35}\left[8 + \frac{11\cdot 19}{9\cdot 5}\right]\left[4c_2^2 + 2b_4\right] + \tau_{\kappa,\textrm{e-e, $\chi$.v.}}^{-1}(T \ll T_\textrm{BG}) ,
\end{gather}
where $\tau_{\sigma,\textrm{e-e, $\chi$.v.}}^{-1}(T\ll T_\textrm{BG})$ and $\tau_{\kappa,\textrm{e-e, $\chi$.v.}}^{-1}(T \ll T_\textrm{BG})$ are the chirality-breaking processes $LR \rightarrow LL$ and $LL \rightarrow RR$ to the electric and thermal conductivities, respectively, extracted from the collision integrals (\ref{appendix_LRLL_result_lt}) and (\ref{appendix_LLRR_result_lt}) for the former, and (\ref{appendix_LRLL_result_lt_thermal_2b}),(\ref{appendix_LLRR_result_lt_thermal_2p}), (\ref{appendix_LLRR_result_lt_thermal_2b}), and (\ref{appendix_LLRR_result_lt_thermal_2p}). The non chirality-violating components in the thermal relaxation times are extracted from the intranode (\ref{ee_intraband_low_T_result}) and internode collision integrals (\ref{appendix_chirality_conserving_low_T_result}).